%% file: ceers_miri_lbg_seds.tex
\documentclass[twocolumn,tighten]{aastex631}

\newif\preprint 
\preprinttrue

\usepackage{amsmath,amstext}
\usepackage[T1]{fontenc}
\usepackage{apjfonts} 
\usepackage{natbib}
\usepackage{multirow}
\usepackage{verbatim}
\usepackage{hyperref}
\hypersetup{
    colorlinks=true,
    linkcolor=blue,
    filecolor=magenta,      
    urlcolor=cyan,
  }
  \usepackage{footmisc}
\graphicspath{{./}{figures/}{spec_plots/}{spec_plots_burst/}{montage_figures/}{1d_plots/}}

\include{defs}

\newcommand{\todo}[1]{\textcolor{aggiemaroon}{\tt #1}}
\newcommand{\addedtwo}[1]{\textbf{#1}}
\renewcommand{\deleted}[1]{}
\renewcommand{\added}[1]{{#1}}
\renewcommand{\replaced}[2]{#2}
\renewcommand{\deleted}[1]{}
\renewcommand{\replaced}[2]{\deleted{#1}\added{#2}}
\ifpreprint
\renewcommand{\added}[1]{{#1}}
\renewcommand{\addedtwo}[1]{{#1}}
\renewcommand{\replaced}[2]{#2}
\renewcommand{\deleted}[1]{}
\fi

\begin{document}


\title{\large \bf CEERS Key Paper IV: Galaxies at $\mathbf{4 < z < 9}$ are Bluer
  than They Appear --- \\ Characterizing Galaxy Stellar Populations from
  Rest-Frame $\mathrm{\sim}$1 micron Imaging}
%


\correspondingauthor{Casey Papovich}
\email{papovich@tamu.edu}

\author[0000-0001-7503-8482]{Casey Papovich}
\affiliation{Department of Physics and Astronomy, Texas A\&M University, College
Station, TX, 77843-4242 USA}
\affiliation{George P.\ and Cynthia Woods Mitchell Institute for
 Fundamental Physics and Astronomy, Texas A\&M University, College
 Station, TX, 77843-4242 USA}

\author[0000-0002-6348-1900]{Justin W. Cole}
\affiliation{Department of Physics and Astronomy, Texas A\&M
  University, College Station, TX, 77843-4242 USA}
\affiliation{George P.\ and Cynthia Woods Mitchell Institute for
  Fundamental Physics and Astronomy, Texas A\&M University, College
  Station, TX, 77843-4242 USA}

\author[0000-0001-8835-7722]{Guang Yang}
\affiliation{Kapteyn Astronomical Institute, University of Groningen, P.O. Box 800, 9700 AV Groningen, The Netherlands}
\affiliation{SRON Netherlands Institute for Space Research, Postbus 800, 9700 AV Groningen, The Netherlands}

\author[0000-0001-8519-1130]{Steven L. Finkelstein}
\affiliation{Department of Astronomy, The University of Texas at Austin, Austin, TX, USA}

\author[0000-0001-6813-875X]{Guillermo Barro}
\affiliation{Department of Physics, University of the Pacific, Stockton, CA 90340 USA}

\author[0000-0003-3441-903X]{V\'eronique Buat}
\affiliation{Aix Marseille Univ, CNRS, CNES, LAM Marseille, France}

\author[0000-0002-4193-2539]{Denis Burgarella}
\affiliation{Aix Marseille Univ, CNRS, CNES, LAM Marseille, France}

\author[0000-0003-4528-5639]{Pablo G. P\'erez-Gonz\'alez}
\affiliation{Centro de Astrobiolog\'{\i}a (CAB), CSIC-INTA, Ctra. de Ajalvir km 4, Torrej\'on de Ardoz, E-28850, Madrid, Spain}

\author[0000-0002-9334-8705]{Paola Santini}
\affiliation{INAF - Osservatorio Astronomico di Roma, via di Frascati 33, 00078 Monte Porzio Catone, Italy}

\author[0000-0001-7755-4755]{Lise-Marie Seill\'e}
\affiliation{Aix Marseille Univ, CNRS, CNES, LAM Marseille, France}

\author[0000-0001-9495-7759]{Lu Shen}
\affiliation{Department of Physics and Astronomy, Texas A\&M
  University, College Station, TX, 77843-4242 USA}
\affiliation{George P.\ and Cynthia Woods Mitchell Institute for
  Fundamental Physics and Astronomy, Texas A\&M University, College
  Station, TX, 77843-4242 USA}

\author[0000-0002-7959-8783]{Pablo Arrabal Haro}
\affiliation{NSF's National Optical-Infrared Astronomy Research Laboratory, 950 N. Cherry Ave., Tucson, AZ 85719, USA}

\author[0000-0002-9921-9218]{Micaela B. Bagley}
\affiliation{Department of Astronomy, The University of Texas at Austin, Austin, TX, USA}

\author[0000-0002-5564-9873]{Eric F.\ Bell}
\affiliation{Department of Astronomy, University of Michigan, 1085 S. University Ave, Ann Arbor, MI 48109-1107, USA}

\author[0000-0003-0492-4924]{Laura Bisigello}
\affiliation{Dipartimento di Fisica e Astronomia "G.Galilei", Universit\'a di Padova, Via Marzolo 8, I-35131 Padova, Italy}
\affiliation{INAF--Osservatorio Astronomico di Padova, Vicolo dell'Osservatorio 5, I-35122, Padova, Italy}

\author[0000-0003-2536-1614]{Antonello Calabr{\`o}} 
\affiliation{INAF - Osservatorio Astronomico di Roma, via di Frascati 33, 00078 Monte Porzio Catone, Italy}

\author[0000-0002-0930-6466]{Caitlin M. Casey}
\affiliation{Department of Astronomy, The University of Texas at Austin, Austin, TX, USA}

\author[0000-0001-9875-8263]{Marco Castellano}
\affiliation{INAF - Osservatorio Astronomico di Roma, via di Frascati 33, 00078 Monte Porzio Catone, Italy}

\author[0000-0003-4922-0613]{Katherine Chworowsky}\altaffiliation{NSF Graduate Fellow}
\affiliation{Department of Astronomy, The University of Texas at Austin, Austin, TX, USA}

\author[0000-0001-7151-009X]{Nikko J. Cleri}
\affiliation{Department of Physics and Astronomy, Texas A\&M University, College Station, TX, 77843-4242 USA}
\affiliation{George P.\ and Cynthia Woods Mitchell Institute for
  Fundamental Physics and Astronomy, Texas A\&M University, College
  Station, TX, 77843-4242 USA}

\author[0000-0001-6820-0015]{Luca Costantin}
\affiliation{Centro de Astrobiolog\'{\i}a (CAB), CSIC-INTA, Ctra. de Ajalvir km 4, Torrej\'on de Ardoz, E-28850, Madrid, Spain}

\author[0000-0003-1371-6019]{M. C. Cooper}
\affiliation{Department of Physics \& Astronomy, University of California, Irvine, 4129 Reines Hall, Irvine, CA 92697, USA}

\author[0000-0001-5414-5131]{Mark Dickinson}
\affiliation{NSF's National Optical-Infrared Astronomy Research Laboratory, 950 N. Cherry Ave., Tucson, AZ 85719, USA}

\author[0000-0001-7113-2738]{Henry C. Ferguson}
\affiliation{Space Telescope Science Institute, 3700 San Martin Dr., Baltimore, MD 21218, USA}

\author[0000-0003-3820-2823]{Adriano Fontana}
\affiliation{INAF - Osservatorio Astronomico di Roma, via di Frascati 33, 00078 Monte Porzio Catone, Italy}

\author[0000-0002-7831-8751]{Mauro Giavalisco}
\affiliation{University of Massachusetts Amherst, 710 North Pleasant Street, Amherst, MA 01003-9305, USA}

\author[0000-0002-5688-0663]{Andrea Grazian}
\affiliation{INAF--Osservatorio Astronomico di Padova, Vicolo dell'Osservatorio 5, I-35122, Padova, Italy}

\author[0000-0001-9440-8872]{Norman A. Grogin}
\affiliation{Space Telescope Science Institute, 3700 San Martin Dr., Baltimore, MD 21218, USA}

\author[0000-0001-6145-5090]{Nimish P. Hathi}
\affiliation{Space Telescope Science Institute, 3700 San Martin Dr., Baltimore, MD 21218, USA}

\author[0000-0002-4884-6756]{Benne W. Holwerda}
\affil{Physics \& Astronomy Department, University of Louisville, 40292 KY, Louisville, USA}

\author[0000-0001-6251-4988]{Taylor A. Hutchison}
\altaffiliation{NASA Postdoctoral Fellow}
\affiliation{Astrophysics Science Division, NASA Goddard Space Flight Center, 8800 Greenbelt Rd, Greenbelt, MD 20771, USA}

\author[0000-0001-9187-3605]{Jeyhan S. Kartaltepe}
\affiliation{Laboratory for Multiwavelength Astrophysics, School of Physics and Astronomy, Rochester Institute of Technology, 84 Lomb Memorial Drive, Rochester, NY 14623, USA}

\author[0000-0001-8152-3943]{Lisa J. Kewley}
\affiliation{Center for Astrophysics | Harvard \& Smithsonian, 60 Garden Street, Cambridge, MA 02138, USA}

\author[0000-0002-5537-8110]{Allison Kirkpatrick}
\affiliation{Department of Physics and Astronomy, University of Kansas, Lawrence, KS 66045, USA}

\author[0000-0002-8360-3880]{Dale D. Kocevski}
\affiliation{Department of Physics and Astronomy, Colby College,
  Waterville, ME 04901, USA}

\author[0000-0002-6610-2048]{Anton M. Koekemoer}
\affiliation{Space Telescope Science Institute, 3700 San Martin Dr., Baltimore, MD 21218, USA}

\author[0000-0003-2366-8858]{Rebecca L. Larson}
\altaffiliation{NSF Graduate Fellow}
\affiliation{Department of Astronomy, The University of Texas at Austin, Austin, TX, USA}

\author[0000-0002-7530-8857]{Arianna S. Long}
\altaffiliation{NASA Hubble Fellow}
\affiliation{Department of Astronomy, The University of Texas at Austin, Austin, TX, USA}

\author[0000-0003-1581-7825]{Ray A. Lucas}
\affiliation{Space Telescope Science Institute, 3700 San Martin Dr., Baltimore, MD 21218, USA}

\author[0000-0001-8940-6768]{Laura Pentericci}
\affiliation{INAF - Osservatorio Astronomico di Roma, via di Frascati 33, 00078 Monte Porzio Catone, Italy}

\author[0000-0003-3382-5941]{Nor Pirzkal}
\affiliation{ESA/AURA Space Telescope Science Institute}
\affiliation{Space Telescope Science Institute, 3700 San Martin Dr., Baltimore, MD 21218, USA}

\author[0000-0002-5269-6527]{Swara Ravindranath}
\affiliation{Space Telescope Science Institute, 3700 San Martin Dr., Baltimore, MD 21218, USA}

\author[0000-0002-6748-6821]{Rachel S. Somerville}
\affiliation{Center for Computational Astrophysics, Flatiron Institute, 162 5th Avenue, New York, NY, 10010, USA}

\author[0000-0002-1410-0470]{Jonathan R. Trump}
\affiliation{Department of Physics, 196 Auditorium Road, Unit 3046, University of Connecticut, Storrs, CT 06269, USA}

\author[0000-0001-8169-7249]{Stephanie M. Urbano Stawinski}
\affiliation{Department of Physics \& Astronomy, University of California, Irvine, 4129 Reines Hall, Irvine, CA 92697, USA}

\author[0000-0001-6065-7483]{Benjamin J. Weiner}
\affiliation{MMT/Steward Observatory, University of Arizona, 933 N. Cherry Ave., Tucson, AZ 85721, USA}

\author[0000-0003-3903-6935]{Stephen M.~Wilkins} %
\affiliation{Astronomy Centre, University of Sussex, Falmer, Brighton BN1 9QH, UK}
\affiliation{Institute of Space Sciences and Astronomy, University of Malta, Msida MSD 2080, Malta}

\author[0000-0003-3466-035X]{L. Y. Aaron\ Yung}
\altaffiliation{NASA Postdoctoral Fellow}
\affiliation{Astrophysics Science Division, NASA Goddard Space Flight Center, 8800 Greenbelt Rd, Greenbelt, MD 20771, USA}

\author[0000-0002-7051-1100]{Jorge A. Zavala}
\affiliation{National Astronomical Observatory of Japan, 2-21-1 Osawa, Mitaka, Tokyo 181-8588, Japan}

%
\begin{abstract}
 We present results from the \textit{Cosmic Evolution Early Release
Survey} (CEERS) on the stellar-population parameters for 28 galaxies
with redshifts $4<z<9$ using imaging data from the \textit{James Webb
Space Telescope} (\jwst) Mid-Infrared Instrument (MIRI) combined with
data from the \textit{Hubble Space Telescope} and the \textit{Spitzer
Space Telescope}. The \jwst/MIRI 5.6 and 7.7 \micron\ data extend the
coverage of the rest-frame spectral-energy distribution (SED) to
nearly 1 micron for galaxies in this redshift range. By modeling the
galaxies' SEDs the MIRI data show that the galaxies have, on average,
rest-frame UV (1600 \AA) -- $I$-band colors 0.4 mag bluer than derived
when using photometry that lacks MIRI. Therefore, the galaxies have
lower \deleted{(stellar)}\added{stellar}-mass--to--light ratios. The
MIRI data reduce the stellar masses by $\langle \Delta\log
M_\ast\rangle=0.25$ dex at $4<z<6$ \deleted{(a factor of 1.8)} and
0.37 dex at $6<z<9$ \deleted{(a factor of 2.3)}.This also reduces the
star-formation rates (SFRs) by $\langle \Delta\log\mathrm{SFR}
\rangle=0.14$ dex at $4<z<6$ and 0.27 dex at $6<z<9$. The MIRI data
also improve constraints on the \textit{allowable} stellar mass formed
in early star-formation. We model this using a star-formation history
that includes both a ``burst' at $z_f=100$ and a slowly varying
(``delayed-$\tau$'') model. The MIRI data reduce the allowable stellar
mass by 0.6 dex at $4<z< 6$ and by $\approx$1 dex at $6<z<9$. Applying
these results globally, this reduces the cosmic stellar-mass density
by an order of magnitude in the early universe
($z\approx9$). Therefore, observations of rest-frame
$\gtrsim$1~\micron\ are paramount for constraining the stellar--mass
build-up in galaxies at very high-redshifts.
\end{abstract}


\section{Introduction}\label{section:introduction}

There is growing evidence that galaxies must have started forming
stars very quickly following the Big Bang.  Theory predicts the first
stars should form at $z \gtrsim 20$
\citep[e.g.,][]{Barkana_2001,Miralda-Escude_2003,Yoshida_2003,Wise_2012,Visbal_2020}.
The ionization from these sources is needed to explain observations
that the hydrogen-neutral fraction of the intergalactic medium (IGM)
was 50\% by $z\sim 8$ (\citealt{Planck_2020}),\footnote{The
\citet{Planck_2020} analysis \deleted{also} suggests a non-zero optical depth of
CMB photons scattering off free electrons at $z \approx 15$, which
implies ionization of the IGM had begun by this epoch.}  \deleted{and to
account for the absorption profile of the 21~cm signal at $z\sim 20$
\citep[e.g.,][]{Bowman_2018}}.   Indeed, early observations from \jwst\
have already identified candidates for galaxies at $z \gtrsim 15$
\citep{Curtis-Lake_2022,Donnan_2022,Finkelstein_2022c,Robertson_2022}.
Spectroscopy from \jwst\ of galaxies at $z\sim 8-9$ shows emission
lines from heavy elements that appear to require metallicities of
$\approx 5-10\%~Z_\odot$
\citep{Arellano-Cordova_2022,Fujimoto_2022,Heintz_2022,Langeroodi_2022,Matthee_2022,Schaerer_2022,Trump_2022,Curti_2023,Katz_2023},
implying these galaxies have experienced at least one (and probably
multiple) generation(s) of previous stars. This is consistent with
earlier detections of metal lines in $z > 6$ galaxies from
ground-based telescopes \citep[e.g.,][]{Stark_2017,Hutchison_2019}.
All of these results point to the fact that star formation began
early and those early generations of stars enriched the universe with
heavy elements. 

It is then important to consider how we may constrain the history of
star-formation  in these early galaxies.   The number of stars\added{,
and therefore the \textit{stellar mass},} \deleted{(and therefore the
\textit{stellar mass})} in galaxies appears to rise rapidly.  The
(co-moving) stellar mass density in galaxies at $z\sim 5-6$ \deleted{(when the
age of the Universe is $\approx 1$~Gyr)} is already 1\% of the present
value  \citep[e.g.,][]{Madau_2014,Finkelstein_2016}\deleted{,}.  \added{
Obviously these stars must form at earlier times, and because the age
of the Universe is only $\approx 1$~Gyr at these redshifts, the time
to form these stars is relatively short. }

%
%
Even early \jwst\ observations find some evidence for massive galaxies
at $z > 7$ \citep{Labbe_2022} with some candidate objects having
masses, $\log M_\ast / M_\odot > 11$ \deleted{)}\added{,} as large as
the stellar mass of the Milky Way \textit{today}
\deleted{,}\added{(}e.g., \citealt{Papovich_2015}).  Such objects would
be in tension with galaxy formation models \citep{Boylan-Kolchin_2022}
where there are not sufficient numbers of massive dark-mater halos to
support these objects, even if \textit{all} the baryons in the halos
are in the form of stellar mass.  However, the uncertainty in these
measurements is in the assumed star-formation histories, the
contributions of emission lines to the photometric measurements from
broad-band data \citep[e.g.,][]{Endsley_2022,
PG_2022,Steinhardt_2022}, and the effects \added{of} young stellar
populations ``outshining'' older stellar populations in the integrated
emission of galaxies \citep[e.g.,][]{Gimenez-Arteaga_2022}.  Clearly
there are unknown systematics in the assumptions of the data analysis,
or missing physics in our theoretical understanding of stellar
populations and galaxy formation, or some combination of all of these
things.  It is therefore crucial to understand constraints on the
stellar masses (which are the integral of the star-formation
histories) as much as possible.  

Motivated by these issues, in this \textit{Paper} we use new data from
\jwst\ to better constrain the stellar masses, star-formation rates
(SFRs), and star-formation \textit{histories}  of galaxies during the
first one and a half billion years after the Big Bang ($z > 4$).  One
of the problems with initial  studies from \jwst\ is that they
currently rely entirely on observations from \jwst's Near-IR Camera
(NIRcam), which only probes to wavelengths $\lsim$5~\micron, or about
6000~\AA\ rest-frame for $z=6-7$ galaxies.  This complicates the
ability to disentangle massive galaxies with older stellar populations
from younger, dusty galaxies or galaxies with emission lines with
extreme equivalent widths (see discussion in
\citealt{Antwi-Danso_2022}, and recent work by
\citealt{Gimenez-Arteaga_2022} who find evidence for older stellar
populations mixed with recent bursts in spatially resolved studies
using \jwst/NIRCam data).   To better constrain the SEDs of these
galaxies requires observations at longer wavelengths.  This is where
\jwst/MIRI is important as it has the sensitivity to detect $z \approx
10$ galaxies at rest-frame 1~\micron\ \citep[see][]{Bisigello_2017}.
Previous work on this subject has been limited to data from the
\spitzer\ \textit{Space Telescope}, which is primarily sensitive to
the emission \replaced{such}{of} distant galaxies at 3.6 to
8.0~\micron. \jwst\ offers immense gains to \spitzer:  \jwst\ has a
collecting area that is 45 times larger than that of \spitzer, and the
larger aperture provides image quality (i.e., angular resolution) that
is improved by a factor of order 10 \citep{Rigby_2022}.  These gains
are especially manifest at longer wavelengths, and  make  \jwst\ 5.6
and 7.7~\micron\ data vastly more sensitive than \spitzer. 

The outline for the \textit{Paper} is as follows.  In
Section~\ref{section:data} we discuss the CEERS dataset and the
ancillary datasets used in this study.   We also discuss the processes
to create (and validate) the flux densities of galaxies in the CEERS
\jwst/MIRI data at 5.6 and 7.7~\micron.     In
Section~\ref{section:sample} we discuss the sample of $4 < z < 9$
galaxies used in this study, and we present the MIRI data for these
objects.   In Section~\ref{section:analysis} we discuss the analysis
methods to derive \addedtwo{constraints} on the galaxy stellar populations.  In
Section~\ref{section:results} we discuss the resulting improvements
that including the MIRI data provide on constraints on the galaxies
stellar populations (specifically their stellar--masses and SFRs)
derived from fitting stellar population models to the observed
photometry.   In Section~\ref{section:bursts} we discuss constraints
on the range of allowed stellar masses in these high redshift galaxies
by allowing for an early (``maximally old'') burst of stars at
$z=100$, and we show that adding the MIRI data improves the limit on
this hypothetical population of $z=100$ stars by a factor of 6 to 10
for galaxies at $4 < z < 9$.   In Section~\ref{section:discussion} we
discuss the implications these constraints have for our understanding
of galaxy colors, stellar populations at these high redshifts, and the
evolution of the galaxy stellar mass density, in particular during the
epoch of reionization, and what this could mean for future studies of
galaxies at higher redshifts (from \jwst).  In
Section~\ref{section:summary} we present our conclusions and prospects
for future studies.

Throughout we use a flat cosmology with $\Omega_{m,0}=0.315$, $H_0 =
67.4$~km s$^{-1}$ Mpc$^{-1}$ \citep{Planck_2020}.    All magnitudes
reported here are on the \textit{Absolute Bolometric} (AB) system
\citep{Oke_1983}.   Throughout we use \added{the} \citet{Chabrier_2003} initial
mass function (IMF) for all stellar masses and SFRs. We denote
magnitudes measured in the MIRI F560W and F770W bands as \mirifive\
and \miriseven, respectively. Similarly, we denote magnitudes measured
in IRAC Channel 1 (3.6~\micron),  Channel 2 (4.5~\micron), Channel 3
(5.8~\micron), and Channel 4 (8.0~\micron) as \mone, \mtwo, \mthree,
and \mfour, respectively. 

\section{Data and Sample}\label{section:data}

\subsection{MIRI Catalog}\label{section:catalog}

We use the data release DR0.5 images produced by the CEERS team for
the MIRI 3 and MIRI 6 fields (see
\citealt{Finkelstein_2022c})\footnote{\url{https://ceers.github.io/releases.html}\label{footnote:ceers_hdr}}.
These data were acquired in 2022 June 21 and 22.  The \deleted{data}
properties \added{of the data} and \added{the data}\deleted{its}
reduction are discussed elsewhere \citep{Yang_2022a}, but we provide a
summary here.  The data were processed using the \jwst\ Calibration
Pipeline (v1.7.2) using the default parameters for stage 1 and 2.  We
then removed the backgrounds with a custom routine that combines
images taken in the same bandpass but from different fields and/or
dither positions\added{, }\deleted{(}rejecting pixels in each image
that contain galaxies\deleted{)}\added{,} in order to create a
``super-background'' image. We then removed this background from each
image and applied an astrometric correction to each image prior to
processing them with stage 3 of the pipeline.  This produced the final
science images (extension \texttt{i2d}), rms images (extension
\texttt{rms}, which account for Poisson, readout, and correlated pixel
noise; see \citealt{Yang_2022a}), and  weight maps (\texttt{wht}) for
each field with a pixel scale of 0.09\arcsec, registered
astrometrically to the existing \hst/CANDELS v1.9 WFC3 and ACS images
(see
\citealt{Koekemoer_2011,Bagley_2022}\footref{footnote:ceers_hdr}).
\added{Our tests find that the MIRI images achieve limiting $5\sigma$
depths in the MIRI images of 26.5 and 27.1 AB mag measured in
0.45\arcsec-diameter apertures}.

For the purpose of this study we are interested in sources detected in
the MIRI data, so we create a catalog of sources derived from these
images.  Prior to object detection we convolved the 5.6~\micron\ image
to match the image quality of the 7.7~\micron\ image. For this step,
we constructed an ``effective'' point source function (ePSF) for each
image by identifying unblended stars using the \texttt{photutils}
(v1.5.0) \texttt{detection} task, and modeling them with the
\texttt{photutils} \texttt{psf} task.  This produced model ePSFs with
measured full-width at half maxima (FWHM) of $0.24\arcsec$
and $0.28\arcsec$, for the F560W and F770W images, respectively.  This
is consistent with the expected image quality, but takes into account
the exact dithering and reduction steps for the CEERS MIRI data.   We
then used \texttt{PyPHER} \citep{Boucaud_2016a} to  construct a
convolution kernel to match the image quality of the model ePSFs.  We
applied these kernels to each F560W image, creating a ``PSF-matched''
image.   Our tests on point sources in the PSF-matched F560W and F770W
images show that we measure the same fraction of light to better than
2\% in fixed circular apertures of radii larger than $0{\farcs}35$. 

We then created a detection image constructed from the sum of the MIRI
F560W and F770W science images (using the extension \texttt{sci})
weighted by the appropriate weight image (using the extension
\texttt{wht}).   We also created a detection weight-map as the sum of
the weights for these images.   We then created F560W and F770W
catalogs using  Source Extractor (\sextractor, version 2.19.5;
\citealt{Bertin_1996}) in ``dual-image'' mode using the detection
image \deleted{)}and its weight map\deleted{)} for object detection, where we measured
photometry in the PSF-matched F560W and F770W image.    We used the
parameters in Table~\ref{table:sextractor}.  We then measured fluxes
and magnitudes  using circular apertures of 0.9\arcsec\ diameter, and
we scaled these to a total aperture (\texttt{MAG\_AUTO}) measured for
each source in the detection image.   Uncertainties for each object
are measured from the rms image in the same apertures, and scaled to a
total magnitude in the same way.  \deleted{Figure~\ref{fig:miri_color_z} shows
the distribution of the MIRI sources with signal-to-noise (SNR) $\geq$
3 in F560W or F770W (compared with our galaxy sample, discussed below
in \ref{section:sample})}.

\begin{deluxetable}{lc}
\tablecolumns{2}
\tablewidth{3in}
\tablecaption{CEERS MIRI F560W and F770W SExtractor Parameter Settings\label{table:sextractor}}
\tablehead{\colhead{SExtractor Parameter} & \colhead{Value} \\
\colhead{(1)} & \colhead{(2)} 
}
\startdata
\texttt{DETECT\_MINAREA} & 10~pixels \\
\texttt{DETECT\_THRESH} & 1.3 \\
\texttt{ANALYSIS\_THRESH} & 1.3 \\
\texttt{FILTER\_NAME} &  gauss\_2.5\_5x5\tablenotemark{a} \\
\texttt{WEIGHT\_TYPE} & MAP\_WEIGHT,MAP\_RMS \\
\texttt{DEBLEND\_NTHRESH} & 32 \\
\texttt{DEBLEND\_MINCONT} & 0.005 \\
\texttt{MAG\_ZEROPOINT} & 25.701\tablenotemark{b} \\
\texttt{PIXEL\_SCALE} & 0.09~arcsec\\ 
\texttt{BACK\_TYPE} & AUTO \\
\texttt{BACK\_FILTERSIZE} & 5~pixels \\ 
\texttt{BACK\_SIZE} & 32~pixels \\
\texttt{BACKPHOTO\_THICK} & 8 \\ 
\texttt{BACKPHOTO\_TYPE} & LOCAL \\ 
\texttt{SEEING\_FWHM} & 0.3~arcsec
\enddata
\tablecomments{\sextractor\ was run using the weighted sum of the
  PSF-matched F560W and F770W images for detection, and using the images separately
  for photometry.  All other \sextractor\ parameters are
  set to the program defaults (for SExtractor v.2.19.5).}
\tablenotetext{a}{This is a Gaussian kernel with $\sigma$=2.5 pixels and
  size $5 \times 5$ pixel$^2$ used to filter the image for
  source detection.} 
\tablenotetext{b}{The AB magnitude zeropoint for the images,
 converting from the \jwst\ default of MJy sr$^{-1}$ to $\mu$Jy
 pixel$^{-1}$ at the $0.09\arcsec$ pixel$^{-1}$ scale.} 
\end{deluxetable}


\begin{figure*}[t]
  \centering
  \includegraphics[width=0.8\textwidth]{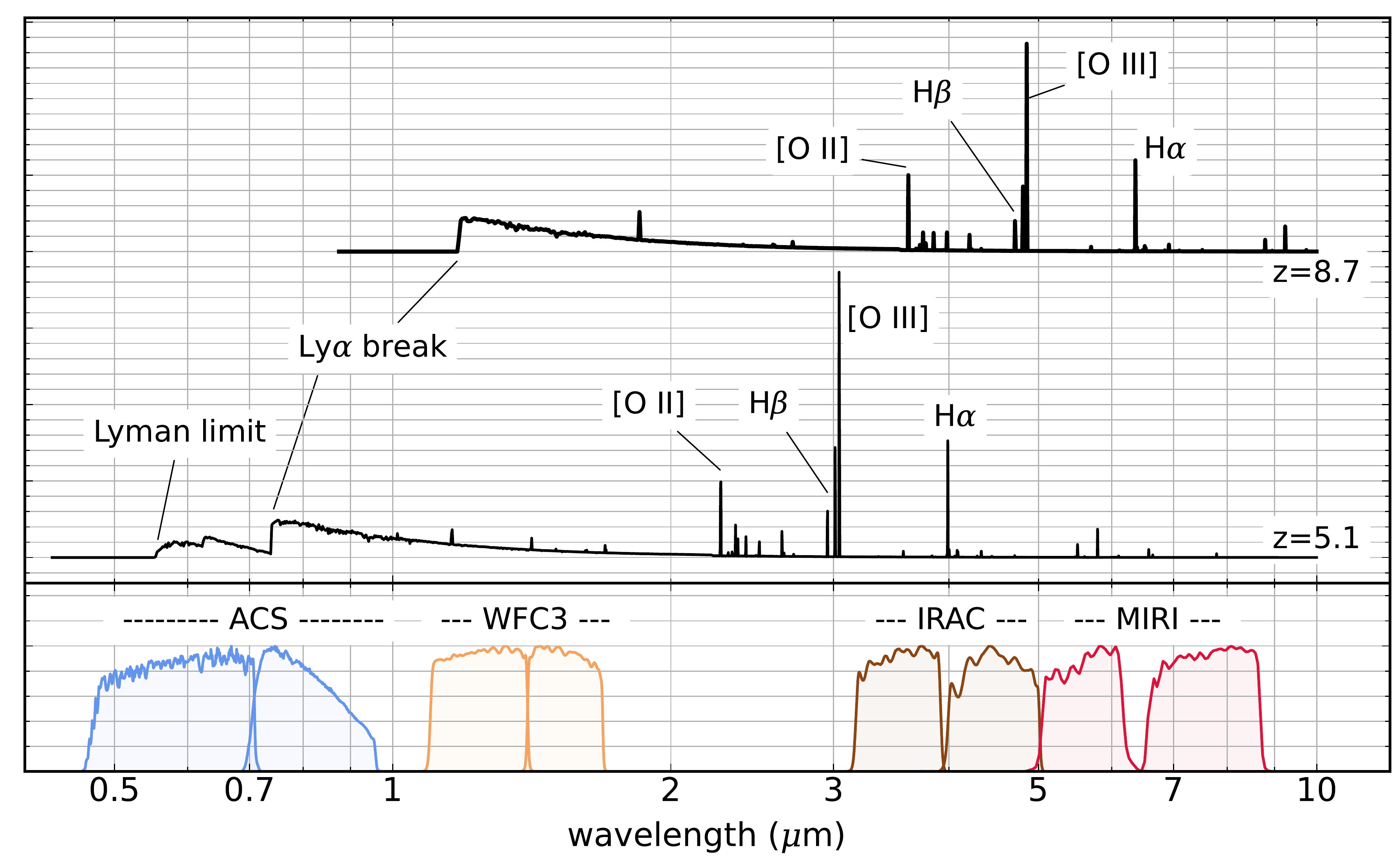}
\caption{ Illustration of galaxy spectra (in relative units of erg~s$^{-1}$
  cm$^{-2}$ \AA$^{-1}$) compared to the broad-band
data for the observations in this work.  The bottom panel shows the
relative transmission functions for the \hst/ACS and WFC3 filters (ACS
F606W, F814 and WFC3 F125W, F160W), \spitzer/IRAC 3.6~\micron\ (Ch1)
and 4.5~\micron\ (Ch2), and \jwst/MIRI F560W and F770W.    The top
panel shows model  spectra of star-forming galaxies at $z=5.1$ and
$z=8.7$\added{,} \deleted{(}which \deleted{coincidentally} match two galaxies with spectroscopic
redshifts in this sample\deleted{)}.   Key emission lines and features are
labeled.   The MIRI data probe the shape of the galaxy spectral energy
distributions to 8000~\AA\ rest-frame, even for galaxies with $z =
9$.\label{fig:filters}}
\end{figure*}

We compared the MIRI flux densities for sources \addedtwo{with} F560W and F770W
\addedtwo{detections} against those for bright objects from existing IRAC 5.8 and
8.0~\micron\ catalogs \added{\citep{Stefanon_2017}}.   For bright objects
($[5.8]$ or $[8.0] \leq 22$~AB mag) in the IRAC data, we measure small
offsets of $\Delta m = \mthree - \mirifive = 0.16$~mag between the
IRAC 5.8 and MIRI 5.6 data,  and $\Delta m = \mfour - \miriseven =
0.07$~mag between the IRAC 8.0 and MIRI 7.7 data (i.e., the MIRI flux
densities are slightly brighter).  Most of these offsets can be
explained by differences in the shape of the MIRI and IRAC passbands
and because of differences in the angular resolution of the
instruments (MIRI has a PSF FWHM smaller by a factor of more than
seven).  These tests are discussed more fully in \citet{Yang_2022a},
but this gives us confidence that the MIRI data are calibrated to
better than $\simeq 0.1-0.2$~mag.


\begin{figure*}[p] 
  \gridline{
    \includegraphics[width=0.85\textwidth]{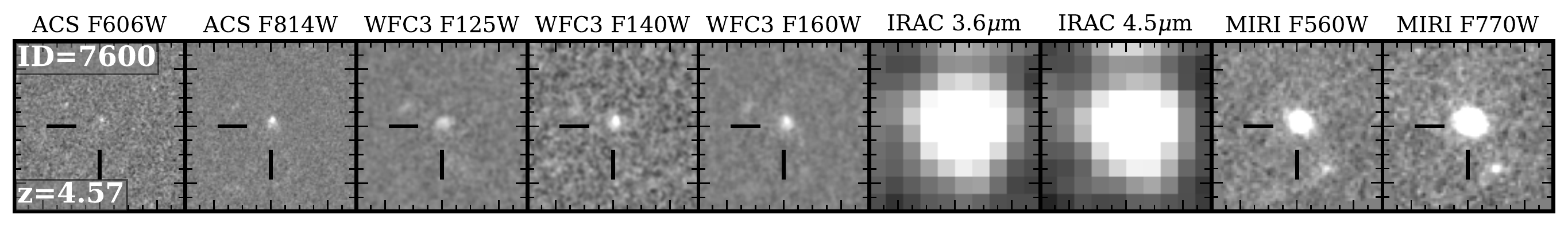}
}
\vspace{-14pt}
  \gridline{
    \includegraphics[width=0.85\textwidth,trim={0pt 0 0 20},clip=true]{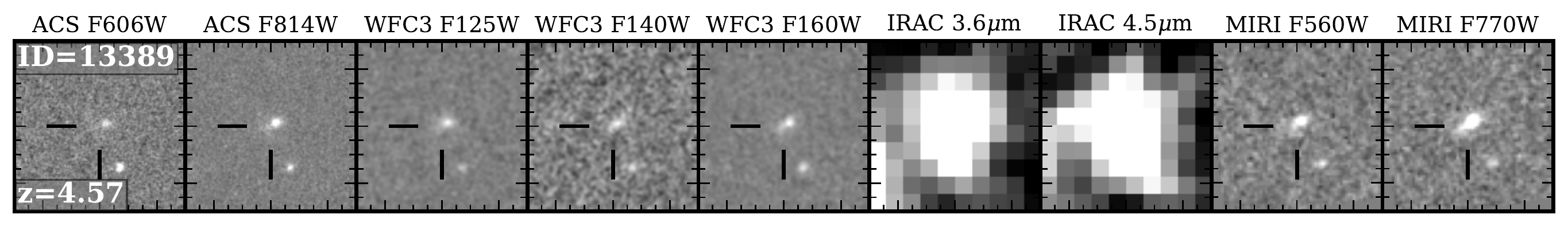}
}
\vspace{-14pt}
 \gridline{
    \includegraphics[width=0.85\textwidth,trim={0pt 0 0 20},clip=true]{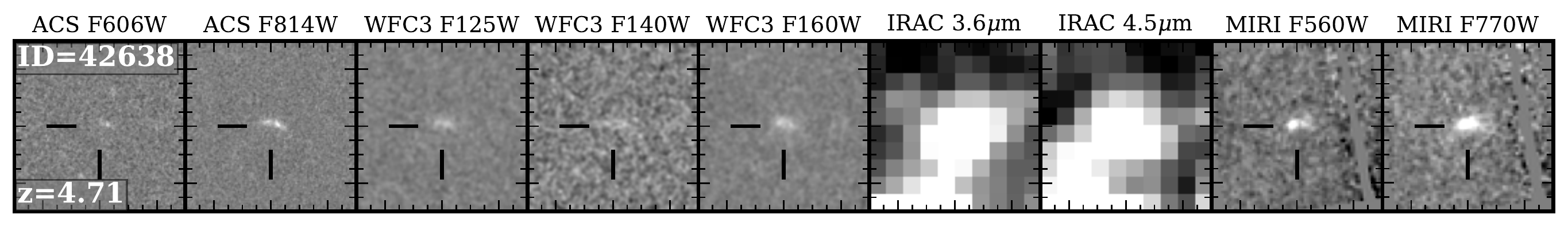}
  }
\vspace{-14pt}
  \gridline{
    \includegraphics[width=0.85\textwidth,trim={0pt 0 0 20},clip=true]{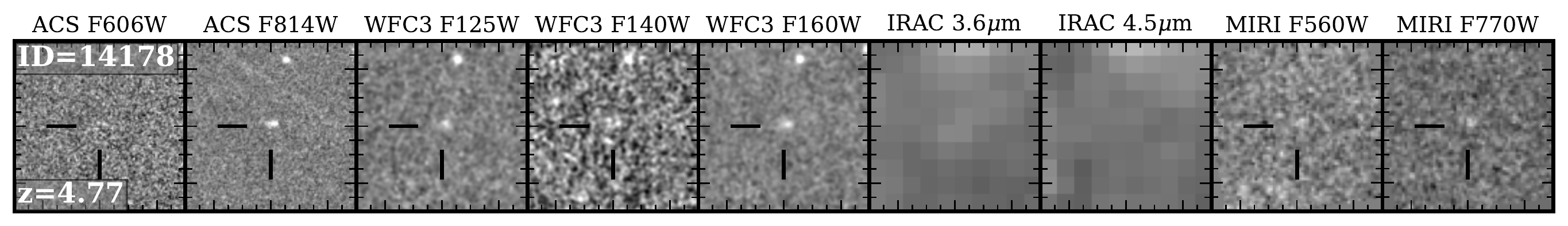}
  }
\vspace{-14pt}
 \gridline{
    \includegraphics[width=0.85\textwidth,trim={0pt 0 0 20},clip=true]{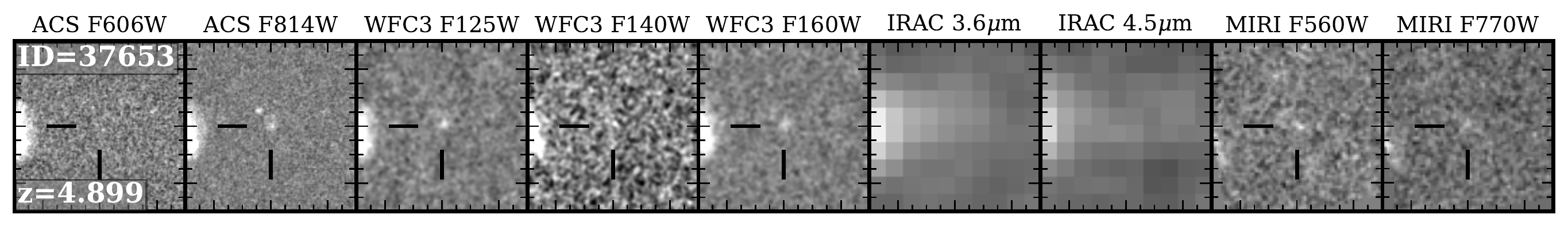}
  }
\vspace{-14pt}
  \gridline{
    \includegraphics[width=0.85\textwidth,trim={0pt 0 0 20},clip=true]{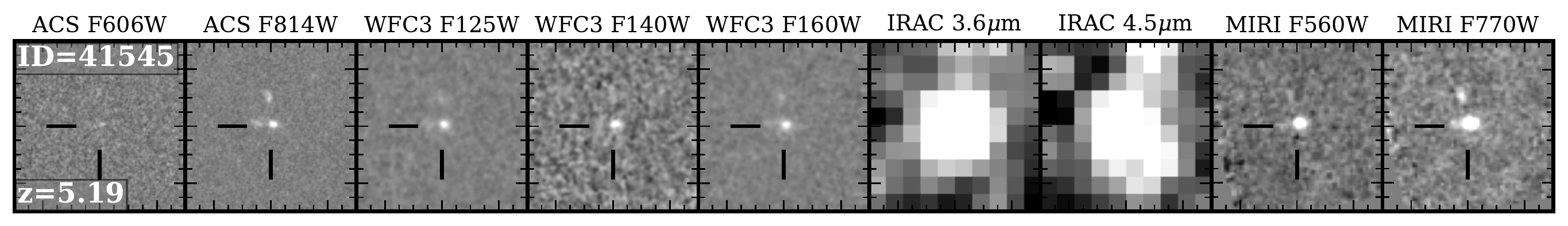}
  }
\vspace{-14pt}
  \gridline{
    \includegraphics[width=0.85\textwidth,trim={0pt 0 0 20},clip=true]{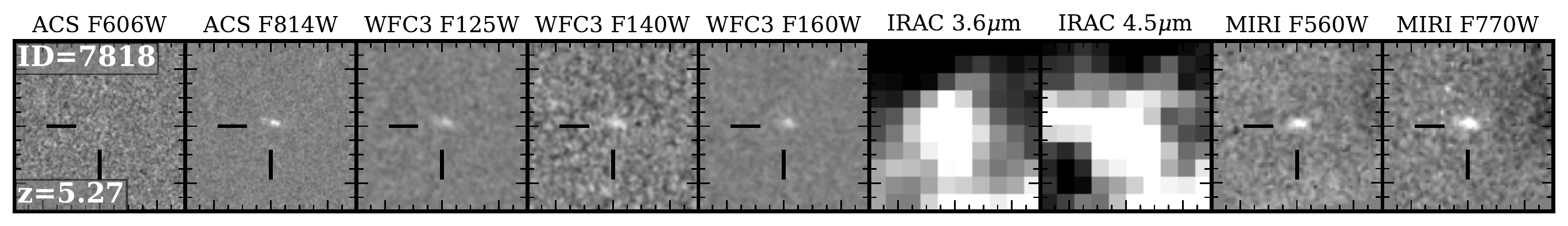}
  }
  \vspace{-14pt}
  \gridline{
    \includegraphics[width=0.85\textwidth,trim={0pt 0 0 20},clip=true]{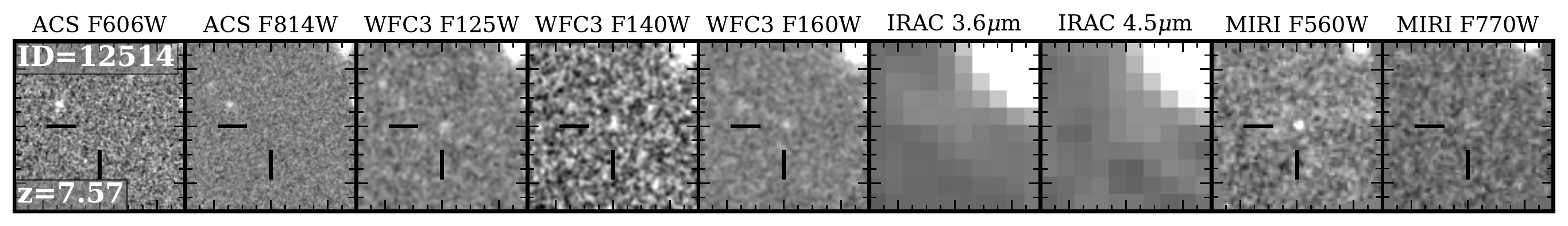}
  }
\vspace{-14pt}
  \gridline{
    \includegraphics[width=0.85\textwidth,trim={0pt 0 0 20},clip=true]{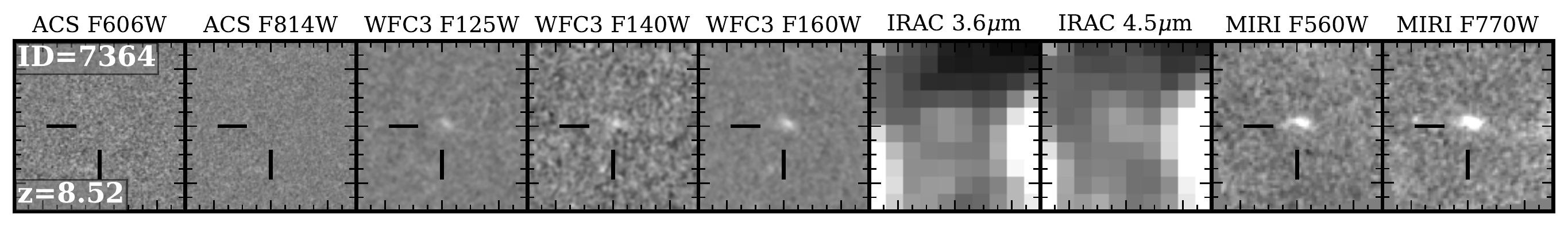}
  }
\vspace{-14pt}
  \gridline{
    \includegraphics[width=0.85\textwidth,trim={0pt 0 0 20},clip=true]{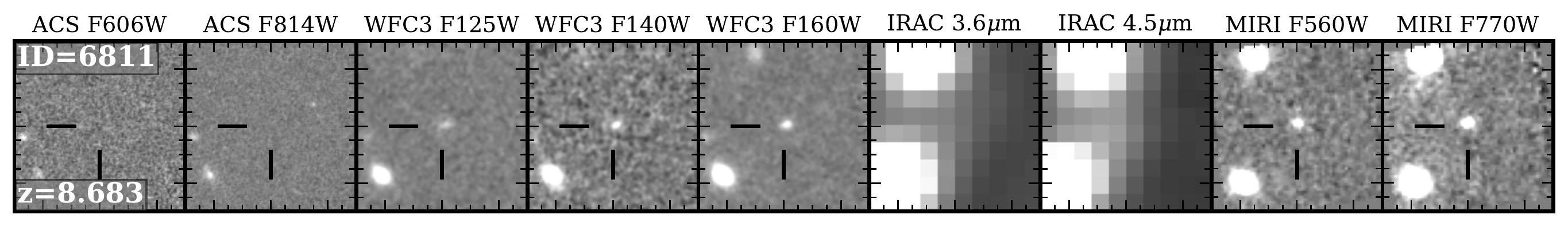}
  }
\vspace{-14pt}
  \gridline{
    \includegraphics[width=0.85\textwidth,trim={0pt 0 0 20},clip=true]{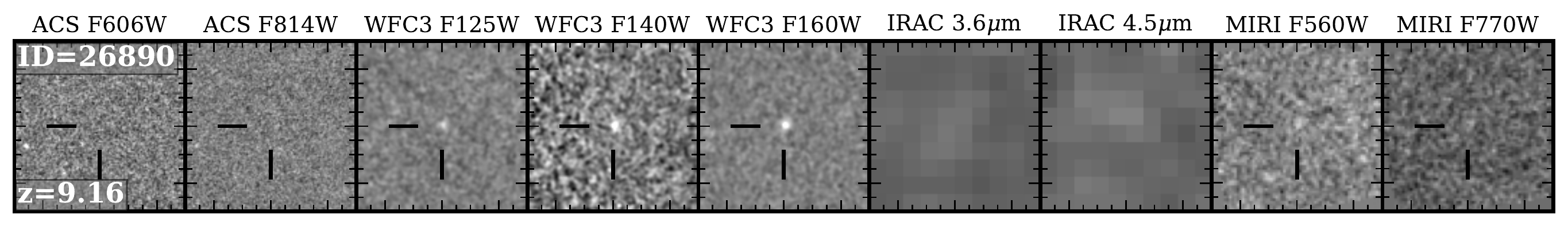}
  }
\vspace{-6pt}
\caption{Montage of images of a subset of the galaxies used in this
study \deleted{(}ordered by increasing redshift\deleted{)}.  Each row shows images for one
galaxy (labeled by galaxy ID and redshift). The images are (left to
right), ACS F606W, F814W, WFC3 F125, F160W, IRAC Ch1 (3.6~\micron) and
Ch2 (4.5~\micron), and MIRI F560W and F770W.   The images above
include the three galaxies with spectroscopic redshifts (ID 37653,
19180, and 6811).  The images are $6\arcsec \times 6\arcsec$ centered
on the  galaxy in each bandpass, as labeled along the top row.   The complete
figure set (28 images) is available online.
  \label{fig:montage}}
\figsetstart
\figsetnum{\thefigure}
\figsettitle{Montage of ACS F606W, F814W, WFC3 F125W, F140W, F160W,
  IRAC 3.6 and 4.5~\micron, and MIRI 5.6 and 7.7~\micron\ images for
  all galaxies in the sample (28 figures). }
\figsetgrpstart
\include{figset_montage.tex}
\figsetgrpend
\figsetend

\end{figure*}
%

\subsection{Galaxy Sample}\label{section:sample}

For this study we use galaxies identified in the CEERS/MIRI first
epoch fields with \added{photometric redshifts $4.3 < z < 10$ from
\citet{Finkelstein_2022a}}.    The lower redshift bound is selected to
ensure the \hst\ photometric data \deleted{(used for galaxy photometric
redshifts)} \deleted{probes} \added{probe} the redshifted Lyman-break.  The upper redshift
limit includes the highest redshift galaxies detectable by \hst/WFC3
data \citep{Bouwens_2019,Finkelstein_2022a}. This is illustrated in
Figure~\ref{fig:filters} which shows that for galaxies around $z\sim
5$ and $z\sim 9$, the \hst/ACS and WFC3 data  constrain this break.
This improves the quality of the sample\added{,}\deleted{(} as compared to, for example,
using galaxies at $z \approx 3$ where the Lyman--break shifts blueward
of the \hst/ACS F606W band.

\added{As mentioned above,} the parent sample for our study is the catalog from
\citet{Finkelstein_2022a}, which uses 
the existing \hst/ACS, WFC3 and \spitzer/IRAC 3.6 and 4.5~\micron\
data to select photometric samples of galaxies at these high
redshifts.  The data include both the imaging from the original CANDELS
survey (\hst/ACS F606W, F814W, WFC3 F125W, F160W) and additional
imaging from WFC3 F140W (see Footnote~\ref{footnote:ceers_hdr}).
\citet{Finkelstein_2022a} then use these data to measure photometric
redshifts and redshift probably distribution functions,
$P(z)$ for each object.

 We \deleted{then} matched objects from the catalog from \citet{Finkelstein_2022a}
with objects in our MIRI catalogs that are detected with S/N
$>$ 3 in either the MIRI 5.6 or 7.7~\micron\ data
(Section~\ref{section:catalog}) using a matching radius of
0.5\arcsec.  This sample includes 29 objects, though one object was
later identified as \added{a} foreground star and removed.
Table~\ref{table:sample} \deleted{which} lists the observed properties of the 28 galaxies
in the sample, their ID numbers from \citet{Finkelstein_2022a}, their
\hst/F160W flux densities the MIRI 5.6 and 7.7~\micron\ flux
densities.  The table includes the photometric redshifts derived by
\citeauthor{Finkelstein_2022a} \deleted{(}including the 16th and 84th-percentile
range from the $P(z)$ used for object selection\deleted{)}.   The Table also
includes the amount of the integrated $P(z)$ contained between $\Delta
z = \pm 0.5$ of the stated redshift, for example,
\begin{equation}
 \mathcal{P}(z=z_c) = \int_{z_c -
  0.5}^{z_c+0.5}\ P(z^\prime)\ 
 dz^\prime
\end{equation}
in bins with central redshifts of $z_c$ = 4, 5, 6, 7, 8, and 9 (see
\citealt{Finkelstein_2022a}).  These integrated probabilities indicate
a likelihood that a given galaxy is within the redshift bin to which
it is assigned.  For our analysis we rederive the photometric
redshifts below (from SED modeling that includes the new 5.6 and
7.7~\micron\ MIRI data) but include this here as we use the $P(z)$ in
Table~\ref{table:sample} as a prior likelihood on the SED fitting
(discussed in Section~\ref{section:analysis} below). 
%
%

Three of the galaxies in our sample have spectroscopic redshifts.
This includes a previously known galaxy (ID 6811 in our catalog) with
$z=8.683$ from \citet{Zitrin_2015}, and two new redshifts obtained by
the CEERS and WERLS collaborations from observations with Keck/DEIMOS
and Keck/LRIS.  The latter two sources are ID 37653 with $z=4.899$
measured by (Stawinski et al.~2023a, in preparation) and ID 19180 with
$z=5.077$  (Stawinski  et al.~2023b, in preparation).
%
%
In all of these  cases the spectroscopic redshifts are
consistent with the photometric redshifts in Table~\ref{table:sample},
and we fix the redshift to the value of the spectroscopic redshift in
our analysis of the spectral energy distributions below.

Figure~\ref{fig:montage} shows the \hst/ACS, \hst/WFC3,
\spitzer/IRAC, and \jwst/MIRI imaging for all the objects in our
sample, with the objects ordered by increasing redshift (the full
Figure Set of all 28 objects is available online).     In all cases the
galaxies show prominent ``Lyman--breaks'' at the location of the
redshifted Lyman-limit and/or Lyman-$\alpha$. In some cases, the
flux density appears to be much brighter in a given passband compared
to the adjacent band (for example, galaxy ID=12514 at $z=7.6$ shows
evidence of enhanced emission at MIRI 5.6~\micron, indicative of
strong redshifted \ha).  Figure~\ref{fig:filters} illustrates how the
bandpasses are sensitive to different features in the SED of galaxies
(using $z=5.1$ and $z=8.7$ as examples as these are similar to two of
the objects with spectroscopic redshifts in our sample).   We will
return to these cases below when we explore constraints on the galaxy
stellar populations by modeling their SEDs
(Section~\ref{section:analysis}).

~\deleted{Figure~1 shows the MIRI [5.6] - [7.7] colors for galaxies}
\deleted{  in our sample, compared to the distribution of MIRI [5.6] --}
\deleted{[7.7] colors for all objects detected in the MIRI images. }
\deleted{The three sources with spectroscopic redshifts are}
\deleted{indicated with larger symbols.}
The \added{$4 < z < 9$ galaxies}\deleted{high-redshift sources} in our
sample have MIRI $\mirifive - \miriseven$ colors largely consistent
with expectations: most objects have relatively flat
($\mirifive-\miriseven \approx 0$~mag) or blue colors ($\mirifive -
\miriseven \lsim 0$ mag).  This implies that in most cases the MIRI
data sample the continuum of galaxies. In some cases the MIRI colors
suggest very blue colors, $\mirifive - \miriseven \lsim -0.5$ to
$-1$~mag. \added{This could indicate the presence of an emission line
  in the F560W band that boosts the brightness in this band.} 
\deleted{, roughly bounded by the bold-dashed line in the}
\deleted{figure. The dashed
  line represents a photoionization model}
\deleted{with a young, metal-poor
  stellar population (10 Myr; 0.1}
\deleted{ $Z_\odot$) driving strong nebular
  emission (set by an ionization}
\deleted{parameter of $\log U = -2$).}  We will
explore evidence for this interpretation below.


\section{Spectral Energy Distribution
  Modelling}\label{section:analysis}

We model the spectral energy distributions (SEDs) of the galaxies in
our sample using stellar population synthesis models.  Our goal is to
test how the inferred properties of the stellar populations in the
high-redshift galaxies change by including the \jwst/MIRI 5.6 and
7.7~\micron\ data, in particular the stellar masses and the
SFRs. Previous work \added{before the launch of \jwst }
\deleted{(pre-\jwst)} showed \added{that MIRI data are} \deleted{the
MIRI data is} able to recover these quantities accurately
\citep{Bisigello_2017}, and here we test how they improve the
constraints on the stellar population parameters.  This is critical
for galaxies at higher redshifts, $z \gtrsim 4$, where the rest-frame
optical features shift to longer wavelengths (rest-frame 4000~\AA\
corresponds to $>$2~\micron), which probes light from longer-lived
stars.

Perhaps more problematic are the
effects of nebular emission lines, which can litter the optical
portion of the SED (see Figure~\ref{fig:filters}).  There are
observations that $z > 2$ galaxies have a higher incidence of
``extreme'' emission lines with rest-frame EWs up to
$\approx$1000~\AA\
\citep[e.g.,][]{vanderWel_2011,Tang_2019,Tran_2020,Boyett_2022,Matthee_2022,PG_2022,Sun_2022},
consistent with inferences made from the $>$3~\micron\ colors of $z >
6$ galaxies
\citep{Smit_2015,Roberts-Borsani_2016,Castellano_2017,Hutchison_2019,Endsley_2021}.
As the EW scales with redshift as $(1+z)$ this implies these lines
have a stronger impact for high-redshift galaxies for bandpasses of
fixed wavelength width \citep[e.g.][]{Papovich_2001,Burgarella_2022}.  

\begin{deluxetable*}{c@{\hskip 25pt}lcc}
\tablecolumns{4}
\tablewidth{0pt}
\tablecaption{Parameter Settings for \bagpipes\label{table:bagpipes}}
\tablehead{\colhead{Model} & \colhead{Parameter} & \colhead{Prior}  & \colhead{Limits} }
  %
%
\startdata
\multirow{3}{*}{\shortstack{Star-Formation History (1): \\
    Delayed-$\tau$, $\Psi \propto
  (t/\tau) \exp(-t/\tau)$}} & $e$-folding timescale, $\tau$ / Gyr & Uniform & (0.01, 10) \\ 
& age, $t$ / Gyr  & Uniform & (0.01,15)  \\
& stellar mass, $\log(M_\ast / M_\odot)$ & Uniform & (5, 12)  \\ \hline
\multirow{2}{*}{\shortstack{Star-Formation History (2): \\  Burst at
    $z_f=100$ and delayed-$\tau$ model from (1)}} & burst age, $t_\mathrm{burst}$ / Gyr & Fixed  &  $t_\mathrm{burst}$ = Age($z$) - Age($z_f=100$)  \\
&burst stellar mass, $\log(M_\ast / M_\odot)$ & Uniform  &  (0, 13) \\ \hline
\multirow{5}{*}{Additional parameters for all models} 
&dust attenuation law & \ldots & \cite{Calzetti_2001} \\ 
&dust attenuation, $A(V)$ / mag & Uniform & (0, 3) \\ 
&metallicity, $Z / Z_\odot$ & Uniform & (0,1) \\ 
&ionization parameter, $\log U$ & Fixed & $-2$ \\ 
&redshift$^\dag$, $z$ & \texttt{EAZY} $P(z)$ & (3, 15)  \\
\enddata
\tablenotetext{$\dag$}{For galaxies with photometric redshifts the
  redshift prior is the posterior from the photometric redshift. For
  galaxies with spectroscopic redshifts the redshift is fixed at the
  spectroscopic redshift.}
%
\end{deluxetable*}
%

 We model each galaxy by fitting \hst/ACS and WFC3, \spitzer/IRAC, and
\jwst/MIRI data with stellar population models using \bagpipes\
\citep{Carnall_2018}.  \bagpipes\ is a Bayesian SED-fitting code that
models the multiband photometry (flux densities) with stellar
population synthesis models formed over a wide range of user-defined
parameters.  The code has flexibility on the type of stellar
population synthesis models, star-formation history, dust attenuation,
and nebular emission.  It has the ability to incorporate prior
knowledge on parameters. The code then computes a probability density
for model parameters (i.e., posteriors) given the data by calculating
a likelihood weighted by priors on the parameters, and samples the
posteriors for the parameters using the \texttt{MultiNest} nested
sampling algorithm \citep[see][]{Feroz_2009,Carnall_2018}.  

Table~\ref{table:bagpipes} lists the range of parameters considered
for the SED fitting in this study.   For all models we use stellar
population synthesis models from \citet{BC_2003} formed with a
Chabrier IMF.     The table defines the parameters and their range of
parameter values we explored.
%
%
In most cases we adopt uniform priors \addedtwo{on these parameters}, as
listed in Table~\ref{table:bagpipes}, with two exceptions.  The first
is related to the nebular ionization parameter, which controls the
strength of the nebular emission features.   Current evidence from
spectroscopy \citep[e.g.,][]{Oesch_2015, Stark_2015, Stark_2017,
LeFevre_2015,  Sanders_2016, Sanders_2020, Laporte_2017,
Backhaus_2022,Papovich_2022},  including recent \jwst\ spectroscopy
\citep{Brinchmann_2022,Schaerer_2022,Trump_2022}, shows that emission
lines are common in star-forming galaxies at $z \gtrsim 1$ (and the
strength appears to increase with increasing redshift). Therefore we
fix the ionization parameter to a high, physically plausible value of
$\log U = -2$ as this is representative of the values used in previous
studies when fitting the SEDs of galaxies (see for example the
discussion in \citealt{Whitler_2022}).   We plan to explore how
variations in the ionization parameter impact the constraints on the
stellar populations using future data that can include spectroscopy,
e.g., from \jwst/NIRSpec.

The other exception is for galaxies with photometric redshifts, where
we use the photometric redshift posterior, $P(z)$, derived from
\texttt{EAZY} as the prior on the redshift (see \citealt{Chworowsky_2022}).
For galaxies with spectroscopic redshifts, we force the fit to the
spectroscopic redshift value listed in Table~\ref{table:sample}.   

For the galaxy star-formation histories (SFHs) we test two
possibilities.  First, we primarily employ ``delayed-$\tau$'' models,
where $\mathrm{SFR} \propto (t/\tau) \times \exp(-t/\tau)$ for age,
$t$, and star-formation $e$-folding timescale, $\tau$.  These
models 
allow SFHs that rise with time (when $t / \tau \ll 1$) (as is
expected for high-redshift galaxies, 
\citealt{Finlator_2011,Papovich_2011}) and for
exponentially declining models (when $t/\tau \gg 1$) and these have
the flexibility to broadly
reproduce the evolution of galaxies over long time periods
\citep[e.g.,][]{Larson_1978, Tinsley_1980,Carnall_2019,Garcia-Argumanez_2022}.

Second, following \cite{Papovich_2001} we also consider more extreme
SFHs that include both the delayed-$\tau$ model (above) \added{and} an early
burst of stars that formed at $z_f=100$.
\added{
  When a burst forms, the stellar
population immediately  begins aging.
A burst at $z_f = \infty$ has
the oldest possible age at any subsequent time, and the smallest
amount of light \deleted{(}at a given mass\deleted{)}, and therefore it would have a maximal
$M/L$ at any observed epoch.  }
\added{As such, an early burst adds the most amount of stellar mass,
with the minimum impact on the observed galaxy SED.} \deleted{(at least for
commonly assumed IMFs, like the Chabrier one assumed here).}   \added{In
contrast, the slowly evolving delayed-$\tau$ model provides a fit to
the light from the more-recently formed stellar populations that
dominate the rest-frame UV and optical light.}  Therefore,
\replaced{the}{it is} \added{conceivably possible to hide significant amounts
of stellar mass formed at earlier times that are}
\deleted{
UV/optical light from the stars formed in the maximally old burst can
be}
``lost in the glare'' of the luminous, more recently formed stars.  This is also referred to as the \deleted{``outshining''}
\added{outshining} effect 
\deleted{\citealt{Conroy_2013}}\added{(e.g., \citealt{Papovich_2001,
Dickinson_2003,
Papovich_2006,Finkelstein_2010,Maraston_2010,Pforr_2012,Conroy_2013}}).

The choice of
$z_f=100$ is motivated by the fact that current models expect stars to
be forming by $z=20-30$ \citealt{Barkana_2001}, \deleted{and
  references in} \added{with galaxy candidates identified at $z\approx
  15$ (see} 
Section~\ref{section:introduction}).  \added{A redshift of $z=100$ is essentially
``immediately'' in the history of the Universe as it corresponds to an
age of only $\simeq$17 Myr after the Big Bang for the assumed
cosmology.}  The time from $z=100$ to $z=20$ spans
less \addedtwo{than} 200 Myr, during which little stellar evolution occurs for the
longer-lived stars that dominate the stellar mass.
Therefore, $z_f = 100$
seems to be a reasonable upper bound to ensure we capture the earliest
time when stars could plausibly form.   \added{This formation redshift allows}
\deleted{By using $z_f=100$
we allow} for a ``maximally old'' stellar population and we constrain
\textit{any} star-formation that may have occurred at the earliest
times which could have the highest possible $M/L$ (for canonical
stellar populations).\footnote{Using a lower formation
redshift,  $z_f < 100$, would lower the upper limit on the stellar
masses that could form in the bursts as these would be younger, with
lower $M/L$. }

By studying the SEDs of galaxies to longer
wavelengths we can constrain the amount of light in this population.
For example, \citet{Papovich_2001} and \citet{Dickinson_2003} found
that using $K$-band data, galaxies at $z\sim 3$ could hide as much as
75--90\% of  their stellar mass in early bursts formed at $z_f=100$.
Indeed, at the risk of foreshadowing, we find that including the MIRI
5.6 and 7.7 micron data reduces the amount of possible stellar mass
formed in such maximally old bursts by up to an order of magnitude
(see Section~\ref{section:bursts}).



  \section{Results}\label{section:results}

  \subsection{Analysis of Galaxy SEDs}\label{section:results:individual}
  
\begin{figure*}
   \gridline{
    \fig{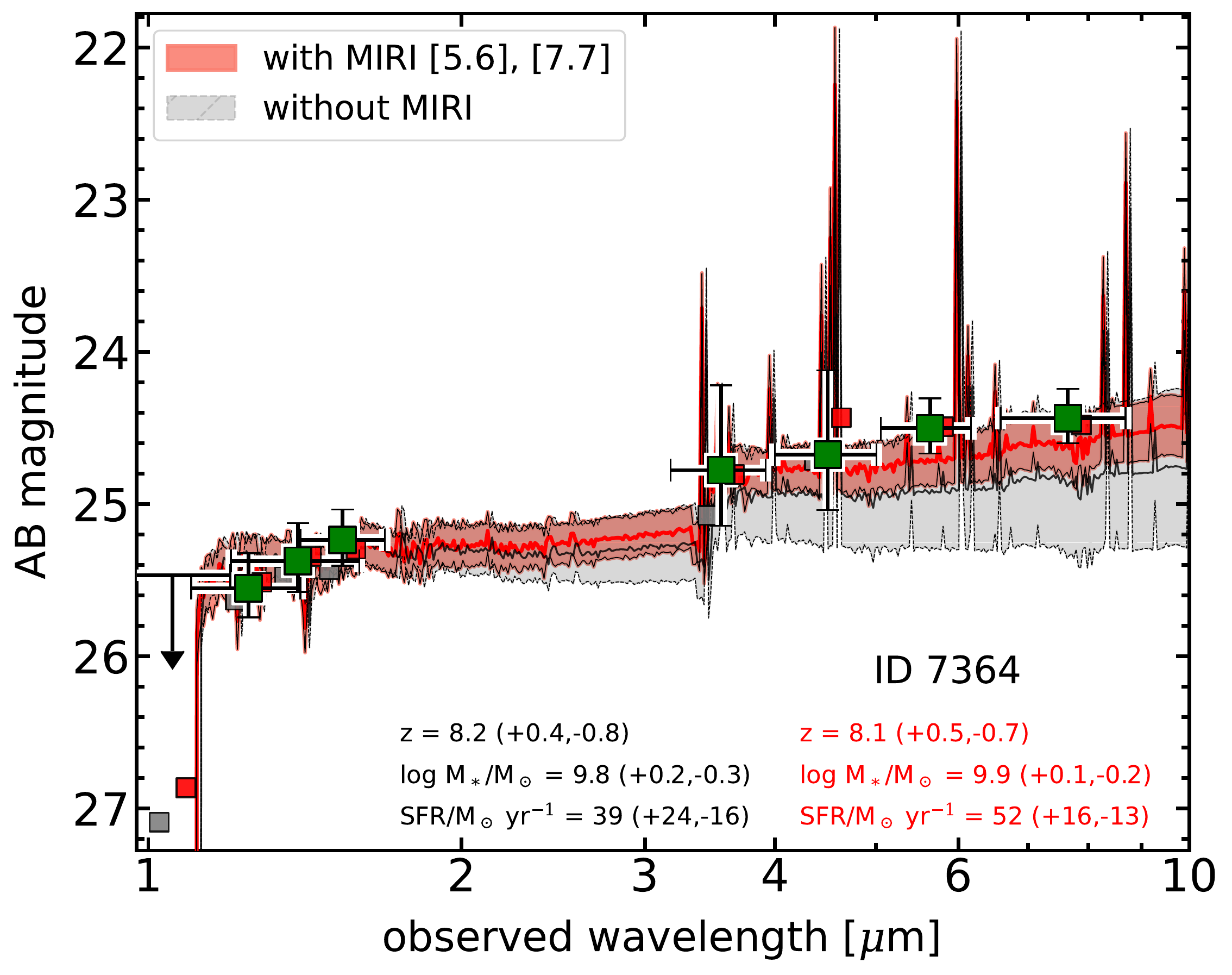}{0.5\textwidth}{}
    \fig{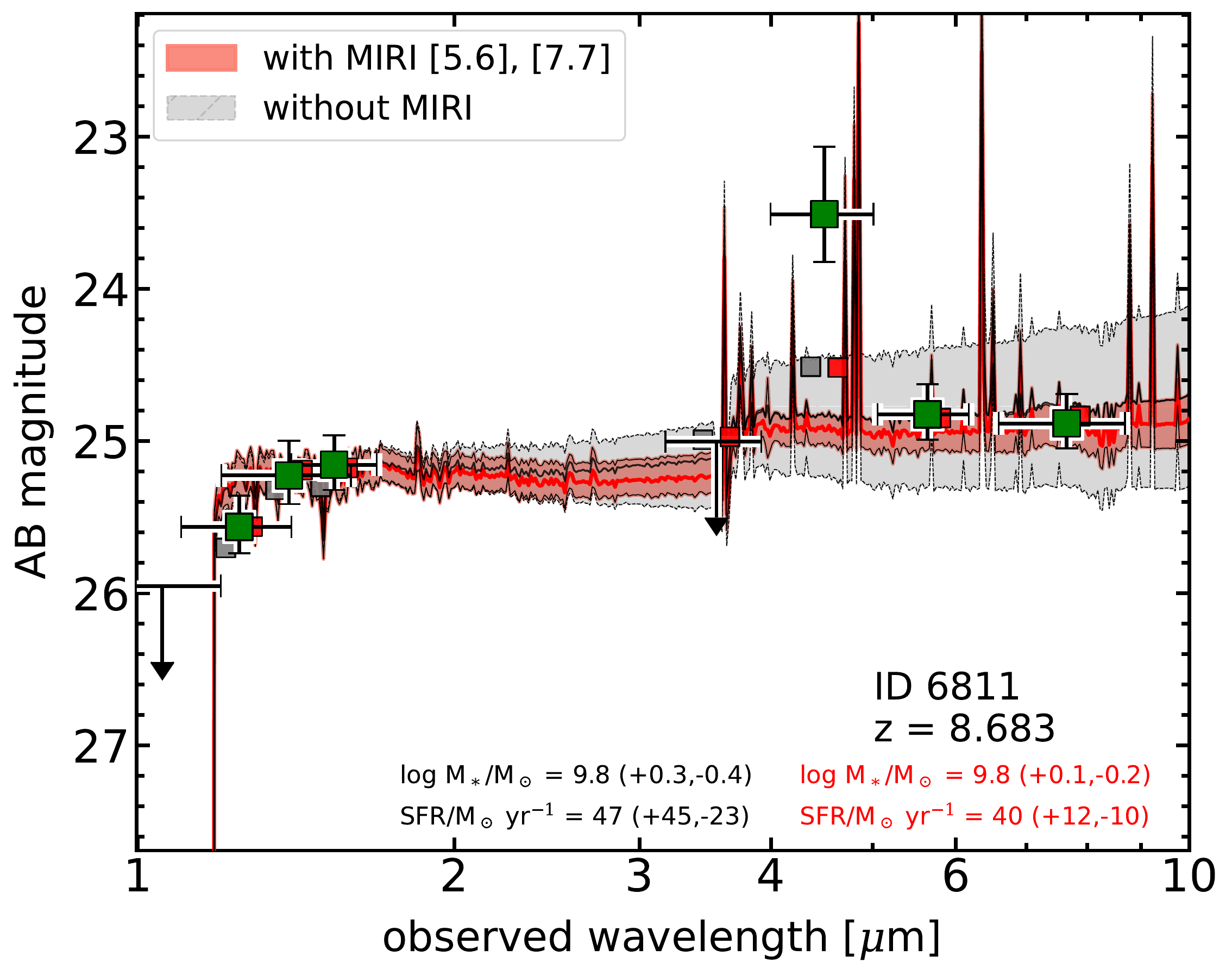}{0.5\textwidth}{}
  }
    \vspace{-24pt}
   \gridline{
    \fig{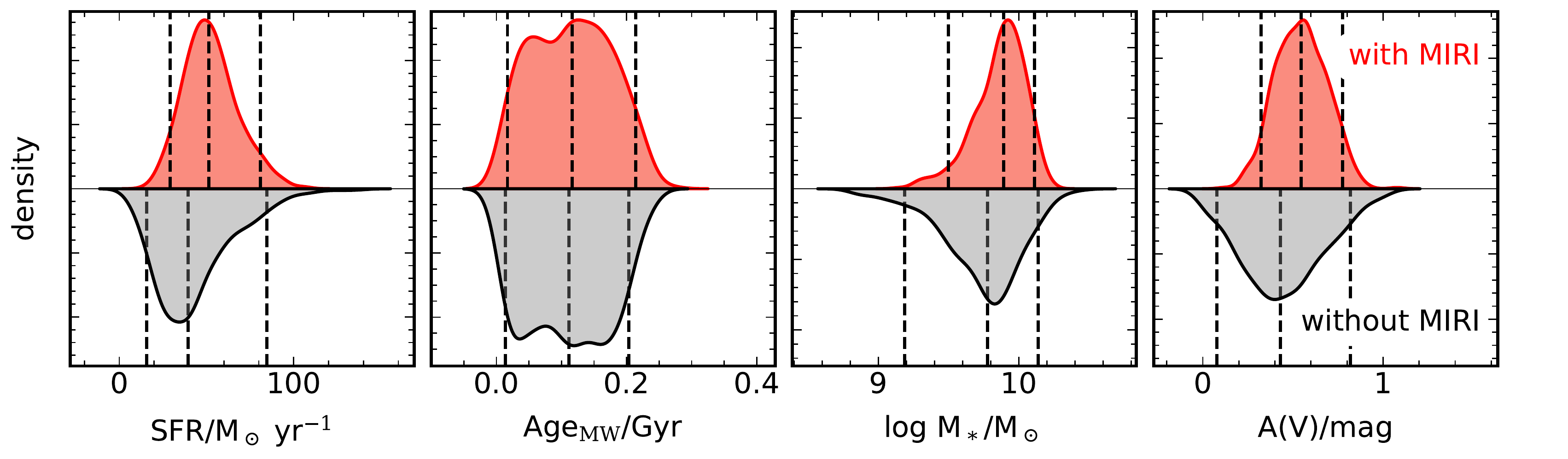}{0.5\textwidth}{(a)}
    \fig{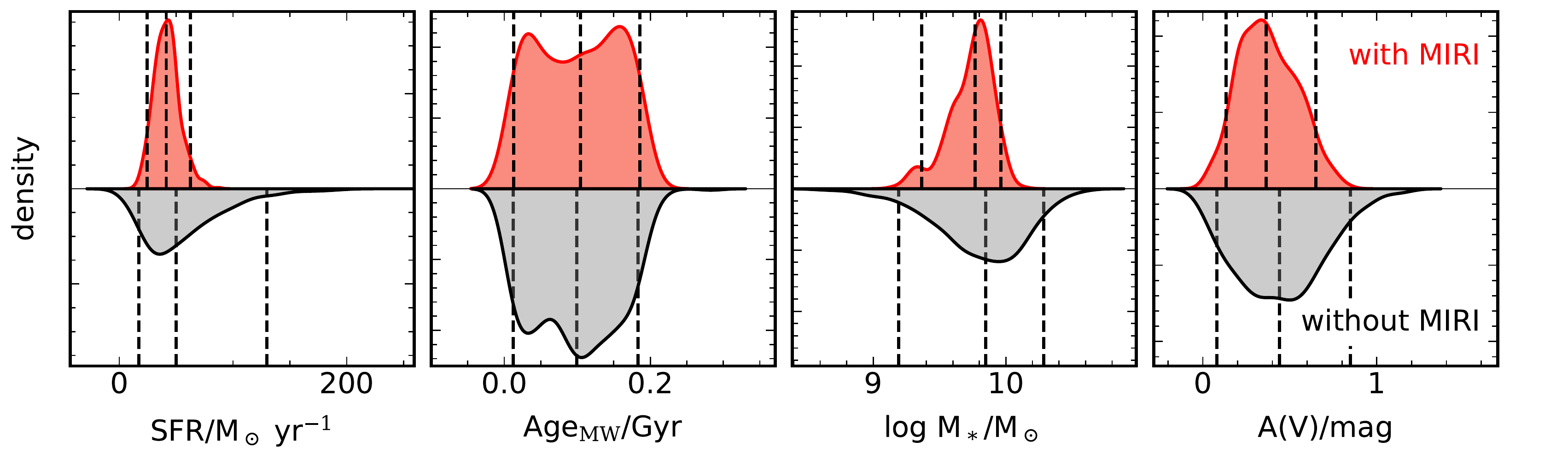}{0.5\textwidth}{(b)}
  }
%
  \caption{ Examples of SED fits for  galaxies in the sample.  The
green data points show the measured flux densities and uncertainties
on the \hst/WFC3, \spitzer/IRAC, and \jwst/MIRI bands.   The
red-shaded regions shows the model fit to all the data points (the
shaded region shows the inner-68\% range of models; the solid red line
shows the median).   The black-shaded region shows the fit to the data
points excluding the MIRI bands.  The small red and grey points show
the median-model photometry.   The inset text gives the median and
68\%-tile uncertainties on the stellar masses and SFRs inferred from
the fits.   Below each SED plot, the panels show the posteriors
(probability density) for the SFR, mass-weighted (MW) age, stellar
mass, and dust attenuation for each galaxy (using MIRI and excluding
MIRI data).  The dashed lines denote the 5, 50, and 95\% intervals.
The examples include galaxies where
adding the MIRI data yields similar stellar masses and SFRs, but with
tighter constraints (panels (a) and (b)).  Other examples show galaxies where
adding the MIRI data greatly reduces both the stellar masses and SFRs.
Typically this results from contamination in the IRAC data or because
of strong emission lines impacting the IRAC data (or both; see panels (c) and (d)), or because the stellar continua appear very blue (see panels (e) and (f)).     The
complete figure set (28 images) is available
online.\label{fig:sed_results}}
\figsetstart
\figsetnum{\thefigure}
\figsettitle{SED fits and 1D posteriors for SFRs, stellar masses,
  mass-weighted ages, and dust attenuation for all galaxies in the sample. }
\figsetgrpstart
\include{figset_spec_plots..tex}
\include{figset_1d_plots..tex}
\figsetgrpend
\figsetend
  \end{figure*}

  Figure~\ref{fig:sed_results} shows the \bagpipes\ SED fits and
  one-dimensional (1D) posterior likelihoods for select parameters of
  the SED fit.   The Figure shows six galaxies as an example.  The
  online version of the \textit{Paper} includes a Figure Set with
  these plots for the full sample.    For each galaxy, the plots compare the SED fits with and
  without the MIRI 5.6 and 7.7 $\mu$m data.    
Tables~\ref{table:bagpipes_wmiri} and
\ref{table:bagpipes_nomiri} provide the medians (50th percentiles),
16th and 84th percentiles derived from these posterior likelihoods for
the stellar masses, SFRs, and photometric redshifts for all galaxies
in the sample, both with and without using the MIRI data,
respectively. \added{Throughout, we take the ``stellar mass'' to be the
``mass formed'', which is equivalent to the integral of the
star-formation history.}

To test the robustness of the stellar masses and SFRs derived from the
\bagpipes\ fits, we have refit all the galaxies in our sample using
several independent SED-fitting codes (
\texttt{CIGALE}, \citealt{Boquien_2019}; \texttt{FAST},
\citealt{Kriek_2009}, and the codes of
\citealt{Santini_2022} and \citealt{PG_2008}).  Comparing the stellar
masses and SFRs, we find they agree in the mean (with bias, $\mu
\simeq 0$~dex) and an inter-method scatter of $\sigma = 0.23$~dex in
stellar mass \deleted{(a factor of 1.7)} and $\sigma=0.27$~dex in SFR \deleted{(a factor
of 1.9)}.  This scatter is typical in comparisons of SED-fitting
results \citep[e.g.,][]{Mobasher_2015}.  We therefore interpret the
scatter as representative of the systematic uncertainties on the
stellar masses and SFRs here.  
  

 \renewcommand{\thefigure}{\arabic{figure} (continued)}
\addtocounter{figure}{-1}

\begin{figure*}
   \gridline{
    \fig{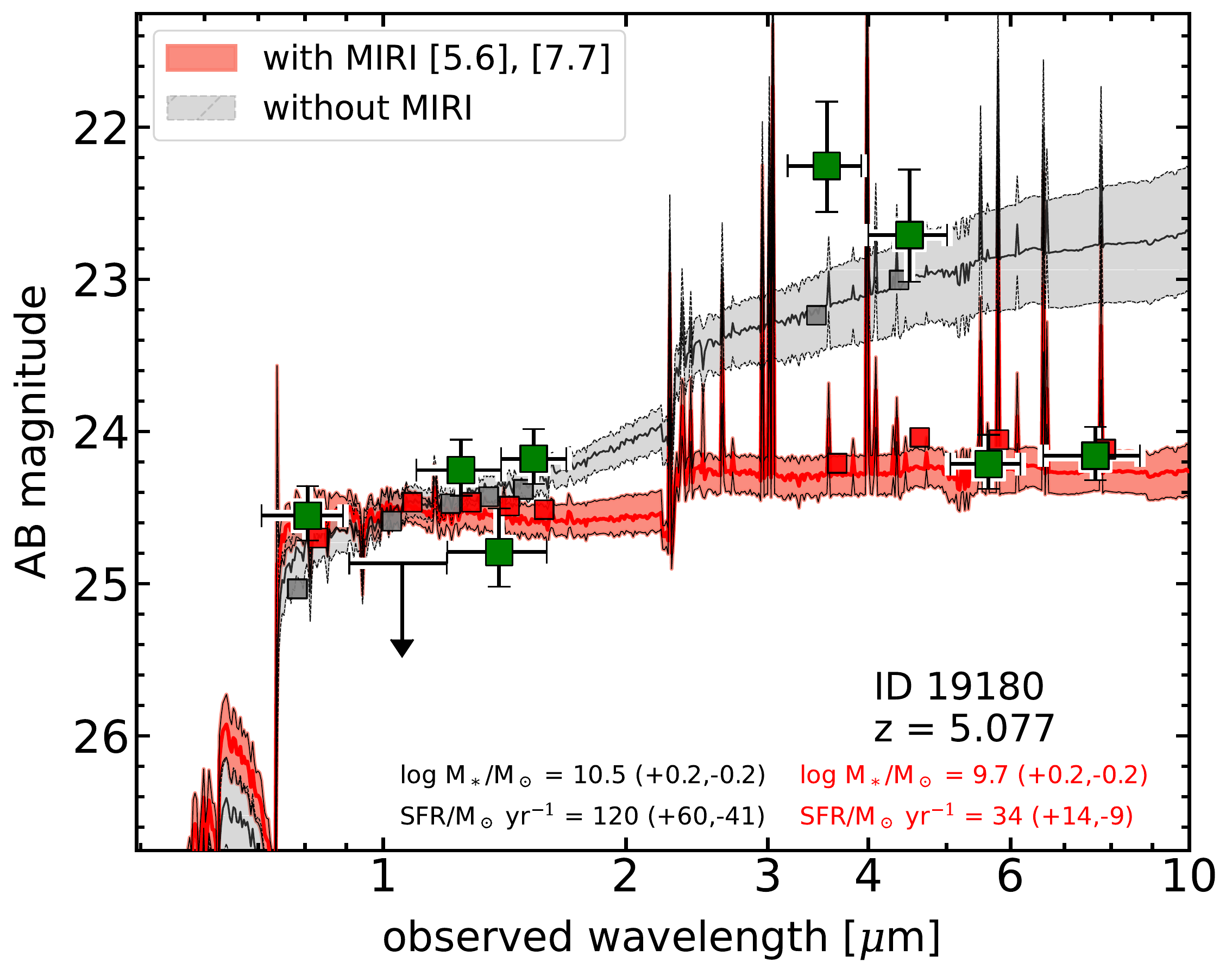}{0.5\textwidth}{}
   \fig{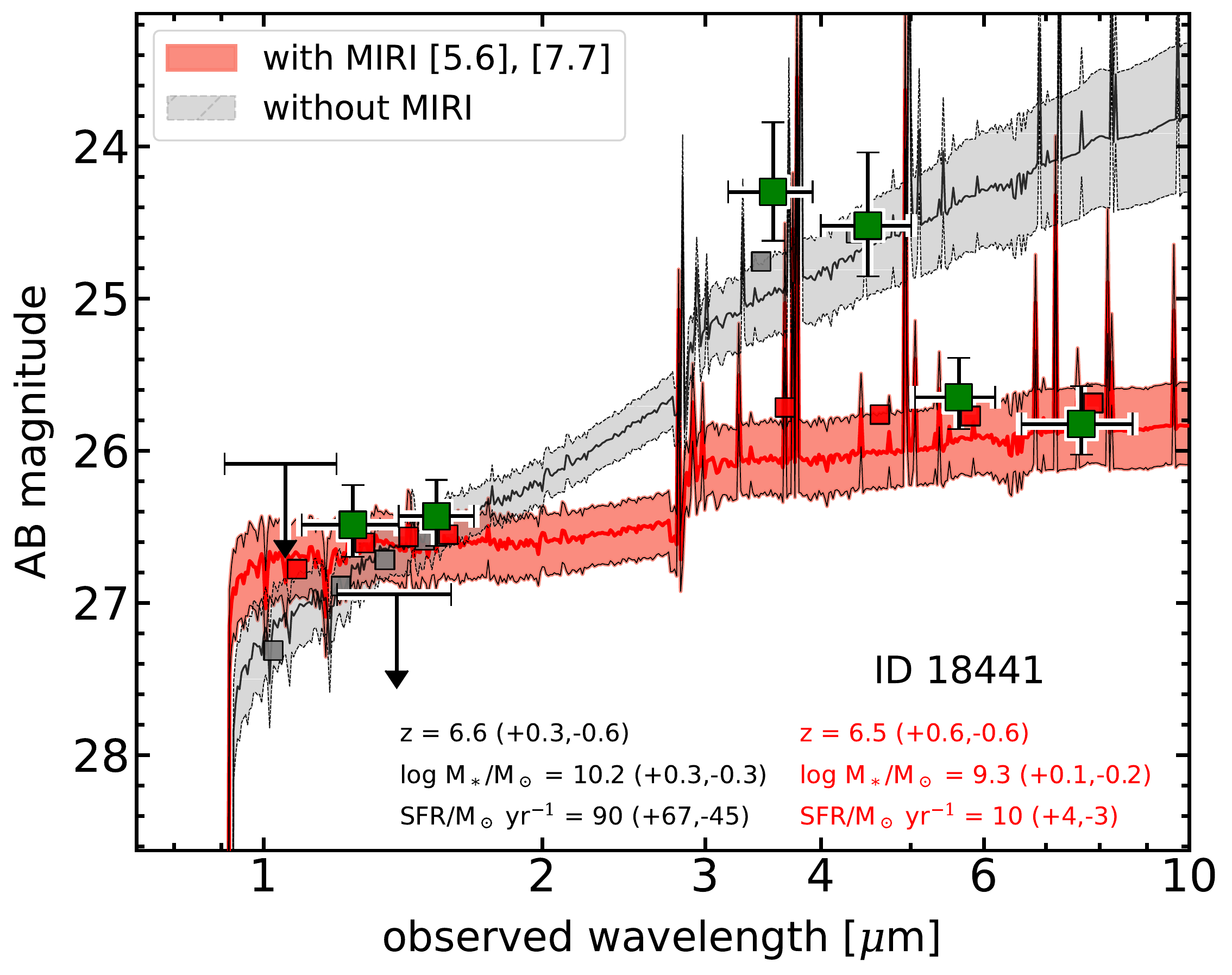}{0.5\textwidth}{}
  }
    \vspace{-24pt}
   \gridline{
    \fig{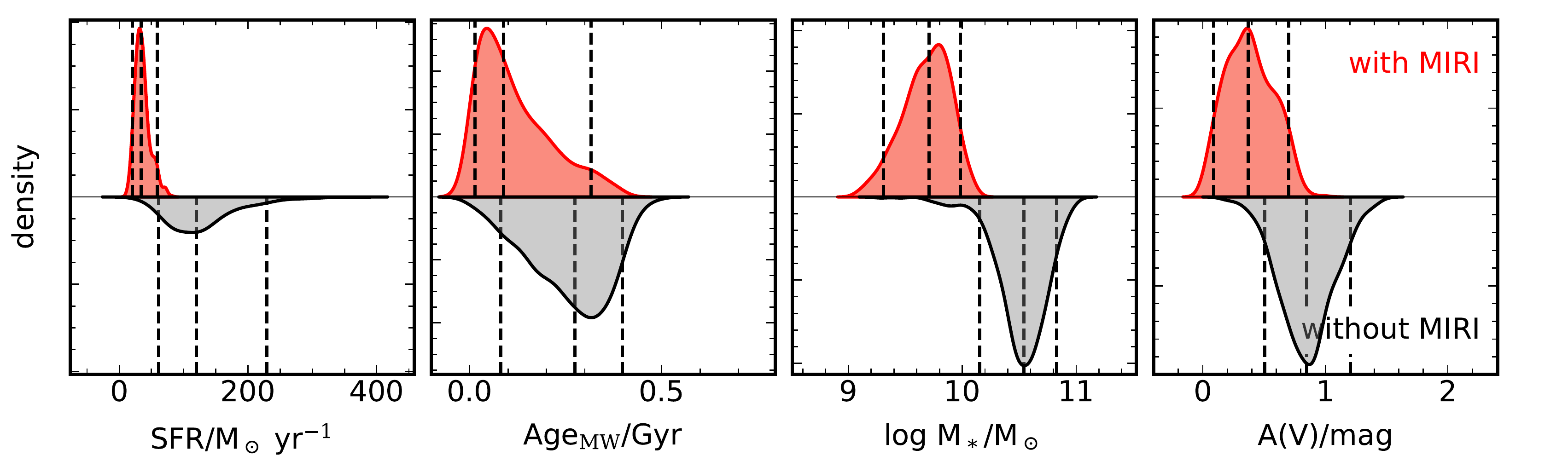}{0.5\textwidth}{(c)}
   \fig{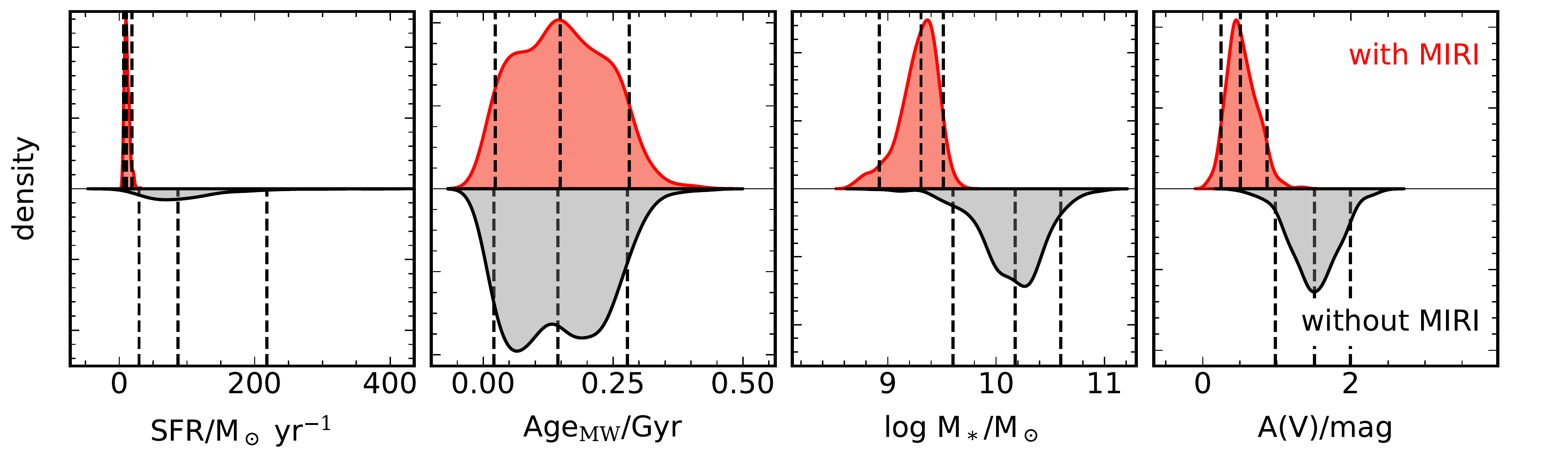}{0.5\textwidth}{(d)}
  }
%
  \caption{ \label{fig:sed_results_b}}
  \end{figure*}
\addtocounter{figure}{-1}

\begin{figure*}
   \gridline{
     \fig{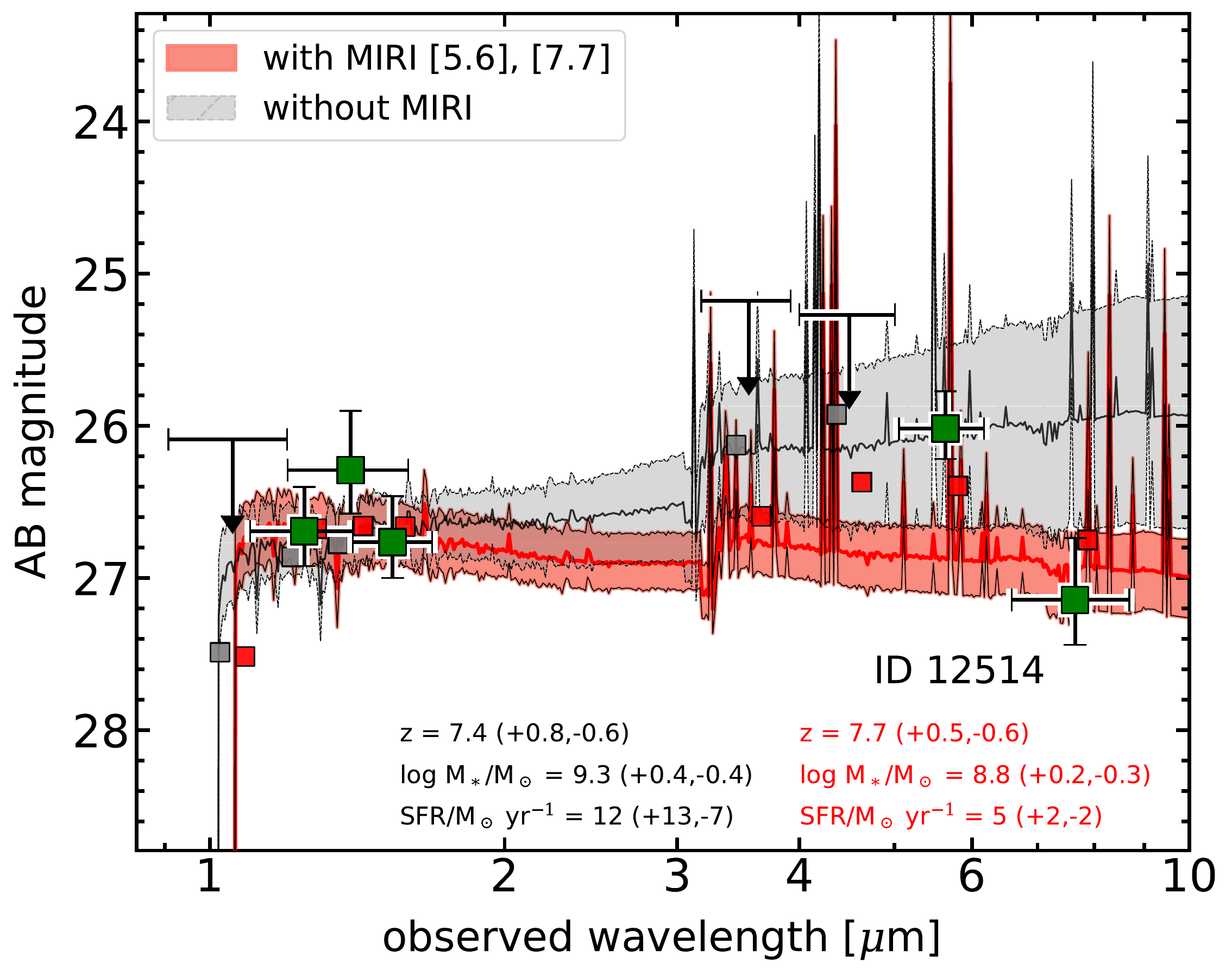}{0.5\textwidth}{}
   \fig{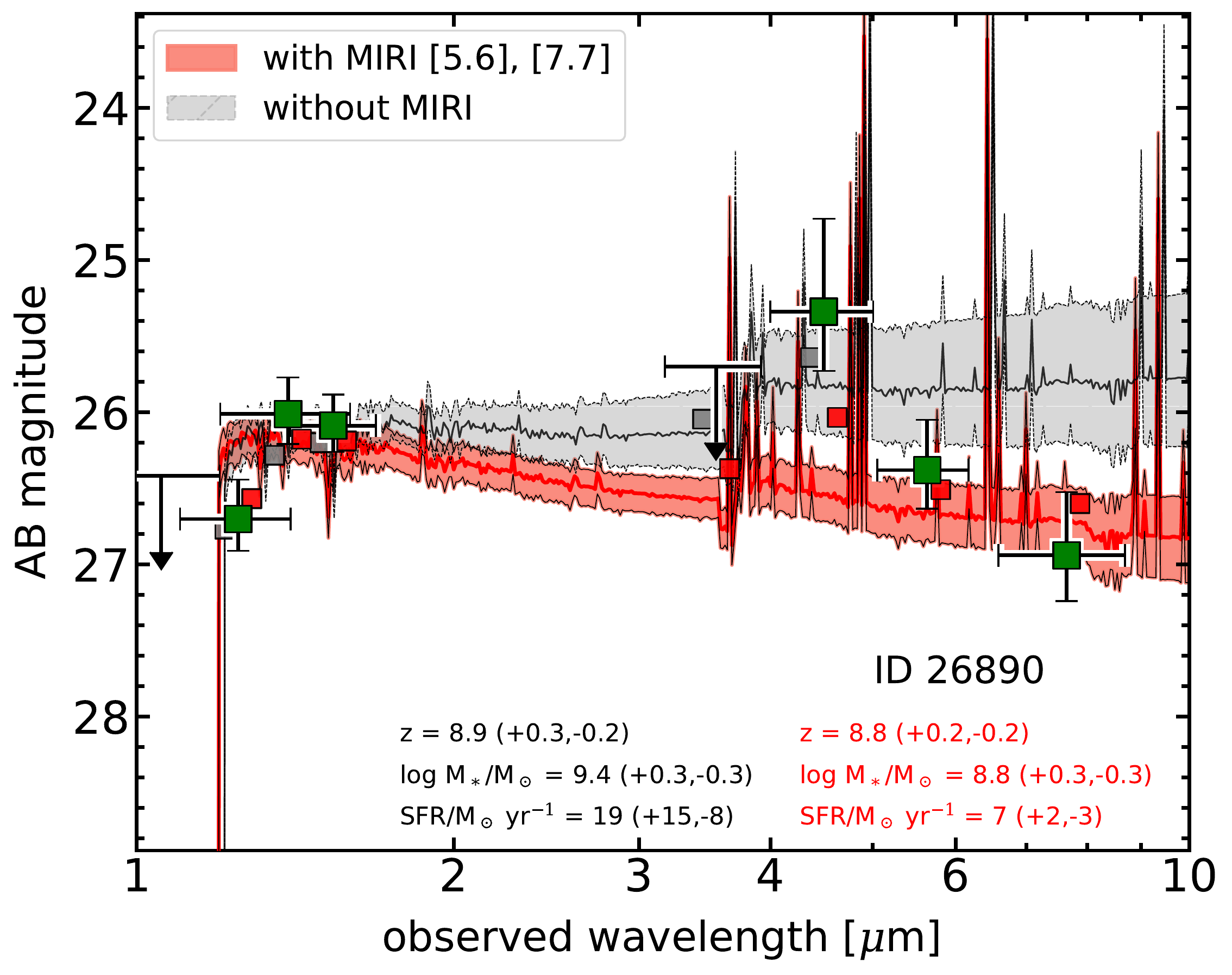}{0.5\textwidth}{}
  }
    \vspace{-24pt}
   \gridline{
    \fig{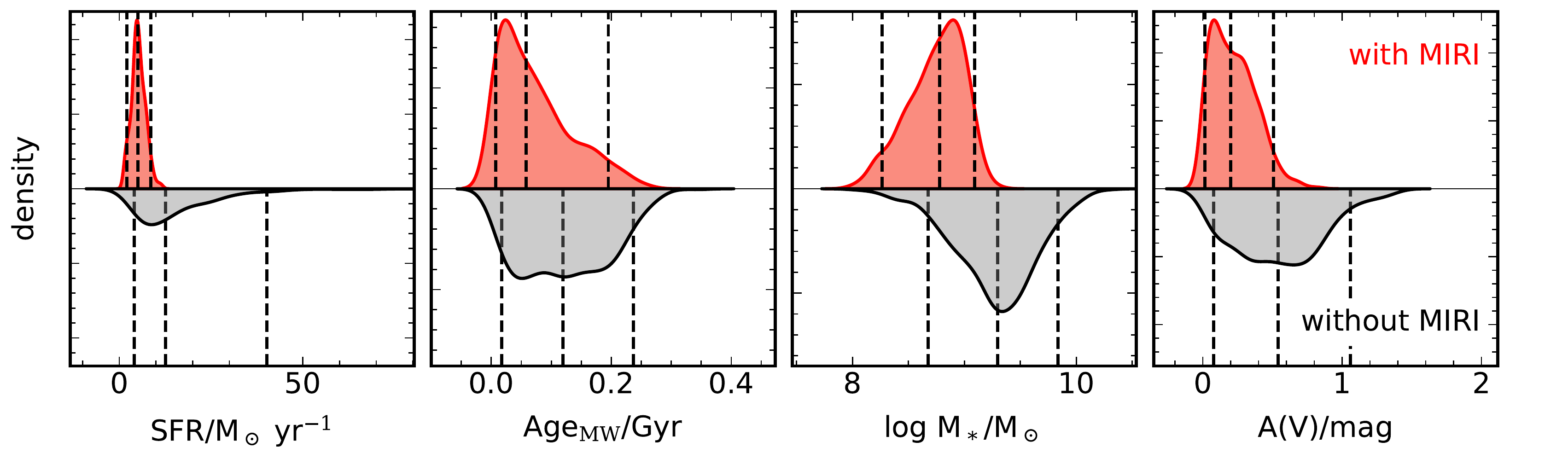}{0.5\textwidth}{(e)}
   \fig{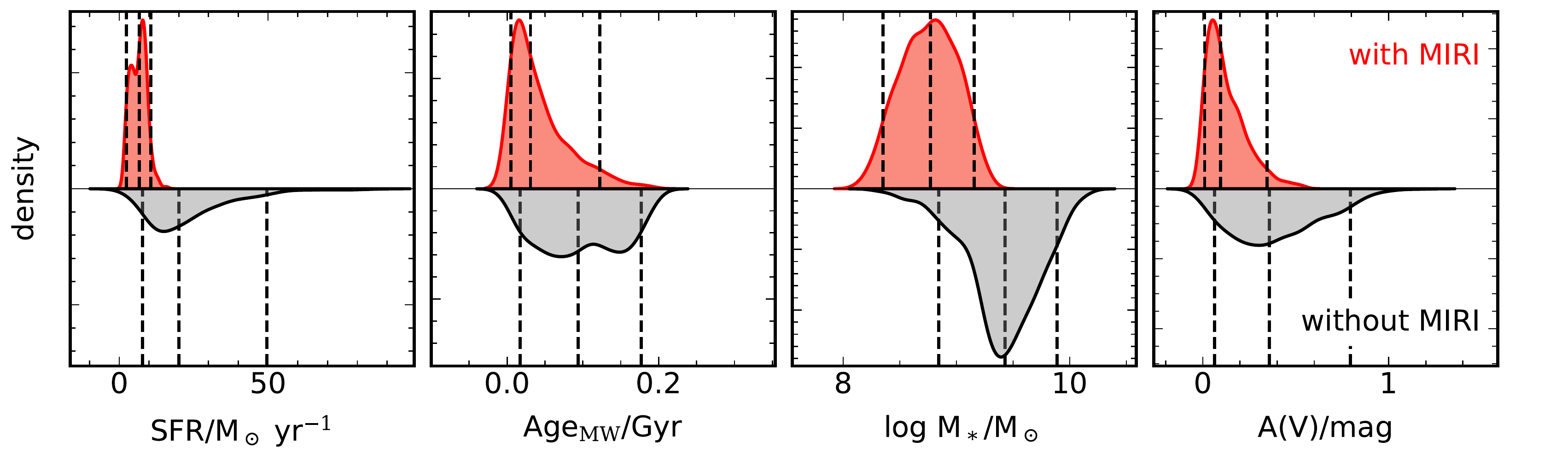}{0.5\textwidth}{(f)}
  }
  \caption{ \label{fig:sed_results_c}}
  \end{figure*}
\renewcommand{\thefigure}{\arabic{figure}}
 
In some cases adding the MIRI data has a small effect on the median
values of the stellar mass and SFR, but it does tighten the allowable
range of these parameters.  Figure~\ref{fig:sed_results} panels (a)
and (b) show  galaxy ID 7364 (at $z=8.1-8.2$) and galaxy ID 6811 (with
$z_\mathrm{sp} = 8.683$, \citealt{Zitrin_2015}).  These have MIRI data
that support the interpretation inferred using only the \hst\ and
\spitzer\ data.  However, in  both of these cases adding the MIRI data
tighten\added{s} the allowed range of models, and thus improve\added{s} the constraints
on the stellar population parameters\deleted{ (this is true for the sample in
general)}.  In both cases shown here, the favored range of stellar mass
and SFR \replaced{are}{is} improved significantly when MIRI data are included (with
improvements in the inter-68\%-tile range by more than a factor of
two).    Below the SED fit for each galaxy,
Figure~\ref{fig:sed_results} shows the posterior probability densities
for the SFR, mass-weighted age (Age$_\mathrm{MW}$), stellar mass, and
dust attenuation.   Adding the MIRI data typically produce narrower
posteriors for SFR and stellar mass.  This is a result of improved
constraints on the dust attenuation ($A(V)$), and this forces the
models to a narrower range of SFR and stellar mass. \added{This is the case
for 25\% of the sample here (7 of the 28 galaxies) based on
our visual inspection of the SEDs and 1D posteriors in
Figure~\ref{fig:sed_results}. }
\deleted{ This is the case for ID 7364 and 6811.}

In other cases adding the MIRI data changes the interpretation of the
galaxy stellar populations dramatically.  Figure~\ref{fig:sed_results}
panels (c) and (d) show galaxies with ID 19180 (with
$z_\mathrm{sp}=5.077$) and ID 18441 (at $z=6.5-6.6$).   In both of
these cases, without MIRI data the \hst\ to IRAC 3.6 and 4.5 $\mu$m
data implied very red rest-UV--to--optical colors, leading to high
dust-attenuation values ($A(V) \gtrsim 1$~mag) with high SFRs
($\gtrsim 90-100$~$M_\odot$ yr$^{-1}$).  The stellar masses in these
cases are also elevated primarily because the higher dust attenuation
increases the mass-to-light ratio ($M/L$) of the models, and therefore
increases the stellar mass.  Including the MIRI data changes the
favored stellar population models to ones with much bluer
rest-UV--to--optical colors.  As a result the dust attenuation, SFR
and stellar mass are decreased, by an order of magnitude in some
cases. \deleted{(t}\added{T}he decrease in stellar mass is more than that of the SFR, implying
the \textit{specific} SFR \deleted{declines}\added{increases} slightly as well\deleted{)}.  \added{This
  impacts  36\% of the sample here (10 of the 28
galaxies}\addedtwo{).}\deleted{, based on
our visual inspection of the SEDs and 1D posteriors in
Figure~\ref{fig:sed_results}. }

Yet in other cases, \added{adding} the MIRI data forces the constraints on the
stellar populations to be bluer than expected based on the \hst\ and
\spitzer\ data.   Figure~\ref{fig:sed_results} panels (e) and (f) show
two galaxies that demonstrate  these situations.  In both the cases
of galaxy ID 12514 (at $z=7.4-7.7$) and ID 26890 (at $z=8.8-8.9$) the
MIRI data favor very blue UV/optical colors.  In the case of galaxy
12514, there are indications that the IRAC and MIRI 5.6~\micron\ data
are boosted by the presence of strong emission lines.  Having the
7.7~\micron\ photometry favors a lower stellar continuum.  As a result
the SFR and stellar mass are reduced (the presence of redshifted
H$\alpha$ in the MIRI F560W band is apparent even in the galaxy image
in Figure~\ref{fig:montage} which shows the 5.6~\micron\ flux density
is noticeably brighter than the 7.7~\micron\ flux density).   For this
reason the SED fit favors a slightly higher photometric redshift \deleted(to
accommodate the \ha\ emission in the F560W bandpass\deleted{)}, see
Section~\ref{section:discussion:redshifts}  below.   \added{This is
  the case for 39\% of the sample here (11 of the 28
galaxies).}

\added{In the latter category,} galaxy  ID 26890 is noteworthy in itself because it is the highest
redshift galaxy in the sample, and because the \hst/WFC3--to--\jwst/MIRI colors
are $H_\mathrm{160} - [5.6] \approx -0.5$~mag, and 
indicative of stellar populations with very low $M/L$ ratios.  The IRAC
4.5~\micron\ emission shows indications of enhancement, possibly
owing to redshifted \hb+\oiii.   The MIRI data reign in the allowable
range of stellar population parameters, favoring models with lower
SFRs than the constraints that lack MIRI data.  

Therefore, the MIRI data favor bluer rest-UV/optical colors compared
to the IRAC data.  In part this may have a physical explanation.   At
the redshifts of the galaxies under consideration, the IRAC data
contain strong emission lines, including \hb+\oiii\ at $5 < z < 8$,
\oii+\neiii\ at $7 < z < 12$, and \ha+\nii\ at $4 < z < 7$
\citep{Labbe_2013,Smit_2015,DeBarros_2019,Roberts-Borsani_2020,Endsley_2021}.
At certain redshifts both the 3.6 and 4.5~\micron\ bands can be \deleted{both}
be affected \citep[e.g.,][]{Smit_2014,Roberts-Borsani_2020}.   Without
the SED fits, the models that fit the data may include both models
with strong emission lines and/or models with redder colors
(indicative of strong dust attenuation) or both.    The models in
Figure~\ref{fig:sed_results} show that the presence of strong emission
lines in the bands augments the flux densities.  This could account
for part of the difference between the SED fits with and without MIRI:
the MIRI data \deleted{are}\added{provide} evidence in these cases that the stellar populations
have bluer colors, and the elevated IRAC flux densities then would
require strong emission line EWs to reproduce them.

However, crowding between sources in the IRAC data may be another reason for
the increased IRAC flux densities.  The lower angular resolution of
the IRAC data can cause blending from bright, nearby galaxies, and
this can lead to additional uncertainties in the flux density.  As the
images in Figure~\ref{fig:montage} illustrate, some galaxies (e.g., ID 19180
and 6811) have bright objects within 3~arcseconds.  The light from
these objects makes deblending more difficult and could potentially
bias the flux-density measurements
\citep[see, e.g.,][]{Laidler_2007,Skelton_2014,Merlin_2016}.   For this
reason the elevated IRAC flux densities may include systematic
measurement uncertainties from blended objects.  In
Appendix~\ref{section:appendix} we test for the effect of source
blending by excluding objects that have any bright neighbor within
3\arcsec\ (we show that source blending does not bias our interpretation for the stellar masses nor SFRs, nor in their
evolution, that we infer for the galaxy population).    However this
does emphasize the benefit of having data with the  enhanced angular
resolution available from \jwst\ imaging\deleted{ (for both NIRCam and MIRI)}.

\subsection{The Impact of MIRI on Redshifts and Implications for
Emission Lines}\label{section:discussion:redshifts}

\begin{figure}
  \begin{center}
    \includegraphics[width=0.45\textwidth]{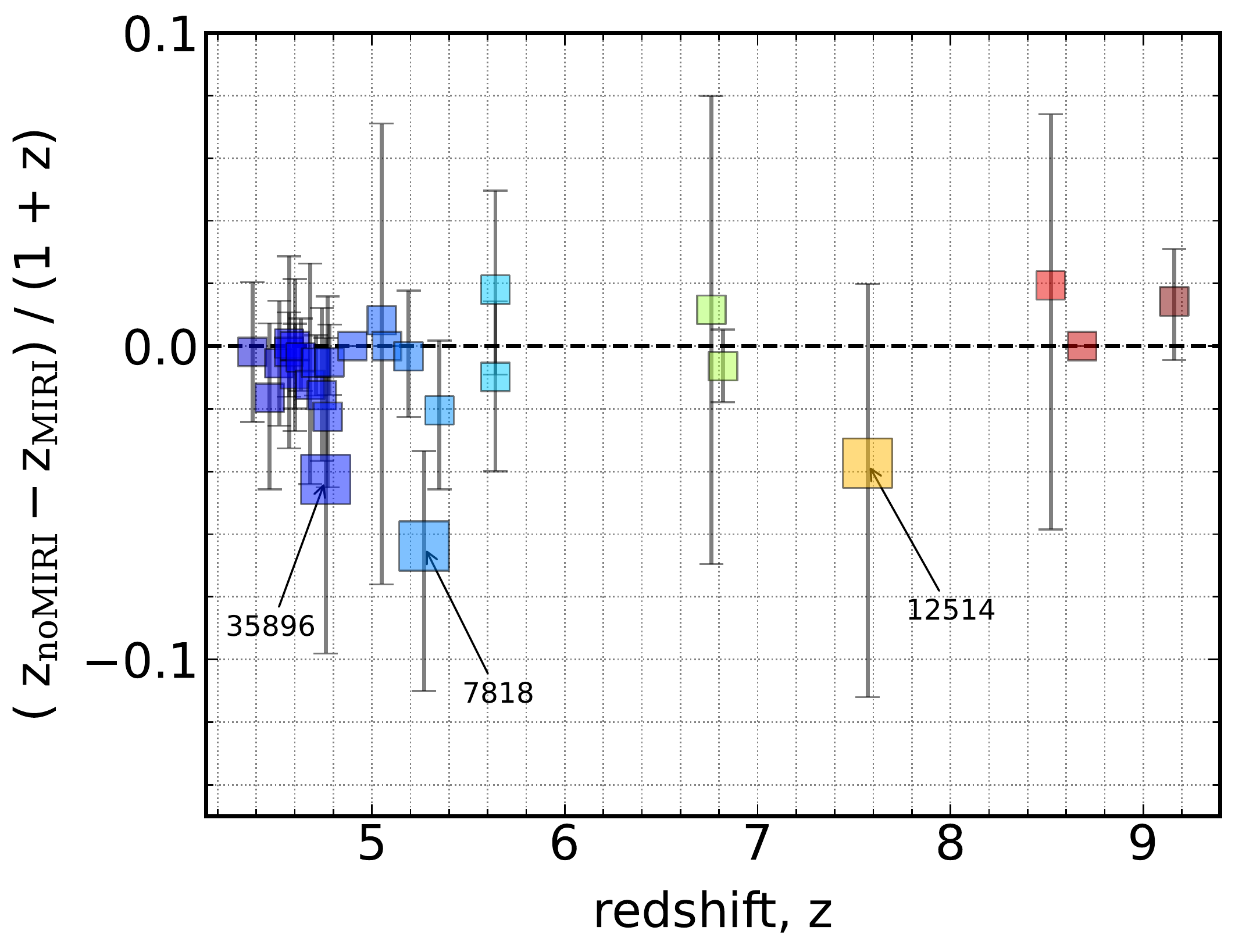}
  \end{center}
  \caption{Comparison of the redshifts for galaxies in our
    sample with and without the MIRI data.  The data points show the
    relative difference between the photometric redshifts (derived
    from our \bagpipes\ fits) for the galaxies excluding MIRI
    ($z_\mathrm{no MIRI}$) and when including MIRI
    ($z_\mathrm{MIRI}$) as a function of the prior photometric
    redshift from \citet{Finkelstein_2022a}.   Annotated points
    indicate objects with $\left| z_\mathrm{no MIRI}-
      z_\mathrm{MIRI}\right| > 0.2$.   These objects indicate shifts
    in their photometric redshift, where at least in part this is because of nebular
    emission in one or more bands. \label{fig:redshifts_miri_vs_nomiri}} 
\end{figure} 

Figure~\ref{fig:redshifts_miri_vs_nomiri} compares the redshifts for
the galaxies in our sample derived from our \bagpipes\ fits for the
galaxies excluding the MIRI data ($z_\mathrm{no MIRI}$) and when
including the MIRI data ($z_\mathrm{MIRI}$) as a function of the prior
photometric redshift from \citet{Finkelstein_2022a}.   For most
galaxies there is good agreement. \deleted{(}For completeness we include all 28
galaxies in this comparison, using the photometric-redshifts even for
galaxies with spectroscopic redshifts.\deleted{) }

    \begin{figure*}
  \begin{center}
    \includegraphics[width=0.31\textwidth]{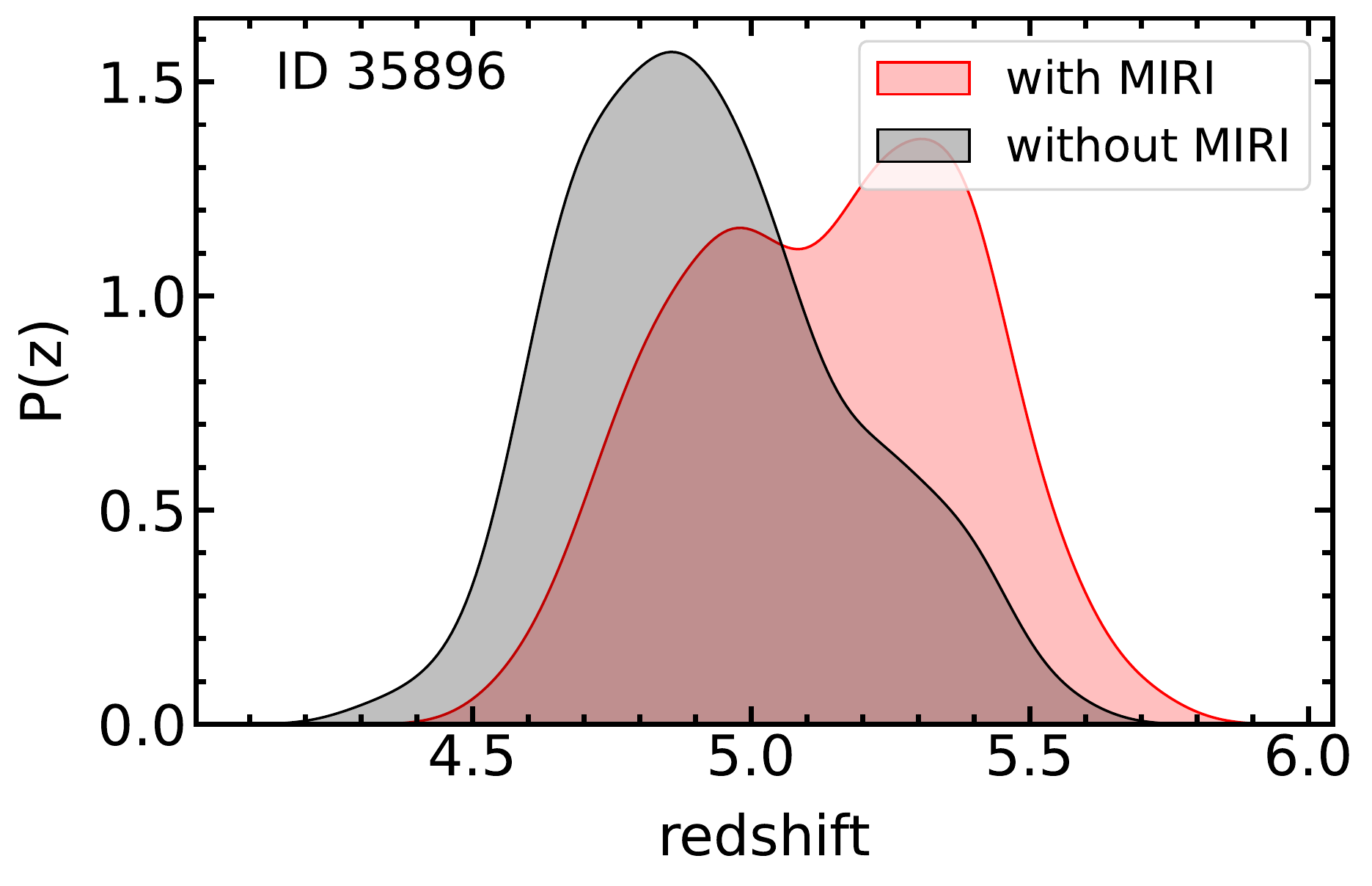}
    \includegraphics[width=0.31\textwidth]{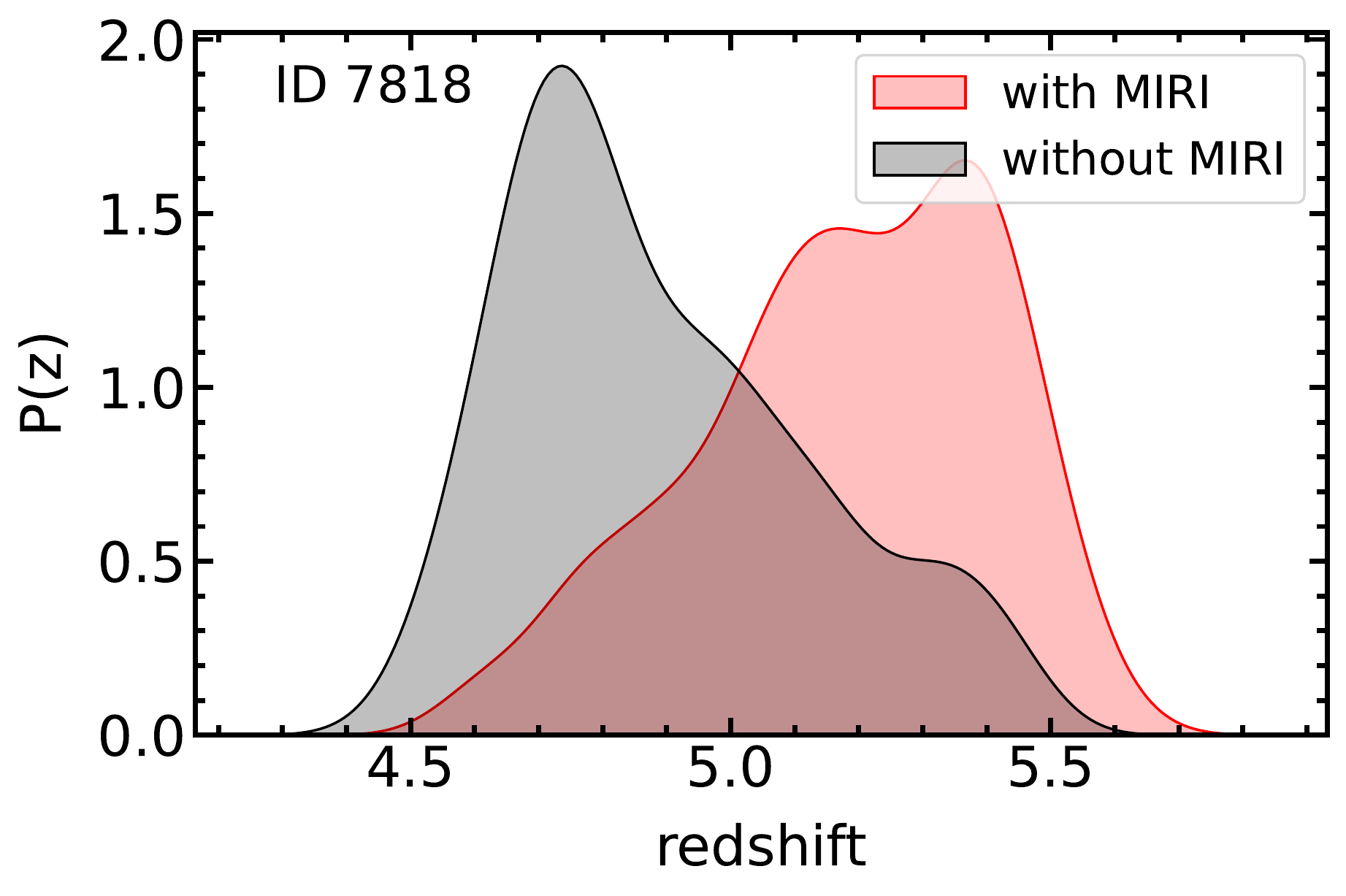}
    \includegraphics[width=0.31\textwidth]{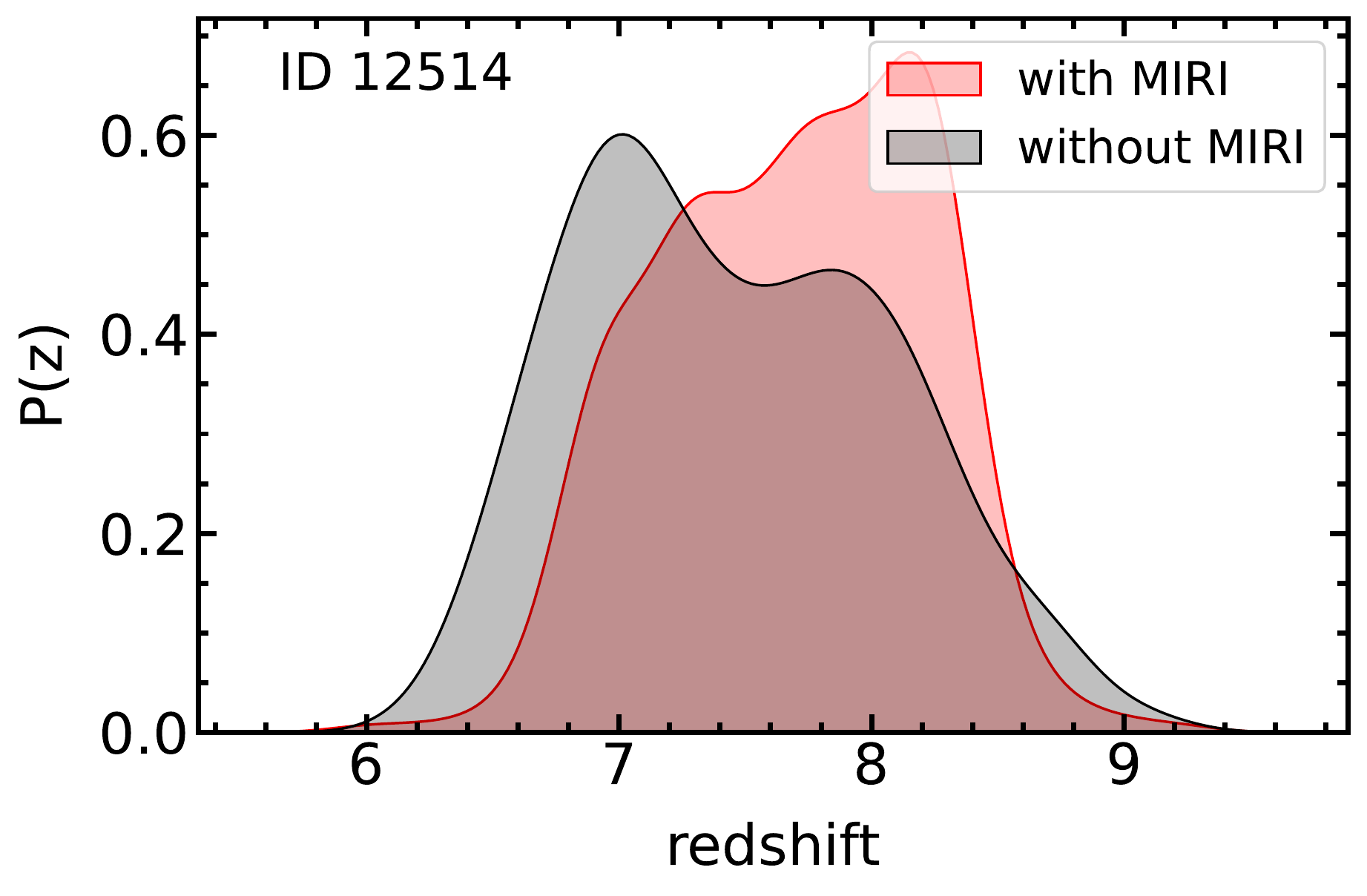}
  \end{center}
  \caption{Comparison of redshift posterior probability densities,
    $P(z)$, derived from the \bagpipes\ fits to three of the galaxies
    in the sample (as labeled). These three galaxies each have
    a change of more than 0.2 between the median redshift when including
    the MIRI F560W and F770W data in the fit versus when they are
    excluded. For the case of galaxy IDs 35896 and 7818, the
    IRAC$-$MIRI colors are blue, indicating redshifted emission lines
    (likely \hb+\oiii\ and \ha) inhabit both IRAC bands \deleted{(} while the
    MIRI data probe the galaxy continua\deleted{)}. In the case of galaxy ID
    12514, the MIRI F560W band clearly shows boosted emission, likely
    from redshifted \ha\ at $z=7.6-8$.\label{fig:zpdfs}}
\end{figure*}

In several instances we see the median redshift from the posterior
shifts appreciably when including the MIRI data \added{(with $\Delta z
  > 0.2$;  these are annotated
in Figure~\ref{fig:redshifts_miri_vs_nomiri}}).\deleted{ shows three objects
where the shift is greater than $\Delta z = 0.2$.}   In all cases the
shift in the redshift probability density appears to be related to the
effects of one or more emission lines in the IRAC or MIRI passbands.
Figure~\ref{fig:zpdfs} illustrates the shift in $P(z)$
from the \bagpipes\ fits for these galaxies.

Two of these galaxies lie at $4 < z < 6$ (galaxy IDs 35896 and
7818). Adding the MIRI data favors having strong nebular emission in
\textit{both} of the IRAC bands.  This occurs at $z \gtrsim 5$ when
\hb+\oiii\ enters the IRAC Ch1 bandpass (at 3.6~\micron) \textit{and}
\ha\ enters the IRAC Channel 2 bandpass (at 4.5~\micron; see
Figure~\ref{fig:filters}).  Inspection of the galaxy SEDs (see
Fig.~Set.~\ref{fig:sed_results}) shows that the IRAC--to--MIRI colors
($[3.6]-[5.6]$ and $[4.5] - [5.6]$) are blue for both of these
galaxies.  The \bagpipes\ fits that include the MIRI data favor models
in which the galaxy has strong nebular emission in the IRAC bands to
account for \deleted{these}\added{the observed} color.  This increases
(decreases) the redshift probability density at higher (lower)
redshifts for these galaxies.

The remaining galaxy, ID 12514, lies at $z \approx 7.7$.
Figure~\ref{fig:zpdfs} shows the redshift probability densities for
this galaxy.  The MIRI F560W image shows
evidence of ``boosted'' flux compared to the MIRI F770W image
\added{(see the images in Fig.~\ref{fig:montage})}, where
the MIRI color is $[5.6] - [7.7] = -1.1 \pm 0.3$~mag \added{(}see
Table~\ref{table:sample}).   The most likely explanation for this
boosted emission appears to be from redshifted \ha+\nii\ at
$z=7.6-8.0$ in the MIRI F560W bandpass (see
Fig.~\ref{fig:sed_results}).  The strength of the flux in the F560W
bandpass decreases the probability density for redshifts with $z \lsim
7$ as these would place \ha\ at wavelengths shorter than covered by
the bandpass (see~Figure~\ref{fig:filters}). 

Therefore, nebular emission lines appear to be responsible for the
\deleted{cases where there are} larger shifts in the photometric
redshifts. \deleted{
solutions,} However, these cases are generally exceptions.  For most
galaxies in the sample the redshift posteriors are consistent, where
Fig.~\ref{fig:redshifts_miri_vs_nomiri}  shows the differences are
negligible in median redshifts with and without including the MIRI
data.   We expect that many (even most) of the galaxies in the sample
also exhibit strong emission lines given the preponderance of evidence
from other galaxies at these redshifts (see
Section~\ref{section:results:individual}) and by inspection of the
SEDs in Figure~\ref{fig:sed_results}.    In these cases adding the
MIRI data supports the redshifts, either because the emission lines
make less of an impact on the redshift likelihoods or because the MIRI
data reinforce them. 

\subsection{The Impact of MIRI on the Stellar Masses, Dust Attenuation, and SFRs}

\begin{figure*}
  \begin{center}
    \gridline{
      \includegraphics[width=0.85\textwidth]{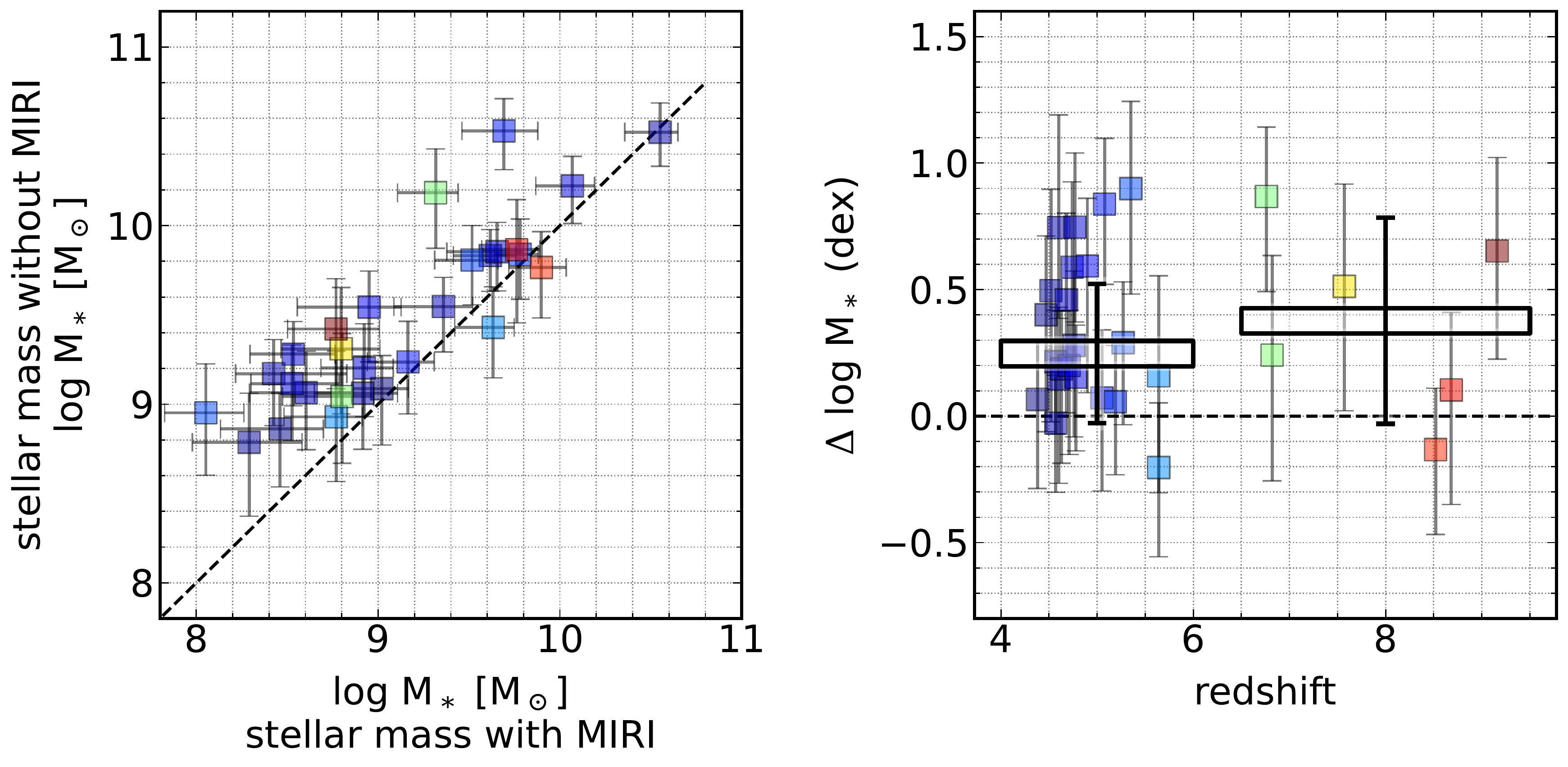}
    }
    \vspace{0pt}
    \gridline{
      \includegraphics[width=0.85\textwidth]{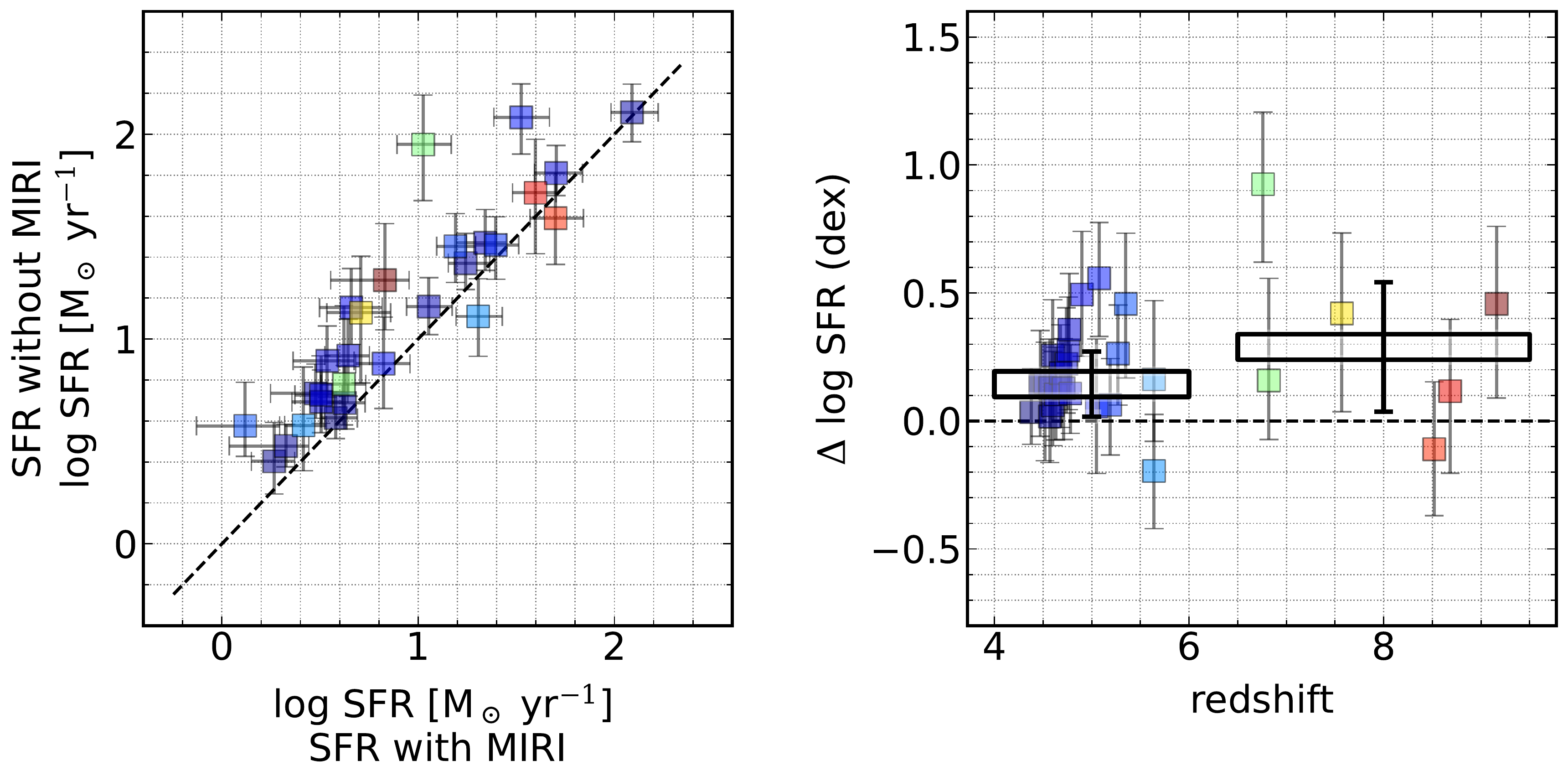}
      }
    \vspace{-24pt}
  \end{center}

\caption{ Comparison of stellar masses and SFRs derived from the SED
modeling for galaxies including the MIRI F560W and F770W data and
without the MIRI data.   The top set of panels show the comparison for
the stellar mass.   The bottom set of panels show the comparison for the SFR.
In each row the left panel shows the direct comparison, where the
dashed line shows the one-to-one relation.  The right panel shows the
logarithmic difference, defined as $\Delta \log
M_\ast^\mathrm{with~MIRI} - \log M_\ast^\mathrm{without~MIRI}$
(similarly for the SFR).   The symbols are color-coded by redshift
(using the right panel).  In the right panel the large rectangular
boxes and error bars show the median value (and scatter)  in two
bins of redshift ($4 < z < 6$ and $6 < z < 10$).  The values for these
are given in Table~\ref{table:offsets}.   Adding the MIRI data
generally lowers the SFRs and stellar masses of these galaxies though
the scatter is significant.  The median offset is larger for galaxies
at higher redshifts.  \label{fig:mass-sfr_miri_vs_nomiri}}
\end{figure*}

Figure~\ref{fig:mass-sfr_miri_vs_nomiri} compares the stellar masses
and SFRs derived for the galaxies in our sample using the simple
delayed-$\tau$ models with \bagpipes\ for the case where we include
the MIRI F560W and F770W data and when we exclude it.    In general,
including the MIRI data \textit{reduces} the stellar masses and SFRs
for the galaxies.

\begin{deluxetable}{l@{\hskip 16pt}cc@{\hskip 16pt}cc}
  \tablecolumns{5}
  \tablecaption{Offsets in
Stellar Mass and SFRs for $4 < z < 10$ galaxies when including the
MIRI F560W and F770W data.\label{table:offsets}}
\tablehead{ \colhead{}
& \multicolumn{2}{c}{Stellar Mass offsets} & \multicolumn{2}{c}{SFR
  offsets} \\ [-2pt]
\colhead{} & \multicolumn{2}{c}{$\langle \Delta \log M_\ast \rangle =$} &
\multicolumn{2}{c}{$\langle$ $\Delta$ log SFR $\rangle =$} \\ [-6pt]
\colhead{} & \multicolumn{2}{c}{$\log
M_\mathrm{\ast,no MIRI} / M_\mathrm{\ast,MIRI}$} &
\multicolumn{2}{c}{log SFR$_\mathrm{no MIRI}$ / SFR$_\mathrm{MIRI}$} \\ [-2pt]
\colhead{Sample} & \colhead{Median}  &
\colhead{Scatter} & \colhead{Median} & \colhead{Scatter} }
\startdata
$4 < z < 6$ & 0.25 dex & 0.28 dex & 0.15 dex & 0.12 dex \\
$6 < z <
10$ & 0.38 dex & 0.44 dex & 0.29 dex & 0.27 dex \\
\enddata
\tablecomments{The quantities with the subscript ``noMIRI'' denote
values derived without the MIRI data.  Quantities with the subscript
``MIRI'' denote the values derived with the 5.6 and 7.7~\micron\ MIRI
data.   The scatter is the inter-68-percentile interval derived from
the median absolute deviation.}
\end{deluxetable} 

We study the offsets in stellar mass and SFR for our galaxies in two
redshift bins, $4 < z < 6$ and $6 < z < 10$, where the median offsets
in stellar mass and SFR are indicated in the
Figure~\added{\ref{fig:mass-sfr_miri_vs_nomiri}} as large rectangles,
and are listed in Table~\ref{table:offsets}.  \added{For this
comparison, we have defined $\Delta \log M_\ast = \Delta \log
M_\ast^\mathrm{with~MIRI} - \log M_\ast^\mathrm{without~MIRI}$ and
$\Delta \log \mathrm{SFR} = \log \mathrm{SFR}^\mathrm{with~MIRI} -
\log \mathrm{SFR}^\mathrm{without~MIRI}$.  We have added their
uncertainties in quadrature, but this assumes the stellar masses and
SFRs derived with and without MIRI are independent.  Strictly speaking
this is not true as both values use the same \hst\ and \spitzer/IRAC
data.  Therefore it is likely that the error bars are
\textit{overestimated} assuming the values are correlated.
Nevertheless this does not have a major impact on our work because we
are interested in the medians and scatter of the sample. } 

Formally, the offsets in stellar mass are $\Delta \log (M_\ast) =
0.25$~dex for $4 < z < 6$ and 0.38~dex for $6 < z < 10$.  The
inter-68\%-tile intervals are  0.28 and 0.44~dex, respectively
(estimated using the median absolute deviation,
$\sigma_\mathrm{NMAD}$).  For the SFRs the offsets are $\Delta \log
(\mathrm{SFR}) = 0.15$ dex for $4 < z < 6$ and 0.29 dex for $6 < z <
10$, with an inter-68~percentile interval of 0.12~dex and 0.27~dex,
respectively.  Because the impact of MIRI is larger on the stellar
mass than the SFR, the specific SFRs will be \deleted{reduced}
\added{increased} by approximately $\Delta \log \mathrm{SFR} -
\Delta \log M_\ast \approx 0.1$ dex (see Table~\ref{table:offsets}).   

The reasons for the offsets are similar to that discussed for the
individual galaxies in Section~\ref{section:results:individual}.
Including the MIRI data generally favors stellar populations with
bluer rest-UV/optical colors, with lower $M/L$ ratios.  This forces
the constraints to lower stellar-mass models.  For the SFRs, because
the colors are bluer, there is less dust attenuation favored in the
models, which lowers the SFRs  compared to models with higher dust
attenuation (at fixed \textit{observed} galaxy luminosity).    The
MIRI data also remove some of the degeneracy between models with
redder stellar populations versus those with strong emission lines
impact select bands.  These have the combined effect of favoring lower
stellar masses and SFRs when including the MIRI data.

There is significant scatter in the offsets of stellar mass and SFR
for individual objects.    In Figure~\ref{fig:mass-sfr_miri_vs_nomiri}
the error bars on the large rectangles show the inter-68-percentile
range (i.e,. the difference between the 16--84 percentiles) for both
the $4 < z < 6$ and $6 < z < 10$. For some galaxies the offsets are
insignificant, with $\Delta \log M_\ast \approx 0$ and $\Delta \log
\mathrm{SFR} \approx 0$.
\deleted{Two of these galaxies are shown in
Figure~\ref{fig:sed_results} (IDs 7364 and 6811; others are available
in Fig.~Set.~\ref{fig:sed_results}).  In these cases the MIRI data
tighten the existing constraints on the derived stellar masses and
SFRs, reducing the uncertainties (by 0.1--0.2~dex in stellar mass and
by 0.3~dex in SFR).}
In other words, \replaced{for some}{in some} cases the MIRI data
support the range of stellar population parameters favored by the fits
to the \hst/ACS + WFC3 and \spitzer/IRAC data, but the MIRI data
improve the accuracy of the measurements, typically by a factor of
order two.  

In \replaced{other}{most} cases the MIRI data
    \deleted{dramatically change the interpretation of
the galaxies. This was noted above for galaxies 18441 and 19180
(Section~\ref{section:results:individual} and
Figure~\ref{fig:sed_results}), where adding the MIRI data}
reduce the stellar mass and SFRs\added{, in some cases substantially.}
Figure~\ref{fig:mass-sfr_miri_vs_nomiri} shows that this is
\replaced{typically}{generally} the case, where the MIRI data decrease
the average stellar mass and SFR for galaxies in our sample, typically
by a factor of order two.  We will explore the implications this has
on the evolution of the galaxy stellar-mass density in
Section~\ref{section:discussion}. 


\subsection{The Impact of MIRI on the Inferred Star-Formation History
(and the Mass Formed in Bursts)}\label{section:bursts}

~\deleted{Arguably, one of the most extreme star-formation histories
imaginable is the case where a galaxy forms in either one burst at
$z_f \rightarrow \infty$, or (slightly less extreme) in a series of
bursts extending back to that time.  When a burst forms, the stellar
population immediately  begins aging.  A burst at $z_f = \infty$ has
the oldest possible age at any subsequent time, and the smallest
amount of light (at a given mass), and therefore it would have a
maximal $M/L$ at any observed epoch.  For all these reasons, it is
\added{conceivable}\deleted{conceivably possible} to hide significant
amounts of stellar mass formed at earlier times (this is the
``outshining'' problem, e.g., \citealt{Papovich_2001, Dickinson_2003,
Papovich_2006,Finkelstein_2010,Pforr_2012,Conroy_2013}).}

~\deleted{One advantage of focusing on galaxies at high redshifts is that the
amount of \textit{time} for discrete episodes of star-formation (i.e.,
many individual bursts) is small given the age of the Universe (the
Universe has an age of 1 Gyr at $z=5.7$ and  500 Myr at $z=9.6$ for
the adopted cosmology).   This is shorter than the lifetimes of stars
of spectral type A and later.   The short age of the Universe,
combined with longer wavelengths probed by the MIRI 5.6 and
7.7~\micron\ data allow us to place tighter constraints on the mass
from earlier bursts in  high redshift galaxies than has been possible
previously.}

\added{Using the results from Section~\ref{section:analysis}, we
  also} explored the possibility that an
early burst \added{of star formation} could contribute
stellar mass to the galaxies in our sample, and how the \jwst/MIRI
data can improve the constraints on this population. \added{For this
  study, w}\deleted{W}e used the SED
fits to the galaxies in our sample using a star-formation history that
\addedtwo{includes} both the delayed-$\tau$ model \deleted{ (as in
Section~\ref{section:analysis} above)} and the burst of stars formed at
$z_f=100$.   The details of the other model parameters are listed in
Table~\ref{table:bagpipes}.

To quantify the amount of allowed stellar mass formed in the burst for
each galaxy, we use the Bayesian Information Criterion
\citep[BIC,][]{Bailer-Jones_2017}.  The BIC provides a criterion for
model selection in the case that one introduces a model with an
additional parameter\added{. I}\deleted{(i}n our case we are selecting between between two
models, one with a star-formation history that has a delayed-$\tau$
model only, and one with both the delayed-$\tau$ model and an early burst at
$z_f=100$\added{.}\deleted{).}     When comparing the two models, the BIC applies a
penalty term for the additional parameter \deleted{(}to determine if the
additional parameter improves the fit, or is overfitting the data\added{.}\deleted{).}

We use the BIC defined as (e.g., \citealt{Buat_2019}),
\begin{equation}\label{equation:bic}
  \mathrm{BIC} = \chi_0^2~ \times~ k \ln N,
\end{equation}
where $\chi_0^2$ is the minimum chi-squared from the
SED-fitting, $k$ is the number of independently fitted parameters (we
use $k=7$) and $N$ is the number of data points ($N = 8$ or 10, depending on if the MIRI [5.6] and [7.7] data are included or not).  
%

\begin{figure*}
 \gridline{
    \fig{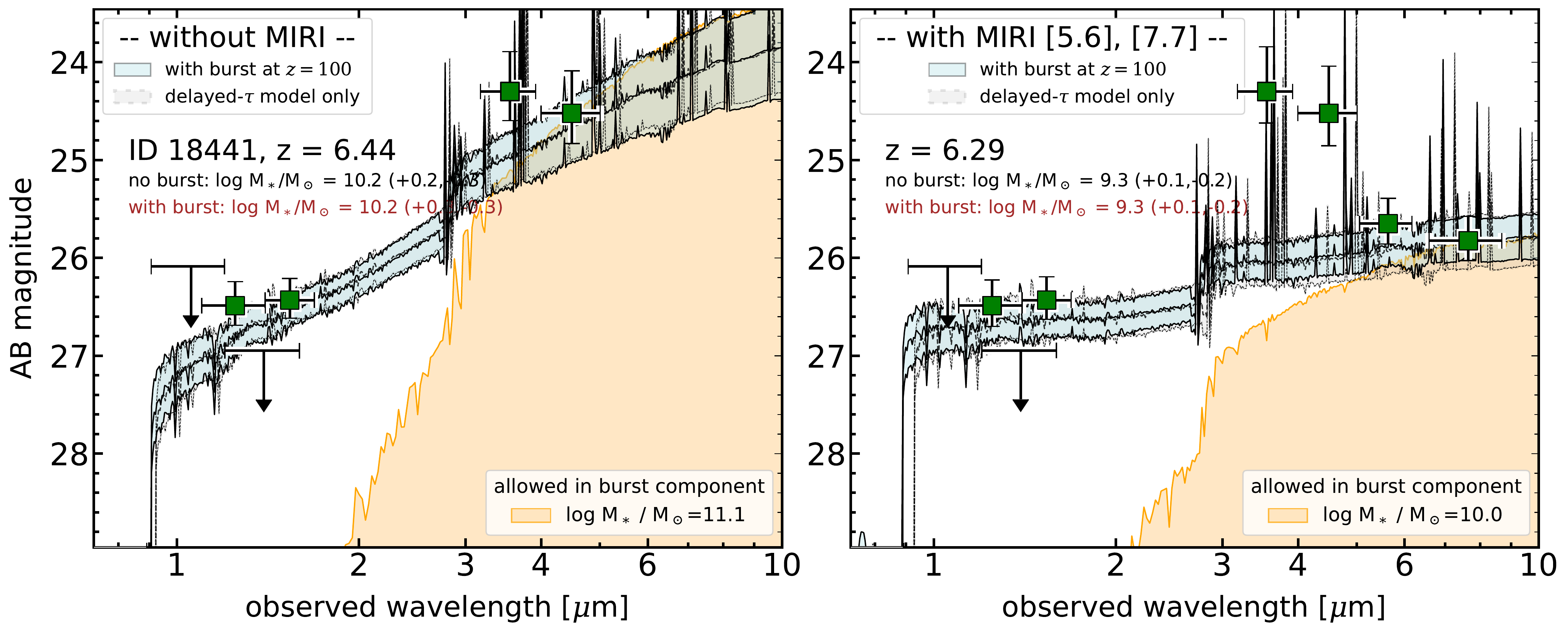}{0.82\textwidth}{}
  }
  \vspace{-18pt}
     \gridline{
    \fig{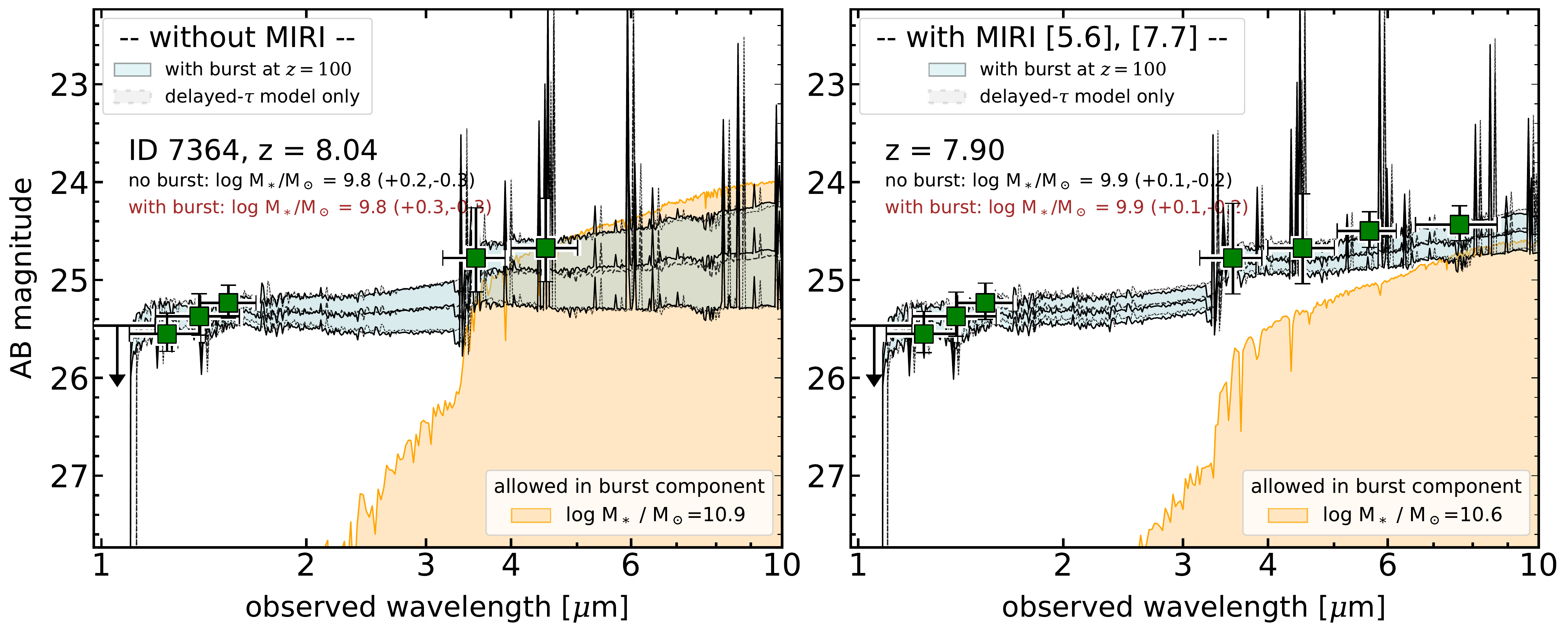}{0.82\textwidth}{}
  }
  \vspace{-18pt}
  \gridline{
    \fig{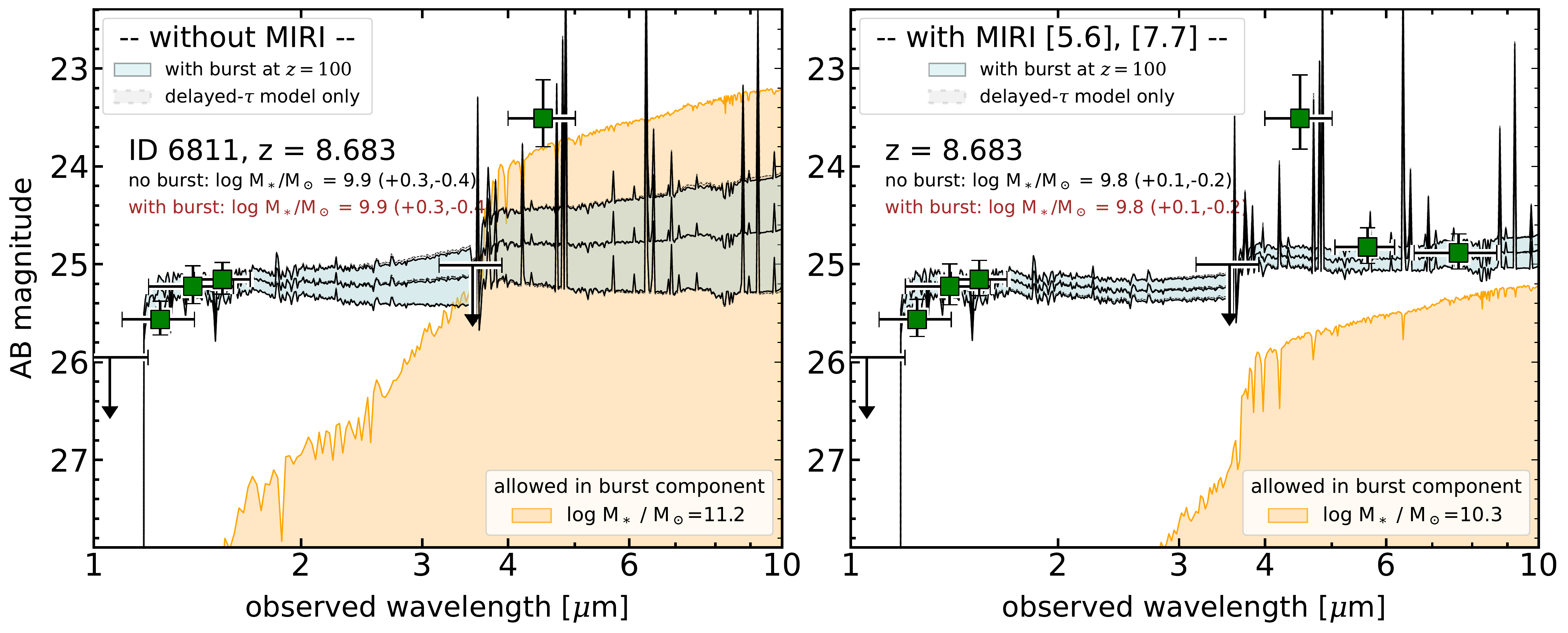}{0.82\textwidth}{}
  }
  \vspace{-12pt}
  \caption{ Example fits to SEDs for the galaxies in our sample,
comparing the results from fits that include an early  burst of
star-formation at $z_f=100$.   Data points and upper limits have the
same definitions as in Figure~\ref{fig:sed_results}.  In each panel,
the shaded regions show the stellar population fit to the SED using
the total model (cyan-shaded = delayed$-\tau$ plus the burst) and
delayed-$\tau$ model only (gray-shaded, these are identical to those
in Figure~\ref{fig:sed_results}; there is almost no difference between
the cyan- and gray-shaded models).  The tan-shaded region shows the
light permitted in the burst component.  The labels indicate the
amount of stellar mass in each component.  Each row shows the results
for one galaxy, where the Left panel shows the results that exclude
the MIRI F560W and F770W data, and the tight panel shows the results
including the MIRI F560W and F770W data. The complete figure set (28
images) is available online.\label{fig:sed_results_burst}}
\figsetstart
\figsetnum{\thefigure}
\figsettitle{SED fits including the MIRI data and excluding the MIRI
  data for all galaxies comparing the fits with and
  without bursts at $z_f=100$. }
\figsetgrpstart
\include{figset_spec_plots_burst.tex}
\figsetgrpend
\figsetend
\end{figure*}

Our goal is to quantify the upper limit on the stellar mass formed in
an early burst that could exist in our galaxies.  To do this, we
select models with bursts ($z_f = 100$) that are not excluded by the
BIC \deleted{criteria}\added{criterion}.   That is, for a given galaxy we identify all models
that satisfy $\chi^2 \leq \mathrm{BIC}$, where the BIC is defined in
Equation~\ref{equation:bic}, and $\chi^2$ is the fit for a given model
(this is similar to the approach adopted by \citealt{Buat_2019}).  For
each galaxy, we select the model with the highest stellar mass in the
$z_f=100$ burst from the subset of models that satisfy the BIC.  We
then take this as \added{an} upper limit on the stellar mass permitted in the
burst.  Comparing this upper limit on the stellar mass formed in
bursts to the range of stellar masses from the fits we find that the
limiting stellar mass from models that satisfy the BIC criteria
corresponds roughly to a 99.7\% upper limit on the mass (i.e., the
values we report correspond approximately to a $3\sigma$ upper limit
on the stellar mass).  Tables~\ref{table:bagpipes_wmiri} and
\ref{table:bagpipes_nomiri} list the upper limit on the stellar mass
in the burst component for all galaxies in the sample for the case
that we include and exclude the MIRI data, respectively. 

Figure~\ref{fig:sed_results_burst}  shows example SED fits for
galaxies in our sample, both with and without the bursts, for \addedtwo{both of the}
the cases that we include and exclude the MIRI 5.6 and 7.7~\micron\
data. The complete figure set (28 images) is available online.  The
examples shown in the Figure include a galaxy where the IRAC data are
bright relative to the MIRI data (ID 18441).  In this case,  without
the MIRI data the allowed stellar mass in the burst can reach $\log
M_\ast/M_\odot$ = 11.0, but adding MIRI reduces this by nearly an
order of magnitude.  For galaxy ID 7364, the MIRI data favor red
IRAC--to--MIRI colors.  Nevertheless, because the MIRI data constrain
the models at longer wavelengths, they also lower the amount of
stellar mass allowed in the burst:  without the MIRI data the burst
can include $\log M_\ast/M_\odot = 10.9$; when MIRI data are included
the mass in the burst declines by 0.3~dex (a factor of two).    For
galaxy ID 6811, the MIRI data show that because they constrain the SED
at longer wavelengths, the amount of stellar mass allowed in the burst
is reduced by a factor of 0.6~dex (nearly a factor of five).


\begin{figure*}
  \begin{center}
         \gridline{
           \includegraphics[width=0.85\textwidth]{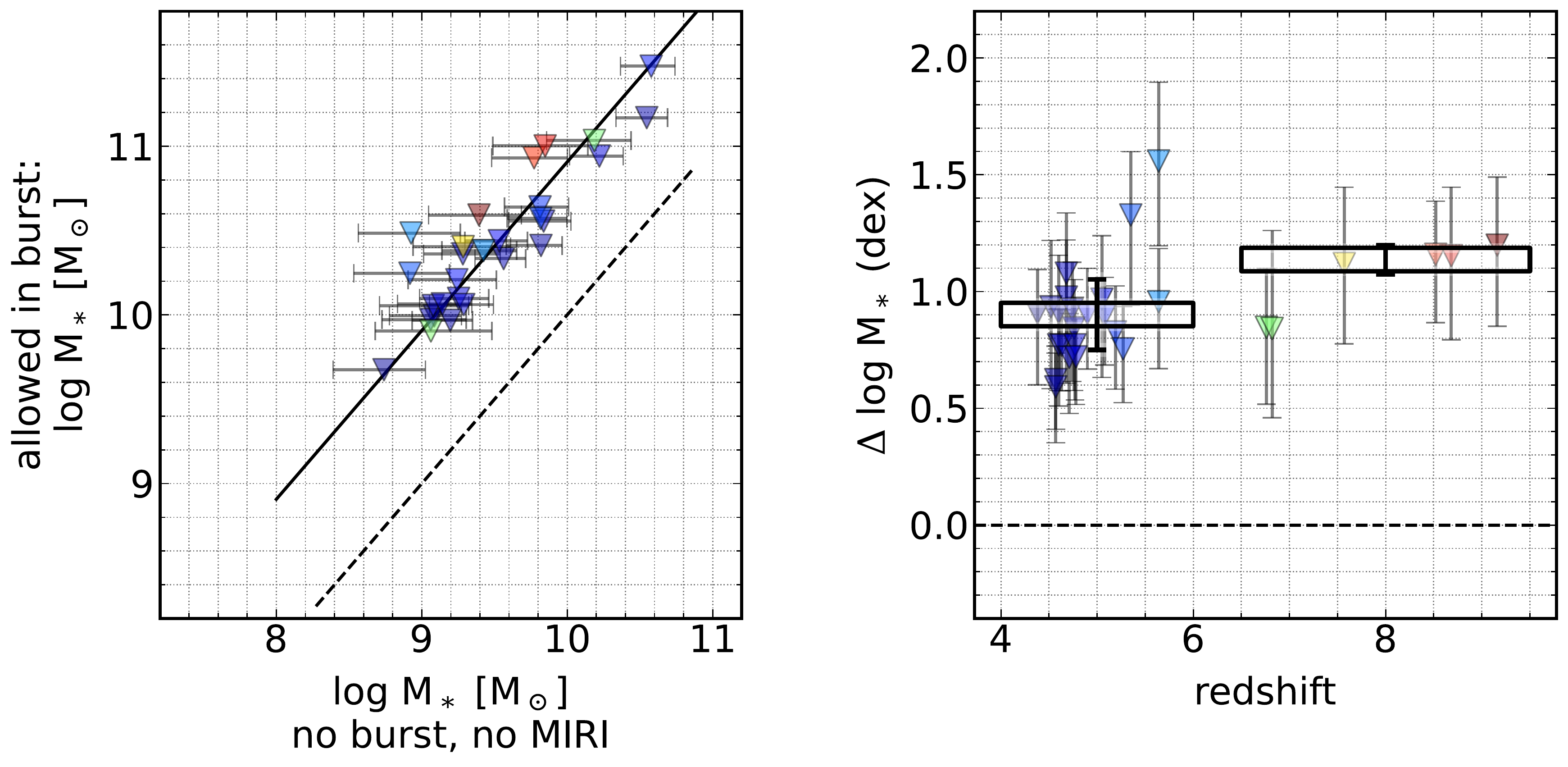}
           }
    \vspace{12pt}
    \gridline{
      \includegraphics[width=0.85\textwidth]{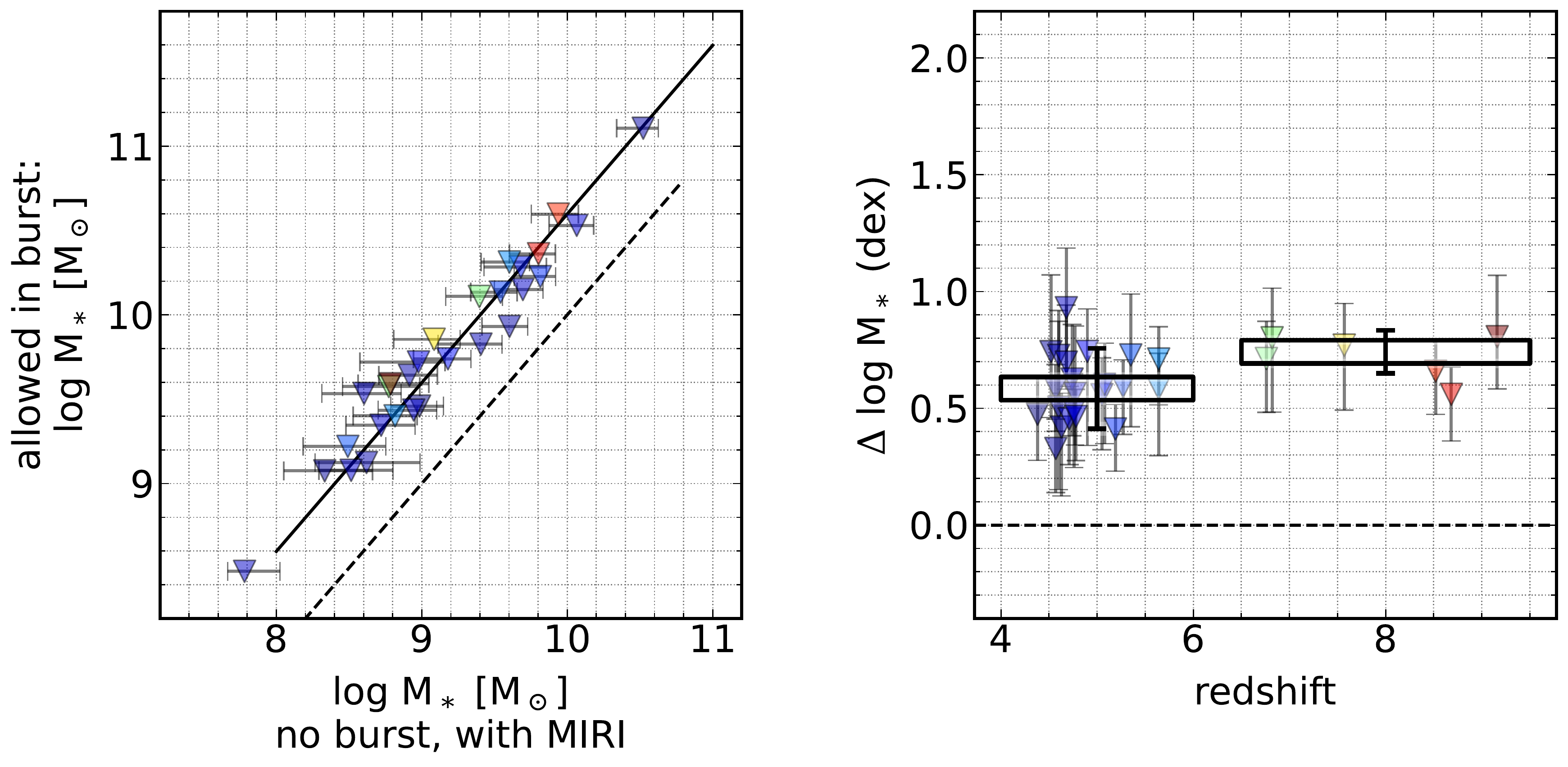}
      }
  \end{center}
\caption{ Comparison of stellar masses derived for galaxies with and
without early bursts (at $z_f=100$). The top row shows the results
that lack MIRI data.  The top-left panel shows the stellar masses
derived from the delay-$\tau$ models only (labeled ``no burst'')
compared to the models that include the burst (labeled ``allowed in
burst'').  The top-right panel shows the difference between the
stellar masses as a function of redshift.   The bottom row shows the
same results for the galaxies including the MIRI data.    In the left
panels, the dashed lines show the one-to-one relation and the solid
lines show the median offsets. In the right panels, the large
rectangles show the medians in two bins of redshift ($4 < z < 6$ and
$6 < z < 9$) these are given in Table~\ref{table:burst_ratios}.
Adding the MIRI data reduces the amount of stellar mass allowed in the
burst components.  \label{fig:burst_noburst}}
\end{figure*}

\begin{deluxetable}{l@{\hskip 24pt}cc@{\hskip 24pt}cc}
  \tablecolumns{5}
  \tablecaption{Ratio of the stellar mass allowed in models that include an
    early burst of star-formation (at $z_f=100$) to those that include
    only a delayed$-\tau$ model.\label{table:burst_ratios}}
\tablehead{ \colhead{}
& \multicolumn{4}{c}{$\log
M_\mathrm{\ast}(with~burst) - \log M_\mathrm{\ast}(no~burst)$}  \\
[-4pt]
\colhead{} &
\multicolumn{2}{c}{with MIRI data}  &
\multicolumn{2}{c}{no MIRI data} \\ [-2pt]
\colhead{Sample} &
\colhead{Median} & \colhead{Scatter~~~~} & \colhead{Median} &
\colhead{Scatter}}
\startdata
$4 < z < 6$ & 0.59 dex & 0.16 dex & 0.87 dex  & 0.23 dex \\
$6 < z < 10$ & 0.69 dex & 0.04 dex & 1.11 dex & 0.13 dex  \\
\enddata
\tablecomments{The values labeled ``with burst'' denote
the upper limit on the stellar mass for models that include bursts.
The values labeled ``no burst'' denote the stellar masses
for models that assume only a delayed-$\tau$ star-formation history.
The values ``with MIRI'' include the 5.6 and 7.7~\micron\ flux
densities from MIRI.}
\end{deluxetable} 

The addition of the MIRI F560W and
F770W data \addedtwo{reduces} the amount of stellar mass that can form in early
bursts. Figure~\ref{fig:burst_noburst} shows the change in stellar
mass for the case that the star-formation histories include only
delayed-$\tau$ models (labeled ``no burst'' in the figure) compared to
when an early burst of star formation at $z_f=100$ is included
(labeled ``allowed in burst'' in the figure).
Figure~\ref{fig:burst_noburst} shows the results for both the case
that the MIRI data are excluded (top row) and when the MIRI data are
included (bottom row).  For the fits that lack MIRI data, the amount
of stellar mass allowed in the burst is nearly an order of magnitude
higher than constrained in the delayed-$\tau$ models:  in this case
the median differences in the log of the stellar mass of the models
with early bursts and those with only delayed-$\tau$ models is 0.9~dex
at $4 < z < 6$ and 1.1~dex at $6 < z < 9$.   For the fits that include
the MIRI data, the amount of stellar mass allowed in the bursts is
significantly reduced:  the median differences in this case \addedtwo{are} 0.6~dex at
$4 < z < 6$ and 0.7~dex at $6 < z < 9$.    These values are listed in
Table~\ref{table:burst_ratios}.



\section{Discussion}\label{section:discussion}

\begin{figure}
  \begin{center}
    \includegraphics[width=0.45\textwidth]{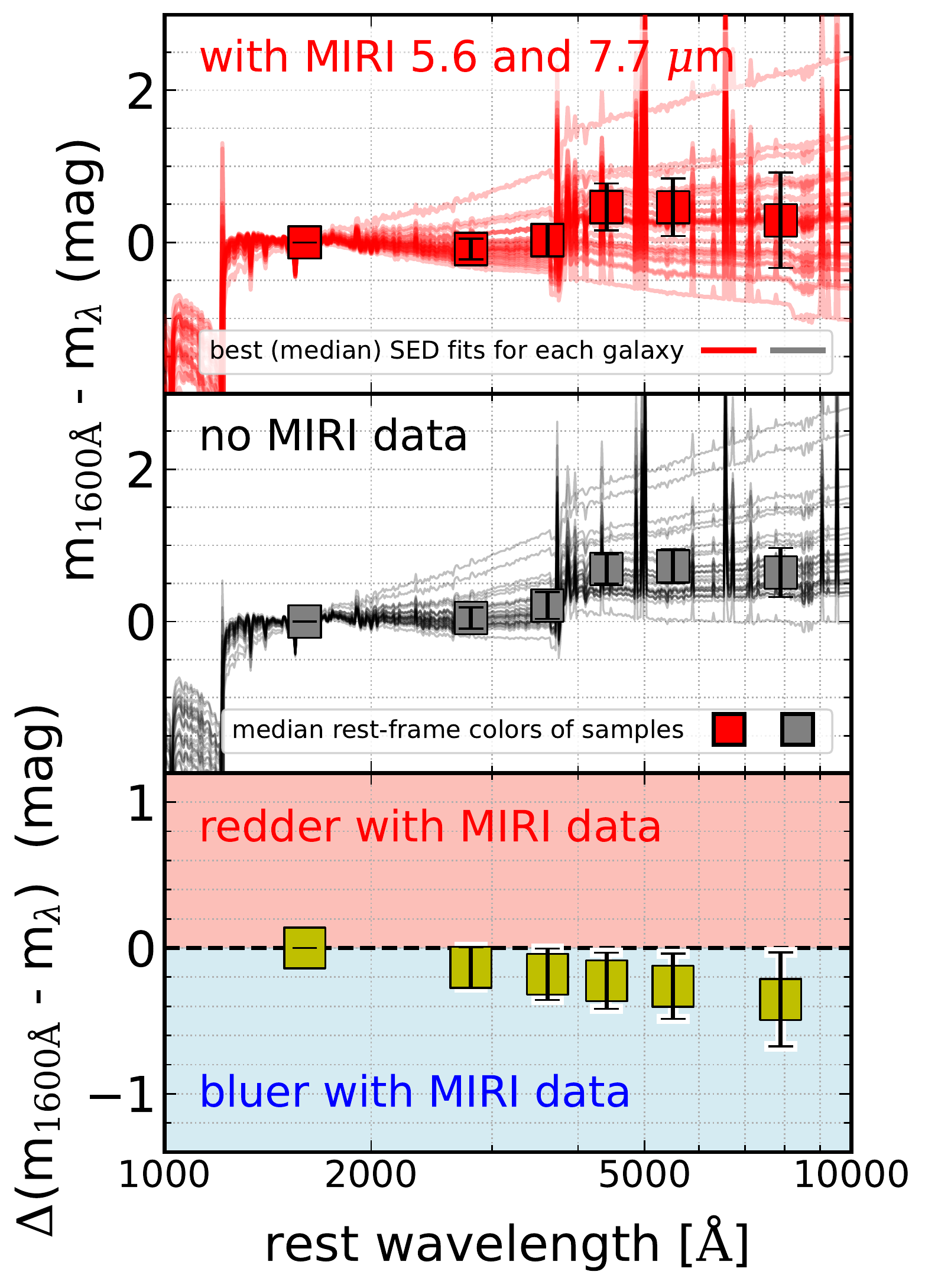}
  \end{center}
\caption{ Comparison of median, relative SED for each galaxy, both for
the case that MIRI data are used (top panel) and when the MIRI data
are excluded (middle panel).  The individual lines are the median SED
model fit to each galaxy in the sample, shifted to the rest-frame.
The large data points show the median rest-frame magnitude at
1600~\AA, 2800~\AA, and $U$, $B$, $V$, and $I$.  The error bars show
the scatter in the sample.  All models have been normalized to the
1600~\AA\ magnitude (which accounts for the lack of scatter at that
wavelength).  The bottom panel shows the difference in color ($\Delta
m$) between the models with and without the MIRI data\added{.  T}\deleted{(t}he error bars
show the range of the 16th-84th percentiles of the sample\added{.}\deleted{).}   The
change in the color at the reddest wavelengths probed (about the
rest-frame $I$-band) corresponds to a $\Delta m \approx 0.4$
mag.\label{fig:rel_colors}}
\end{figure}

\subsection{On the Colors, Stellar Masses, and Nebular Emission in
  Early Galaxies}\label{section:rest-frame_colors}

One of main findings in this paper is that the MIRI data favor bluer
colors in galaxies at $4 < z < 9$.  Figure~\ref{fig:rel_colors} shows
this by comparing the relative SED for each galaxy, both in the case
of including the MIRI data and without the MIRI data.  Including the
MIRI data reduces the derived rest-frame $I$-band light by
approximately $\Delta m_\mathrm{1600~\AA} - I \approx 0.4$ mag.
Inspecting Figure~\ref{fig:rel_colors} this appears to result from
many galaxies favoring bluer SEDs when the MIRI data are included.  In
other words, without the MIRI data, the SED is unconstrained at longer
wavelengths, and this allows for a greater range of SED shape (where
the median favors a solution which on average is redder).  Adding the
MIRI data shifts the likelihood to bluer populations for many
galaxies.  This has a major impact on the implied $M/L$, as the blue
rest-frame colors \added{imply}\deleted{implies} younger ages, lower
dust attenuation, or both. The fact that adding the MIRI data makes
the galaxies bluer largely explains the differences in the derived
stellar masses and SFRs observed in
Figure~\ref{fig:mass-sfr_miri_vs_nomiri}, where adding the MIRI data
\deleted{lower}\added{lowers} the stellar masses and SFRs compared
\added{to}\deleted{what the models favor} when the MIRI data are excluded.

Therefore, our interpretation of the MIRI data is that galaxies at
high redshifts ($z > 4$) are bluer than inferred from previous
studies.   This adds to other studies that find that galaxies at high
redshifts must have (very) blue colors.  Studies from the pre-\jwst\
era argued that galaxies at $z > 4$ show indications of declining
(i.e., steepening) UV spectral slopes with increasing redshift
\citep[e.g.,][]{Bouwens_2012a,Finkelstein_2012,Wilkins_2016,Bhatawdekar_2021}. These
conclusions have been reinforced by early \jwst\ imaging that shows
very blue colors among UV-selected galaxies
\citep[e.g.,][]{Nanayakkara_2022,Topping_2022}.  We find here that
including the MIRI 5.6 and 7.7~\micron\ data
\deleted{show}\added{provides} strong evidence that the stellar
populations are very blue in their rest-frame UV--to--$I$-band colors,
seemingly more so than inferred from these previous studies (as in
some cases the galaxies in our sample are identical to those in these
other studies).  Therefore, the MIRI data favor lower stellar masses
\textit{and} lower SFRs in distant galaxies. 

Adding the MIRI data also changes the interpretation of the strength
of nebular emission in the galaxies in the sample.  Many of the
galaxies show red colors between the \hst/WFC3 and \spitzer/IRAC data
(Figure~\ref{fig:sed_results}).   In the absence of MIRI data, the SED
fitting interprets these colors as a combination of higher dust
obscuration, older ages, \textit{and} strong nebular emission.  This
has been reported previously for individual galaxies and for stacked
samples at $z > 6$ \citep[e.g.,][]{Castellano_2017,Stark_2017,
DeBarros_2019,Hutchison_2019,Endsley_2021,Stefanon_2022}.  The fact
that \textit{without} the MIRI data the models allow for higher dust
obscuration and older stellar populations increases the $M/L$ and the
stellar masses.  The higher dust obscuration also leads to higher
SFRs.  However, including the MIRI data changes this interpretation
for the galaxy population.  Nebular emission lines appear to be the
primary explanation for the red \hst/WFC3 to \spitzer/IRAC colors
(while there are some galaxies where the MIRI data does not change
[e.g., galaxy IDs 6811 and 7364 in Figure~\ref{fig:sed_results}] this
does not change the main result \textit{in general}).   This means (1)
the nebular emission lines for high redshift galaxies must be strong
and (2) the stellar populations are blue.   Early \jwst\ results show
emission lines remain strong in high-redshift galaxies and impact the
reddest ($4-5$~\micron) bandpasses in NIRCam
\citep[e.g.,][]{Endsley_2022,Gimenez-Arteaga_2022,Santini_2022b,Topping_2022,Whitler_2022}.
The MIRI data appear necessary to extend the rest-frame
wavelength coverage to $>$7000~\AA, past the strongest of the nebular
emission features in the rest-frame optical.  

\subsection{Implications for Early Star-Formation and Stellar Masses
  in Galaxies} 

The difference between the delayed-$\tau$ star-formation history and
the model that includes bursts, while simplistic, arguably
\added{spans}\deleted{span} the gamut of available star-formation
histories of galaxies.  Simulations show that galaxies experience many
discrete bursts, but when averaged over long time baselines the
evolution is mostly smooth
\citep[e.g.,][]{Diemer_2017,Iyer_2019,Leja_2019}. Therefore the
smoothly evolving delayed-$\tau$ model represents the slowly evolving
evolution of galaxy star-formation histories.  This can reproduce the
bluest colors (and lowest $M/L$) of the stellar population.  Of
course, galaxies can experience a host of stochastic bursts through
changes in gas accretion or events that can \added{cause} sudden changes in the star
formation, possibly as a result of mergers (e.g., \citealt{Kartaltepe_2022}), or
\added{from} the \deleted{sudden} onset of strong feedback from an AGN
(e.g.,
\citealt{Wagner_2016}).  ``Bursts'' of star-formation from these
events will add stellar mass, but if these occur at $z < 100$ then \added{they}
will be younger, and will have $M/L$ lower than a model with a burst
at $z_f=100$.  As such, the models with a burst at $z_f=100$ have the
maximum $M/L$ and the oldest possible ages for a stellar population at
the observed redshift of each galaxy.  \replaced{Therefore}{For these
reasons,} the models with this burst represent a maximum stellar mass
possibly formed in these galaxies.

\added{An advantage of focusing on galaxies at high redshifts is that
the amount of \textit{time} for discrete episodes of star-formation
(i.e., many individual bursts) is small given the age of the Universe} \added{is}
\deleted{(the Universe has an age of} 1 Gyr at $z=5.7$ and  500 Myr at $z=9.6$
for the adopted cosmology\deleted{)}.   \added{This is shorter than the lifetimes of
stars of spectral type A and later.   The short age of the Universe
limits the maximum $M/L$ for stellar populations in these galaxies.
Combined with longer wavelengths probed by the MIRI 5.6 and
7.7~\micron\ data, this allows us to place tighter constraints on the
mass from earlier bursts in  high redshift galaxies than has been
possible previously. }

Using the MIRI data, we find that the maximum stellar mass from
bursts for the galaxies in our sample is \addedtwo{lower by 0.6 dex at $4
  < z < 6$ and 0.7~dex at $6 < z < 9$} (see
Table~\ref{table:burst_ratios}).  Furthermore, o\deleted{O}ur results
show that including MIRI reduces the amount of stellar mass allowed in
these models, by an order of magnitude in some cases.  In itself this
is interesting as nearly all galaxies show no direct evidence for such
early star-formation. Comparing the \textit{median} stellar masses of
galaxies when fit by the delayed-$\tau$ models only and those with the
delayed-$\tau$ models and the early burst at $z_f=100$ shows they
differ by $\mathrm{median}(\Delta M_\ast) \approx 0.1$~dex\added{, and the
16th-to-84th percentile range also is similar} (see the plots in
Figure~\ref{fig:sed_results_burst}).  We find no convincing cases in
our sample where the galaxies \textit{require} a burst at $z_f=100$ to
better fit their SEDs.   This would imply that galaxies do
\textit{not} experience early bursts of star-formation (or at least
such bursts do not form sufficient mass that we require them).

\added{We can compare our results for $4 < z < 9$ galaxies to those at lower
redshifts.  For example, \citet{Papovich_2001} found that with
$K$-band data for $z\sim 2-3.5$ galaxies, early bursts could contain
between 3 and 8 times the stellar mass of the younger populations that
dominated the SED in those galaxies.  The $K$-band is roughly the same
rest-frame wavelength for galaxies at $z\sim 2$ as MIRI is for
galaxies at $z\sim 8$.   With MIRI we find that the stellar mass in an
early burst can be $\sim4-5\times$ the stellar mass of the younger
population in $4 < z < 9$ galaxies.  The early burst mass is modestly
lower than the results from \citet{Papovich_2001}.   The reason for
this is a combination of effects.  At $4 < z < 9$, the
galaxies are blue, so they can hide more stellar mass in
an early burst.   However, because the universe is younger at $4 < z <
9$, any early burst of stars is also younger, with a lower $M/L$ and
therefore a lower stellar mass.  These effects offset, leading to only
a modest difference in the amount of stellar mass that could be
contained in bursts.  Nevertheless, this highlights the importance of
probing the SEDs of galaxies at rest-frame wavelengths longer than
$\sim$5000-7000~\AA\ to constrain these populations.}

Part of our findings could be impacted by biases \added{and
limitations}.  First there is selection bias: all the galaxies studied
here were selected in \hst/WFC3 data, and therefore required some
rest-frame UV emission above the \hst/WFC3 detection limit.
\replaced{It will be important to test for objects in, for example,
future \jwst/NIRCam-selected populations show evidence for early
bursts (especially \jwst-selected samples that lack \hst\
counterparts}{ It will be important to test if there is evidence for
early episodes of star-formation in galaxies with future \jwst\
observations.  This will be potentially very important for red
galaxies, including \jwst/NIRCam-selected galaxies that lack \hst\
counterparts} \citep{Glazebrook_2022,PG_2022}.  Second, very recent
work shows evidence for older stellar populations in the spatially
resolved colors of $6 < z < 9$ galaxies \citep{Gimenez-Arteaga_2022}.
As these issues become better constrained, then the ages of the
stellar populations in distant galaxies could begin to inform us about
when galaxies first form stars (see \citealt{Whitler_2022}).

\begin{figure}[t]
  \begin{center}
    \includegraphics[width=0.5\textwidth]{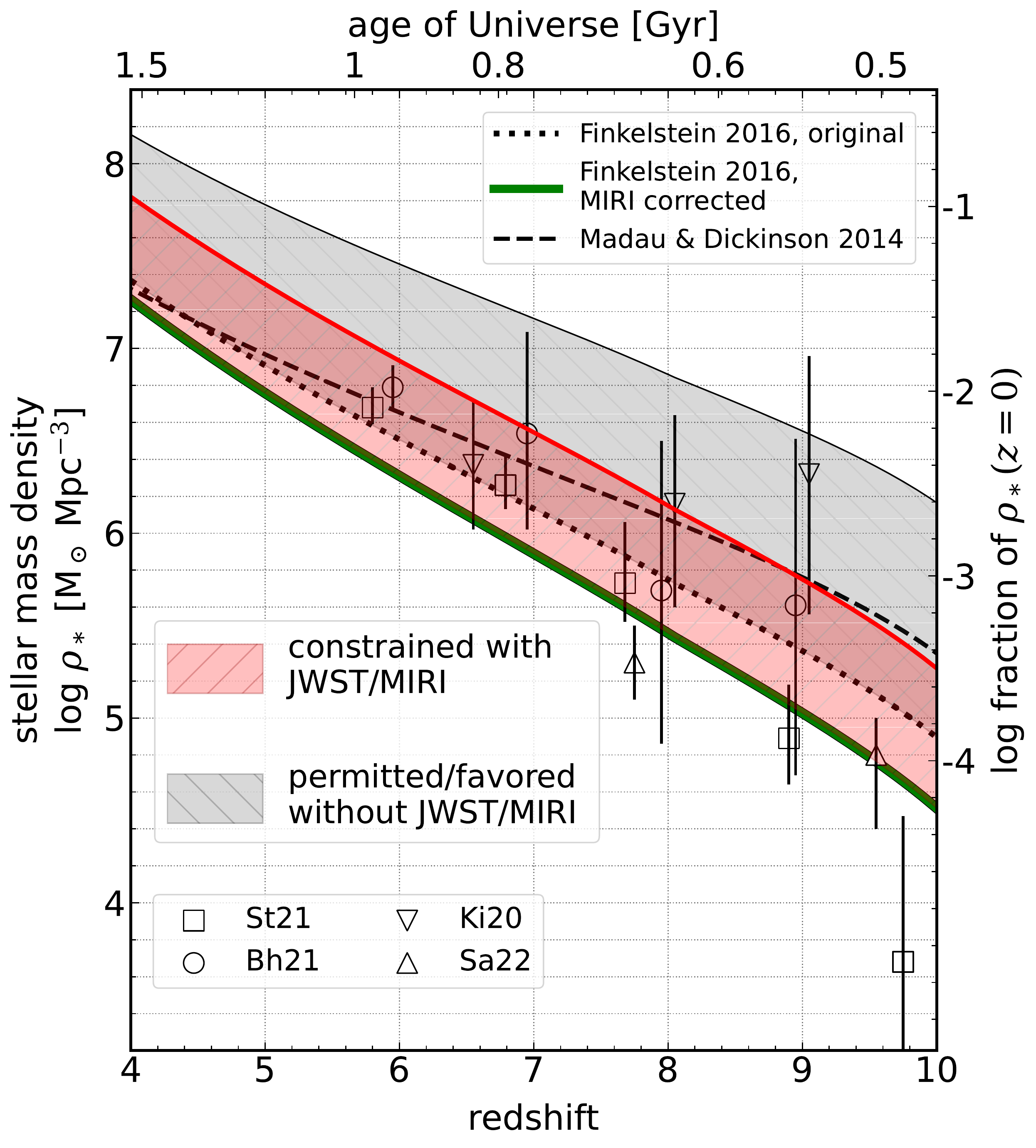}
  \end{center}
\caption{ Evolution of the cosmic stellar mass density, $\rho_\ast$,
in galaxies from $4 < z < 10$.  The lines show pre-{\it JWST}
constraints from \citet{Finkelstein_2016} and \citet{Madau_2014}.
The data points show recent measurements of $\rho_\ast$ at $z > 6$
from the literature (\citealt{Kikuchihara_2020} [Ki20], \citealt{Bhatawdekar_2021}
[Bh21], \citealt{Stefanon_2021} [St21], \citealt{Santini_2022}
[Sa22]), which largely follow the pre-\jwst\ constraints.
The shaded regions show maximally allowable stellar mass density
assuming galaxies experience a burst at $z=100$ followed by ``normal''
star-formation.   Constraints lacking {\it JWST}/MIRI coverage to
rest-frame 1~\micron\ allow for a stellar mass density that is up to
0.8 dex higher at $z=4$ and 1.4 dex higher at $z=10$.  Including MIRI
5.6 and 7.7~\micron\ data lowers the maximum allowed by up to a factor
of 5.  \label{fig:smd}}
\end{figure}

\subsection{Implications for Galaxy Growth} 

The question of how much mass is contained in galaxies is related to
the integral of the galaxies' star-formation histories.  This is
important because it contains the integrated record of how rapidly
galaxies acquire their baryons, and how efficiently they convert these
into stars.  This has already been discussed as an impossibly early
galaxy problem, where galaxies may have acquired too much stellar
mass:  in typical \jwst\ surveys galaxies should have less stellar
mass than a few times $10^{11}$~\msol\
\citep{Behroozi_2018,Boylan-Kolchin_2022}.

The effect of adding the MIRI data already \deleted{show
  that}\added{is that the} stellar masses
and SFRs derived for galaxies tend to be overestimated \added{when such data
are excluded}.  The median offsets for
the galaxies in our sample are 0.15 dex at $4 < z < 6$ and rise to
$\approx$0.3 dex at $6 < z < 9$ (Figure~\ref{fig:mass-sfr_miri_vs_nomiri}  and
Table~\ref{table:offsets}).  Assuming that these (median) offsets
apply to previous estimates of galaxy stellar masses, then it implies that measurements of the
cosmic SFR density (SFRD, which is the average SFR in all galaxies per
co-moving volume element) are similarly biased to higher values at
these redshifts.

Figure~\ref{fig:smd} shows the impact of these lower SFRs on the
cosmic \textit{stellar-mass density}, $\rho_\ast$.  Firstly, the
figure shows $\rho_\ast$ derived from the integral of the SFRD for two
empirical models calibrated against measurements from the literature
\citep{Madau_2014,Finkelstein_2016}.  \citet{Finkelstein_2016} shows
that these models are consistent with a compilation of measurements of
$\rho_\ast$ from the literature at $4 < z < 10$ prior to the launch of
\jwst, \citep{Oesch_2014,Duncan_2014,Grazian_2015,Song_2016}.  The
thick line in Figure~\ref{fig:smd} labeled ``MIRI corrected'' shows
the empirical model of \citeauthor{Finkelstein_2016} corrected by the
offsets of the SFRs derived in Table~\ref{table:offsets}.  To derive
these corrections we have interpolated the results from
Table~\ref{table:offsets} assuming median redshifts of $z=5$ and 8 for
the derived offsets in the two redshift bins.
\replaced{Parenthetically, a}{A}lthough we use the offsets derived
from the SFRs, using those for the stellar masses changes the results
by $\approx$0.1 dex.  We note, however, that because the MIRI data
imply offsets in the SFRs of galaxies, \deleted{the}\added{they}
similarly lower the values of the cosmic SFRD at $z > 4$. 

Secondly, the MIRI data improve the constraints \added{on} the amount of stellar
mass possible in early bursts of star-formation
(Section~\ref{section:bursts}).  This is illustrated by the shaded
regions in Figure~\ref{fig:smd}.   To \added{derive}\deleted{derived} the area in the shaded
swaths, we applied the ratio between the mass permitted in early
\added{bursts}\deleted{burst} at $z_f=100$ \deleted{given} to the fiducial value (listed in
Table~\ref{table:burst_ratios}), \added{and} interpolated \added{them over}\deleted{in} redshift as above.
Therefore, the effects of adding the MIRI data both lower SFR (and
stellar masses)  and limit the total stellar mass allowed in early
bursts.  The combination of these effects reduces the upper bound on
the total cosmic stellar mass density allowed by the data by 0.4 dex
at $z=4$ and by 1.0 dex at $z=9$.    As illustrated in
Figure~\ref{fig:smd}, this implies that the \jwst/MIRI data have
constrained the stellar mass in galaxies at $z=9$ to be less than
0.1\% that of the present-day value ($\rho_\ast(z=0)$).

\subsection{\added{Implications for the Nature of Early Galaxies}\deleted{Random Musings}}

The fact that galaxies have \added{bluer colors than inferred
from}\deleted{bluer than} previous constraints has other consequences
(i.e., they are ``bluer than they appear''). Our findings
dovetail\deleted{and likely dovetails} with other recent results from
\jwst\ NIRCam imaging. The results in this \textit{Paper} are likely
only the first foray into the properties of distant galaxies using the
longer wavelength data available from MIRI.   Future studies will be
able to combine both NIRCam and MIRI imaging,
providing \jwst--quality data from
$0.8-10$~\micron, with NIRSPec spectroscopy,  which will improve the
constraints \deleted{in this \textit{Paper}}\added{presented here}.
Already there are \deleted{indicates}\added{indications that} using
\jwst/NIRCam data only that the number density of luminous galaxies at
$z > 10$ may be much higher than predictions
\citep[e.g.,][]{Bouwens_2022,Donnan_2022,Finkelstein_2022c,Harikane_2022,Naidu_2022,Robertson_2022}.
One \replaced{explanation for}{interpretation of} these discoveries
\replaced{could be}{is} that the UV--luminosity per unit stellar mass
(the UV ``efficiency'') may be higher than \deleted{our} models
predict.  This could be a result of changes in the stellar populations
\deleted{(a shift toward bluer/harder ionizing spectra) or a change in
the stellar IMF that is weighted toward higher-mass stars} such
that they produce more ionizing photons per unit mass.  This is
expected both for the case that the galaxies contain high-mass, very
metal-poor stars \citep{Olivier_2022} or if their stellar populations
have an IMF weighted toward high-mass stars \citep{Raiter_2010}.
Importantly, either of these effects would further \textit{decrease}
the $M/L$ ratio of the stellar poplation, and thus lead to even lower
stellar masses than we have measured with the MIRI data.  Therefore,
it will be important to constrain the nature of the stars in these
galaxies.  In the meantime, the MIRI data have better constrained the available light in stars at
these early epochs, finding that \addedtwo{they} contain at least three
times less ``light'' at rest-frame $0.7-1$~\micron\ than previously
known.

~\deleted{However, a change in the IMF or in the UV efficiency does not change
the conclusion that the light from older stars could be lost in the
glare of the more recently formed stars, nor would it change the
conclusion that there is less light in general at longer wavelengths
from these galaxies. Even if the lower-mass
cutoff of the IMF is higher, e.g., $>50 M_\odot$, \citep{Raiter_2010}
then after $\sim$10~Myr the \textit{mass} left in such stars would be
effectively zero.   Therefore, any of these effects would further
lower the galaxy $M/L$ values, and therefore lower the stellar masses
by even more than what we have measured with the MIRI data.  (The only
way to \textit{add} more stellar mass in galaxies at this epoch is if
there is a substantial population of galaxies at $z > 6$ that are
undetected in \hst, which would be an important discovery for \jwst).
Regardless, the MIRI data have better constrained the available light in stars at these
early epochs and shown that galaxies contain more than three times
less ``light'' at rest-frame $0.7-1$~\micron\ than previously known.}


\section{Conclusions}\label{section:summary}

In this \textit{Paper} we have presented  results from CEERS  on the
stellar population parameters for 28 galaxies with redshifts $4 < z <
9$ using new imaging data from \jwst/MIRI at 5.6 and 7.7~\micron.  Our
galaxy sample was detected in deep data from \hst/WFC3 and ACS and
has observations from \textit{Spitzer}/IRAC at 3.6 and 4.5~\micron.
The MIRI 5.6 and 7.7~\micron\
data extend the coverage of the rest-frame spectral energy
distribution to nearly 1 micron for galaxies in this redshift range.
We use these data to study the improvements in the stellar masses and
SFRs of the galaxies at these redshifts when the MIRI data are
included.    Our main results are the following.

\begin{enumerate}

\item Galaxies at $4 < z < 9$ have bluer rest-frame UV--$I$-band
colors ($m_\mathrm{1600~\AA} - I$) when using the MIRI \addedtwo{data}
compared to when \addedtwo{they} are excluded.  Using the MIRI data we model the
SEDs using stellar population synthesis models (with \bagpipes).  When
we compare the average galaxy SED (Figure~\ref{fig:rel_colors}) we
find that models that include the MIRI data are (on average) $\Delta
(m_\mathrm{1600~\AA} - I) \approx 0.4$ mag bluer in their rest-frame
colors compared to models that exclude the MIRI data.

\item Galaxies generally have lower stellar masses and SFRs when the
MIRI data are included.  For the majority of the galaxies
(Figure~\ref{fig:mass-sfr_miri_vs_nomiri})   adding the MIRI data
\textit{reduces} the derived stellar masses by 0.25~dex at $4 < z < 6$
\deleted{(a factor of 1.8)}  and by 0.38~dex at $6  < z < 9$ \deleted{(a factor of 2.4)}.
Similarly including the MIRI data \textit{reduces} the SFRs by 0.15
dex at $4 < z < 6$ \addedtwo{Because the impact is larger
on the stellar
masses than on the SFRs, the specific SFRs will be increased by
approximately $\Delta \log \mathrm{SFR} - \Delta \log M_\ast \approx
0.1$ dex when the MIRI data are included.}

 There are multiple reasons the stellar masses and SFRs are lowered
when we include the MIRI data.  The first reason is that the galaxies
are blue,  and the fits favor models with lower dust attenuation and
models with lower $M/L$ in general.  The second reason is that in many
cases the IRAC 3.6 and 4.5~\micron\ data probe the rest-frame optical,
and these show indications of containing light from strong emission
lines (e.g., redshifted \hb\ + \oiii, \ha+\nii, \oii, etc.).  These
boost the flux in these bands.  In the absence of MIRI data the models
can not determine if the red rest-frame UV-optical colors are a result
of dust attenuation, older stellar populations, or strong emission
lines (or all of them).  The parameter constraints then give more
weight (\deleted{probably}\added{probability} density) to models with
higher stellar masses and SFRs.  When the MIRI data are included,
\deleted{then}\added{they} probe more of the stellar continuum at
$>$7000~\AA\ rest-frame.  The model fits that include
\added{the}\deleted{he} MIRI data then show the dominant effect in the
rest-frame optical are strong nebular emission lines.  This problem
will persist for models that use NIRCam data as it also is limited to
wavelengths less than 5~\micron, but this can be tested with
forthcoming datasets. 

\item The amount of stellar mass that could have formed in early
bursts is lower.   We estimated the amount of stellar mass formed by
using a star-formation history that includes an early burst (at
$z_f=100$) in addition to a smoothly evolving component.   A stellar
population formed at this early time would fade and redden with time,
and it would have the highest $M/L$ at any subsequent time and
therefore \deleted{representations}\added{represents} an upper limit
on the amount of mass that could exist in these galaxies. The MIRI
data \added{improve}\deleted{improves} the constraint on this stellar
population by probing longer wavelengths \deleted{(}where the impact from this
stellar population is more pronounced\deleted{)}.
Figure~\ref{fig:burst_noburst} shows that without the MIRI data, the
amount of stellar mass in this population can be as much as 0.9~dex
higher at $4 < z <6$ \deleted{(a factor of 7)} and as much as 1.1~dex higher at
$6 < z < 9$ \deleted{(a factor of $>$10)}.  Including the MIRI data, these drop
to 0.6~dex at $4 < z < 6$ \deleted{(a factor of 4)} and 0.7~dex at $6 < z < 9$
\deleted{(a factor of 5)}.  Therefore, adding the MIRI reduces the amount of
mass in early bursts by a factor of order 2 \deleted{(}compared to when no MIRI
data are used\deleted{)}.

  \item Our analysis of the MIRI 5.6 and 7.7~\micron\ therefore
provides  evidence that there is \textit{less} star-formation in
distant galaxies than found in previous studies, \deleted{(}because
the SFRs and stellar masses are lowered\deleted{)}.  The MIRI data
also reduce the limits on the amount of stellar mass possibly formed
at early times.  The combination of these results has implications for
the evolution of the cosmic stellar-mass density, $\rho_\ast$.  We
showed (Figure~\ref{fig:smd}) that applying our results \added{lowers}\deleted{to the galaxy
population shows that} the amount of stellar mass density in galaxies
at $z=9$ \added{to be}\deleted{is} less than 0.1\% of the present day, $z=0$, value. This is
an order of magnitude lower than implied by previous studies
(i.e,. pre-\jwst).

\end{enumerate}

\begin{acknowledgements}

We wish to thank everyone that brought \jwst\ to fruition.    We also
thank our other colleagues in the CEERS collaboration for their
hard work and valuable contributions on this project.    CP thanks Marsha and Ralph
Schilling for generous support of this research.   Portions of this
research were conducted with the advanced computing resources provided
by Texas A\&M High Performance Research Computing (HPRC,
\url{http://hprc.tamu.edu}).  This work benefited from support from
the George P. and Cynthia Woods Mitchell Institute for Fundamental
Physics and Astronomy at Texas A\&M University.    This work
acknowledges support from the NASA/ESA/CSA James Webb Space Telescope
through the Space Telescope Science Institute, which is operated by
the Association of Universities for Research in Astronomy,
Incorporated, under NASA contract NAS5-03127. Support for program
No. JWST-ERS01345 was provided through a grant from the STScI under
NASA contract NAS5-03127.

\end{acknowledgements}

\begin{deluxetable*}{ccccccccccccccccccc}
  \tabletypesize{\footnotesize}
  \rotate
\tablecolumns{19}
\tablewidth{0pt}
\tablecaption{Observed Properties of the Galaxy Sample}. \label{table:sample}
\tablehead{\colhead{ID} & \colhead{R.A.\ (J2000)} & \colhead{Decl.\ (J2000)} &
  \colhead{F160} & \colhead{E160} & \colhead{F560} & \colhead{E560} &
  \colhead{F770} & \colhead{E770} & \colhead{$z_\mathrm{phot}$} &
\colhead{$z_\mathrm{16}$} &
  \colhead{$z_\mathrm{84}$} & \colhead{$z_\mathrm{spec}$} & \colhead{$\mathcal{P}(z=4)$} &
  \colhead{$\mathcal{P}(z=5)$} & \colhead{$\mathcal{P}(z=6)$} & \colhead{$\mathcal{P}(z=7)$} &
  \colhead{$\mathcal{P}(z=8)$} & \colhead{$\mathcal{P}(z=9)$} \\ [-8pt]
\colhead{ } & \colhead{(deg)} & \colhead{(deg)} & \colhead{$\mathrm{nJy}$} & \colhead{($\mathrm{nJy}$)} & \colhead{($\mathrm{nJy}$)} & \colhead{($\mathrm{nJy}$)} & \colhead{($\mathrm{nJy}$)} & \colhead{($\mathrm{nJy}$)} & \colhead{ } & \colhead{ } & \colhead{ } & \colhead{ } & \colhead{ } & \colhead{ } & \colhead{ } & \colhead{ } & \colhead{ } & \colhead{ }}
\startdata
5090 & 215.04973 & 52.89656 & 102.0 & 12.5 & 158.8 & 17.9 & 148.0 & 14.4 & 4.38 & 4.18 & 4.56 & -1.000 & 0.68 & 0.27 & 0.00 & 0.00 & 0.00 & 0.00 \\
11329 & 215.04086 & 52.90623 & 49.4 & 6.9 & 63.5 & 20.4 & 44.5 & 15.7 & 4.47 & 4.15 & 4.65 & -1.000 & 0.55 & 0.39 & 0.00 & 0.00 & 0.00 & 0.00 \\
34813 & 214.97597 & 52.920297 & 82.2 & 16.2 & 67.0 & 16.7 & 43.9 & 19.4 & 4.52 & 4.17 & 4.59 & -1.000 & 0.52 & 0.36 & 0.00 & 0.00 & 0.00 & 0.00 \\
7600 & 215.04415 & 52.898731 & 564.7 & 29.4 & 3283.1 & 32.4 & 4101.7 & 24.8 & 4.57 & 4.13 & 4.65 & -1.000 & 0.62 & 0.38 & 0.00 & 0.00 & 0.00 & 0.00 \\
13389 & 215.03793 & 52.909329 & 453.9 & 29.7 & 578.3 & 26.9 & 661.0 & 22.0 & 4.57 & 4.43 & 4.61 & -1.000 & 0.38 & 0.61 & 0.00 & 0.00 & 0.00 & 0.00 \\
37703 & 214.99094 & 52.924279 & 116.7 & 11.8 & 278.9 & 30.4 & 61.0 & 17.5 & 4.60 & 4.47 & 4.70 & -1.000 & 0.22 & 0.78 & 0.00 & 0.00 & 0.00 & 0.00 \\
15445 & 215.02689 & 52.907215 & 124.5 & 14.4 & 187.8 & 20.2 & 116.6 & 14.9 & 4.60 & 4.21 & 4.84 & -1.000 & 0.24 & 0.62 & 0.00 & 0.00 & 0.00 & 0.00 \\
41564 & 214.98708 & 52.912734 & 266.4 & 13.4 & 481.6 & 20.8 & 276.8 & 15.7 & 4.63 & 4.56 & 4.71 & -1.000 & 0.03 & 0.97 & 0.00 & 0.00 & 0.00 & 0.00 \\
45145 & 215.00878 & 52.919869 & 101.2 & 15.2 & 108.6 & 29.6 & 69.0 & 19.3 & 4.68 & 4.37 & 5.20 & -1.000 & 0.15 & 0.74 & 0.03 & 0.00 & 0.00 & 0.00 \\
14913 & 215.02081 & 52.901523 & 78.3 & 14.1 & 72.8 & 17.8 & 9.4 & 8.9 & 4.68 & 4.47 & 4.94 & -1.000 & 0.19 & 0.81 & 0.00 & 0.00 & 0.00 & 0.00 \\
42638 & 214.97762 & 52.90349 & 547.0 & 31.8 & 711.4 & 28.5 & 596.3 & 23.4 & 4.71 & 4.57 & 4.77 & -1.000 & 0.02 & 0.90 & 0.00 & 0.00 & 0.00 & 0.00 \\
41375 & 215.00283 & 52.924312 & 109.2 & 15.4 & 47.2 & 18.1 & 119.1 & 14.8 & 4.74 & 4.60 & 5.13 & -1.000 & 0.07 & 0.92 & 0.01 & 0.00 & 0.00 & 0.00 \\
35896 & 214.98522 & 52.924266 & 96.3 & 10.2 & 114.5 & 14.8 & 110.0 & 15.3 & 4.76 & 4.35 & 5.33 & -1.000 & 0.06 & 0.73 & 0.07 & 0.00 & 0.00 & 0.00 \\
13179 & 215.04260 & 52.911968 & 736.4 & 23.7 & 1355.2 & 26.8 & 1519.1 & 21.5 & 4.78 & 4.72 & 4.91 & -1.000 & 0.00 & 1.00 & 0.00 & 0.00 & 0.00 & 0.00 \\
19180 & 215.03170 & 52.919632 & 773.3 & 35.6 & 749.1 & 26.0 & 787.7 & 19.0 & 4.93 & 4.86 & 4.95 & 5.077$^\dag$ & 0.00 & 1.00 & 0.00 & 0.00 & 0.00 & 0.00 \\
37653 & 214.99190 & 52.925053 & 103.1 & 12.5 & 154.0 & 22.2 & 108.4 & 18.5 & 4.95 & 4.72 & 5.59 & 4.899$^\ddag$ & 0.06 & 0.71 & 0.22 & 0.00 & 0.00 & 0.00 \\
18449 & 215.02314 & 52.912683 & 69.0 & 10.0 & 185.0 & 24.2 & 165.3 & 17.6 & 5.05 & 4.30 & 5.57 & -1.000 & 0.14 & 0.59 & 0.20 & 0.00 & 0.00 & 0.00 \\
41545 & 215.00312 & 52.924103 & 384.9 & 18.3 & 749.2 & 21.5 & 656.6 & 17.5 & 5.19 & 5.02 & 5.44 & -1.000 & 0.00 & 0.90 & 0.10 & 0.00 & 0.00 & 0.00 \\
7818 & 215.02758 & 52.887744 & 289.9 & 21.4 & 421.4 & 24.5 & 399.5 & 20.9 & 5.27 & 4.74 & 5.46 & -1.000 & 0.02 & 0.86 & 0.12 & 0.00 & 0.00 & 0.00 \\
12773 & 215.03057 & 52.902606 & 56.4 & 12.2 & 17.0 & 7.5 & 71.0 & 14.1 & 5.35 & 4.87 & 5.43 & -1.000 & 0.00 & 0.90 & 0.09 & 0.00 & 0.00 & 0.00 \\
24007 & 214.95185 & 52.928275 & 47.1 & 4.9 & 107.8 & 18.8 & 56.7 & 12.6 & 5.64 & 5.16 & 5.79 & -1.000 & 0.00 & 0.46 & 0.51 & 0.00 & 0.00 & 0.00 \\
49365 & 215.00994 & 52.910669 & 143.9 & 14.1 & 408.5 & 26.8 & 541.0 & 20.6 & 5.64 & 5.03 & 5.82 & -1.000 & 0.01 & 0.38 & 0.50 & 0.00 & 0.00 & 0.00 \\
18441 & 215.03209 & 52.918972 & 97.2 & 11.5 & 199.6 & 28.2 & 169.9 & 22.0 & 6.76 & 5.88 & 7.24 & -1.000 & 0.00 & 0.01 & 0.36 & 0.48 & 0.07 & 0.00 \\
39096 & 214.98901 & 52.919652 & 64.1 & 8.7 & 159.5 & 17.0 & 57.7 & 11.2 & 6.82 & 6.67 & 7.08 & -1.000 & 0.00 & 0.00 & 0.02 & 0.97 & 0.01 & 0.00 \\
12514 & 215.03716 & 52.906712 & 71.6 & 13.2 & 142.3 & 17.8 & 50.4 & 13.6 & 7.57 & 6.90 & 8.41 & -1.000 & 0.00 & 0.00 & 0.02 & 0.39 & 0.44 & 0.13 \\
7364 & 215.03561 & 52.892208 & 292.6 & 17.5 & 575.7 & 28.4 & 611.4 & 23.1 & 8.52 & 7.47 & 8.68 & -1.000 & 0.00 & 0.00 & 0.00 & 0.17 & 0.52 & 0.31 \\
6811 & 215.03538 & 52.890666 & 314.5 & 13.3 & 426.8 & 21.6 & 404.0 & 16.9 & 8.93 & 8.60 & 9.05 & 8.683$^\ast$ & 0.00 & 0.00 & 0.00 & 0.00 & 0.08 & 0.92 \\
26890 & 214.96754 & 52.932966 & 133.3 & 8.9 & 101.9 & 21.4 & 60.8 & 16.8 & 9.16 & 8.46 & 9.24 & -1.000 & 0.00 & 0.00 & 0.00 & 0.00 & 0.06 & 0.83 \\
 \enddata
%
%
\tablecomments{References for spectroscopic redshifts:
  $^\ddag$ Stawinski, S., et al., in prep; $^\dag$ \todo{WERLS}; $^\ast$\cite{Zitrin_2015}.  }
\vspace{-6pt}
\end{deluxetable*}

\begin{deluxetable*}{lccc|ccc|ccc|c}
  \tabletypesize{\footnotesize}
\tablecolumns{11}
\tablewidth{6in}
\tablecaption{Derived Stellar masses, SFRs, and Redshifts including
  the MIRI [5.6] and [7.7] data. } \label{table:bagpipes_wmiri}
\tablehead{
  \colhead{} & \multicolumn{3}{c}{redshift}  &
  \multicolumn{3}{c}{$\log M_\ast / M_\odot$ (dex)} &
  \multicolumn{3}{c}{$\log \mathrm{SFR} / M_\odot~\mathrm{yr}^{-1}$ (dex)} &
  \colhead{``Burst'' Mass} \\[-6pt]
  \colhead{ID} & \colhead{$z_{50}$} & \colhead{$z_{16}$} &
    \colhead{$z_{84}$} &
    \colhead{$\log M_{50}$}  & \colhead{$\log M_{16}$} &
    \colhead{$\log M_{84}$} &
    \colhead{log SFR$_{50}$} & \colhead{log SFR$_{16}$} & \colhead{log
      SFR$_{84}$} &
    \colhead{$\log M_\ast/M_\odot$ (dex)}\\[-6pt]
  \colhead{(1)} & \colhead{(2)} & \colhead{(3)} &
    \colhead{(4)} &
    \colhead{(5)}  & \colhead{(6)} &
    \colhead{(7)} &
    \colhead{(8)} & \colhead{(9)} & \colhead{(10)} &
    \colhead{(11)}
}
\startdata
5090 & 4.33 & 4.19 & 4.45 & 9.01 & 8.84 & 9.16 & 0.59 & 0.50 & 0.69 & 9.55 \\
6811 & \ldots & \ldots & \ldots & 9.76 & 9.57 & 9.88 & 1.60 & 1.49 & 1.72 & 10.34 \\
7364 & 8.07 & 7.35 & 8.55 & 9.91 & 9.72 & 10.04 & 1.70 & 1.57 & 1.82 & 10.61 \\
7600 & 4.41 & 4.22 & 4.56 & 10.54 & 10.36 & 10.65 & 2.09 & 1.97 & 2.20 & 10.93 \\
7818 & 5.19 & 4.93 & 5.39 & 9.51 & 9.30 & 9.64 & 1.19 & 1.10 & 1.29 & 10.23 \\
11329 & 4.45 & 4.31 & 4.59 & 8.48 & 8.14 & 8.73 & 0.27 & 0.17 & 0.37 & 9.17 \\
12514 & 7.69 & 7.03 & 8.21 & 8.78 & 8.47 & 9.02 & 0.70 & 0.53 & 0.85 & 9.45 \\
12773 & 5.08 & 4.91 & 5.24 & 8.01 & 7.78 & 8.24 & 0.08 & -0.18 & 0.30 & 8.87 \\
13179 & 4.79 & 4.73 & 4.86 & 10.07 & 9.86 & 10.20 & 1.71 & 1.60 & 1.83 & 10.57 \\
13389 & 4.53 & 4.45 & 4.60 & 9.61 & 9.42 & 9.75 & 1.25 & 1.15 & 1.37 & 10.16 \\
14178 & 4.89 & 4.76 & 5.08 & 8.54 & 8.29 & 8.79 & 0.54 & 0.35 & 0.67 & 9.16 \\
15445 & 4.59 & 4.46 & 4.71 & 8.92 & 8.67 & 9.11 & 0.63 & 0.54 & 0.73 & 9.36 \\
18441 & 6.54 & 5.90 & 7.11 & 9.32 & 9.12 & 9.46 & 1.03 & 0.90 & 1.17 & 10.00 \\
18449 & 4.97 & 4.43 & 5.32 & 9.17 & 8.93 & 9.32 & 0.81 & 0.67 & 0.95 & 9.83 \\
19180 & \ldots & \ldots & \ldots & 9.72 & 9.48 & 9.90 & 1.54 & 1.38 & 1.68 & 10.14 \\
24007 & 5.36 & 5.17 & 5.52 & 8.76 & 8.48 & 8.91 & 0.41 & 0.32 & 0.53 & 9.59 \\
26890 & 8.80 & 8.61 & 8.97 & 8.79 & 8.46 & 9.04 & 0.84 & 0.50 & 0.97 & 9.45 \\
34813 & 4.38 & 4.27 & 4.50 & 8.26 & 7.97 & 8.62 & 0.33 & 0.02 & 0.46 & 8.97 \\
35896 & 5.13 & 4.82 & 5.40 & 8.92 & 8.67 & 9.08 & 0.63 & 0.51 & 0.76 & 9.44 \\
37653 & \ldots & \ldots & \ldots & 8.95 & 8.58 & 9.13 & 0.66 & 0.53 & 0.83 & 9.57 \\
37703 & 4.63 & 4.53 & 4.71 & 8.41 & 8.20 & 8.86 & 0.47 & 0.23 & 0.65 & 9.21 \\
39096 & 6.78 & 6.68 & 6.87 & 8.81 & 8.48 & 8.98 & 0.63 & 0.52 & 0.75 & 9.41 \\
41375 & 4.79 & 4.66 & 4.96 & 8.54 & 8.27 & 8.77 & 0.51 & 0.34 & 0.61 & 9.15 \\
41545 & 5.13 & 5.00 & 5.25 & 9.80 & 9.62 & 9.89 & 1.38 & 1.29 & 1.49 & 10.32 \\
41564 & 4.66 & 4.59 & 4.74 & 9.36 & 9.08 & 9.52 & 1.05 & 0.93 & 1.17 & 9.78 \\
42638 & 4.70 & 4.64 & 4.76 & 9.65 & 9.39 & 9.80 & 1.34 & 1.20 & 1.45 & 10.12 \\
45145 & 4.72 & 4.55 & 4.96 & 8.59 & 8.28 & 8.83 & 0.51 & 0.34 & 0.61 & 9.52 \\
49365~~~ & 5.31 & 5.14 & 5.52 & 9.63 & 9.43 & 9.75 & 1.29 & 1.18 & 1.42 &
10.23 
\enddata
%
%
\tablecomments{(1) Galaxy ID;  (2)--(4) redshift median (50\%-tile),
  and 16th and 84th-percentiles;  galaxies with no redshift use the
  spectroscopic redshift in Table~\ref{table:sample};  (5)--(7) stellar mass (50th
  percentile), and 16th and 84th--percentiles in the delayed-$\tau$
  model; (8)--(10) SFR median (50th percentile), and 16th and 84th
  percentiles (all SFRs are averaged over the past 100 Myr) in the
  delayed-$\tau$ model; (11) maximum stellar mass allowed in the
  ``burst'' formed at $z_f=100$, these satisfy the BIC criteria (equation~\ref{equation:bic} in Section~\ref{section:bursts}) and correspond approximately to a 3$\sigma$ upper limit.  }
\vspace{-6pt}
\end{deluxetable*}

\begin{deluxetable*}{lccc|ccc|ccc|c}
  \tabletypesize{\footnotesize}
\tablecolumns{11}
\tablewidth{6in}
\tablecaption{Derived Stellar masses, SFRs, and Redshifts excluding
  the MIRI data. } \label{table:bagpipes_nomiri}
\tablehead{
  \colhead{} & \multicolumn{3}{c}{redshift}  &
  \multicolumn{3}{c}{$\log M_\ast / M_\odot$ (dex)} &
  \multicolumn{3}{c}{$\log \mathrm{SFR} / M_\odot~\mathrm{yr}^{-1}$ (dex)} &
  \colhead{``Burst'' Mass} \\[-6pt]
  \colhead{ID} & \colhead{$z_{50}$} & \colhead{$z_{16}$} &
    \colhead{$z_{84}$} &
    \colhead{$\log M_{50}$}  & \colhead{$\log M_{16}$} &
    \colhead{$\log M_{84}$} &
    \colhead{log SFR$_{50}$} & \colhead{log SFR$_{16}$} & \colhead{log
      SFR$_{84}$} &
    \colhead{$\log M_\ast/M_\odot$ (dex)}\\[-6pt]
  \colhead{(1)} & \colhead{(2)} & \colhead{(3)} &
    \colhead{(4)} &
    \colhead{(5)}  & \colhead{(6)} &
    \colhead{(7)} &
    \colhead{(8)} & \colhead{(9)} & \colhead{(10)} &
    \colhead{(11)}
}
\startdata
5090 & 4.32 & 4.17 & 4.46 & 9.08 & 8.75 & 9.26 & 0.61 & 0.50 & 0.74 & 10.00 \\
6811 & -1.00 & -1.00 & -1.00 & 9.85 & 9.47 & 10.12 & 1.70 & 1.43 & 1.98 & 11.17 \\
7364 & 8.23 & 7.42 & 8.61 & 9.77 & 9.48 & 9.97 & 1.60 & 1.38 & 1.80 & 10.85 \\
7600 & 4.42 & 4.26 & 4.57 & 10.53 & 10.32 & 10.68 & 2.10 & 1.97 & 2.24 & 11.17 \\
7818 & 4.83 & 4.67 & 5.17 & 9.83 & 9.57 & 10.02 & 1.45 & 1.28 & 1.61 & 10.51 \\
11329 & 4.40 & 4.24 & 4.56 & 8.82 & 8.50 & 9.05 & 0.40 & 0.24 & 0.55 & 9.84 \\
12514 & 7.38 & 6.77 & 8.13 & 9.30 & 8.92 & 9.64 & 1.10 & 0.77 & 1.44 & 10.41 \\
12773 & 4.97 & 4.83 & 5.15 & 8.96 & 8.55 & 9.22 & 0.58 & 0.41 & 0.78 & 10.23 \\
13179 & 4.76 & 4.71 & 4.81 & 10.25 & 10.03 & 10.38 & 1.81 & 1.70 & 1.96 & 10.89 \\
13389 & 4.52 & 4.44 & 4.59 & 9.82 & 9.65 & 9.97 & 1.37 & 1.24 & 1.51 & 10.41 \\
14178 & 4.78 & 4.65 & 4.92 & 9.27 & 8.99 & 9.45 & 0.90 & 0.72 & 1.07 & 10.08 \\
15445 & 4.58 & 4.45 & 4.70 & 9.03 & 8.70 & 9.25 & 0.68 & 0.55 & 0.85 & 10.07 \\
18441 & 6.61 & 6.07 & 6.87 & 10.18 & 9.90 & 10.42 & 1.94 & 1.66 & 2.16 & 11.11 \\
18449 & 5.00 & 4.43 & 5.41 & 9.24 & 8.92 & 9.47 & 0.90 & 0.64 & 1.10 & 10.13 \\
19180 & -1.00 & -1.00 & -1.00 & 10.52 & 10.31 & 10.70 & 2.09 & 1.90 & 2.25 & 11.52 \\
24007 & 5.28 & 5.06 & 5.47 & 8.93 & 8.59 & 9.27 & 0.58 & 0.38 & 0.88 & 10.40 \\
26890 & 8.92 & 8.67 & 9.17 & 9.39 & 9.10 & 9.67 & 1.28 & 1.05 & 1.51 & 10.53 \\
34813 & 4.34 & 4.23 & 4.47 & 8.78 & 8.40 & 9.05 & 0.48 & 0.39 & 0.59 & 9.64 \\
35896 & 4.88 & 4.66 & 5.20 & 9.23 & 8.94 & 9.47 & 0.93 & 0.72 & 1.12 & 10.08 \\
37653 & -1.00 & -1.00 & -1.00 & 9.55 & 9.26 & 9.75 & 1.18 & 1.00 & 1.36 & 10.39 \\
37703 & 4.58 & 4.48 & 4.68 & 9.16 & 8.90 & 9.35 & 0.75 & 0.62 & 0.88 & 9.94 \\
39096 & 6.73 & 6.59 & 6.84 & 9.04 & 8.68 & 9.46 & 0.80 & 0.57 & 1.21 & 10.03 \\
41375 & 4.73 & 4.60 & 4.87 & 9.15 & 8.79 & 9.34 & 0.73 & 0.58 & 0.91 & 10.19 \\
41545 & 5.12 & 4.98 & 5.26 & 9.85 & 9.61 & 10.03 & 1.45 & 1.31 & 1.60 & 10.49 \\
41564 & 4.64 & 4.57 & 4.71 & 9.56 & 9.34 & 9.72 & 1.15 & 1.04 & 1.29 & 10.18 \\
42638 & 4.68 & 4.62 & 4.75 & 9.85 & 9.62 & 10.01 & 1.47 & 1.32 & 1.64 & 10.55 \\
45145 & 4.66 & 4.52 & 4.86 & 9.08 & 8.71 & 9.30 & 0.69 & 0.53 & 0.88 & 9.98 \\
49365~~~ & 5.43 & 5.23 & 5.63 & 9.45 & 9.14 & 9.65 & 1.09 & 0.89 & 1.29 & 10.31
\enddata
%
%
\tablecomments{(1) Galaxy ID, (2)--(4) redshift median (50\%-tile),
  and 16th and 84th-percentiles;  galaxies with no redshift use the
  spectroscopic redshift in Table~\ref{table:sample};  (5)--(7) stellar mass (50th
  percentile), and 16th and 84th--percentiles in the delayed-$\tau$
  model; (8)--(10) SFR median (50th percentile), and 16th and 84th
  percentiles (all SFRs are averaged over the past 100 Myr) in the
  delayed-$\tau$ model; (11) maximum stellar mass allowed in the
  ``burst'' formed at $z_f=100$, these satisfy the BIC criteria (equation~\ref{equation:bic} in Section~\ref{section:bursts}) and correspond approximately to a 3$\sigma$ upper limit.  }
\vspace{-6pt}
\end{deluxetable*}

\software{ AstroPy \citep{Astropy_2013}, \bagpipes\ \citep{Carnall_2018}, matplotlib \citep{Hunter_2007},
  NumPy \citep{vanderWalt_2011}, \textit{photutils} \citep{Bradley_2020}, \textit{PyPHER} \citep{Boucaud_2016b}, \sextractor\ \citep{Bertin_1996}, SciPy \citep{scipy20}, Seaborn \citep{Waskom_2021}.}


\appendix

\section{Impact of Crowded Sources in IRAC data}\label{section:appendix}.

One source of potential bias relates to the photometry of our sources
in the \spitzer/IRAC data.  As illustrated in the images
(Figure~\ref{fig:montage}) some objects have bright neighbors.  In the
case of the IRAC images, the light from the wings of these objects can
blend with that for our sources.  There is a large body of literature
on the subject of performing crowded source photometry
\citep{Laidler_2007,Labbe_2013,Merlin_2015,Merlin_2016}. We have used
the catalog from \added{\citet{Finkelstein_2022a}}\deleted{\citep{Finkelstein_2022a}} who used the \hst/F160W
image as a prior for the locations of sources and neighbors. Source
photometry is then carried out using \texttt{TPHOT}
\citep{Merlin_2015}, which estimates the source flux from objects
simultaneously when measuring photometry.    While this method is
theoretically robust, residuals from poorly modeling ePSFs and changes
in galaxy morphology with wavelength (the ``Morphological
$K$-correction'', \citealt{Papovich_2005}) can lead to systematic
uncertainties in source photometry. 

To test if our results are impacted by blended sources in the IRAC
bands, we did the following.  We first searched around each of the
galaxies in our sample and identified galaxies that \addedtwo{have} neighbors in the
MIRI 5.6~\micron\ catalog within a radius of $r \leq 3\arcsec$ and a
magnitude of $[5.6] \leq 26.7$~mag (near the flux limit).  We selected
neighbors in the MIRI 5.6~\micron\ image as the central wavelength is
closest to that of IRAC for our dataset (see
Figure~\ref{fig:filters}).   The IRAC ePSF has a FWHM of
$\approx$2\arcsec, so any source within 3\arcsec\ in the MIRI
data could therefore have IRAC light blended with our source of
interest.

\begin{figure*}
  \begin{center}
    \gridline{
      \includegraphics[width=0.45\textwidth]{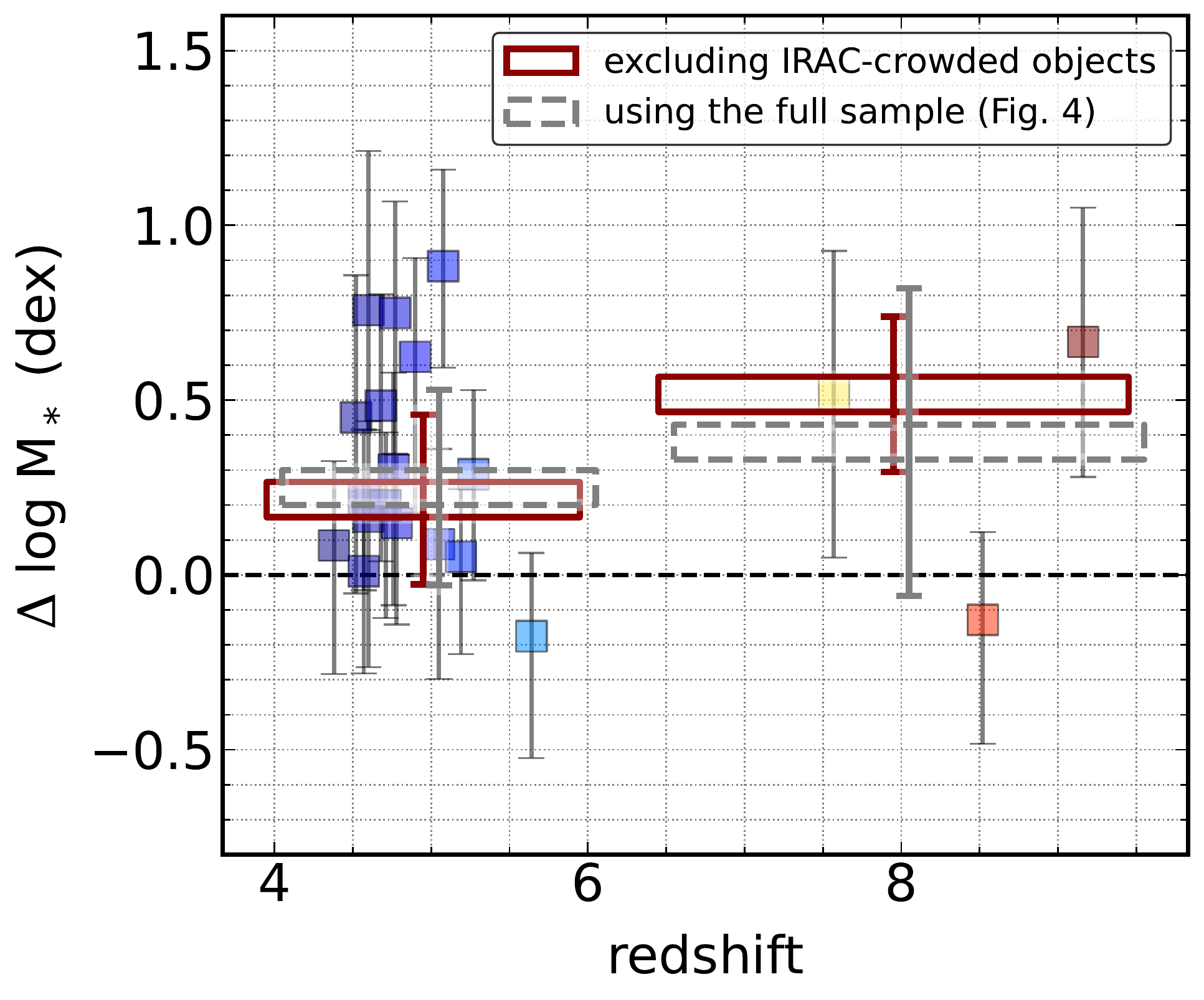}
      \includegraphics[width=0.45\textwidth]{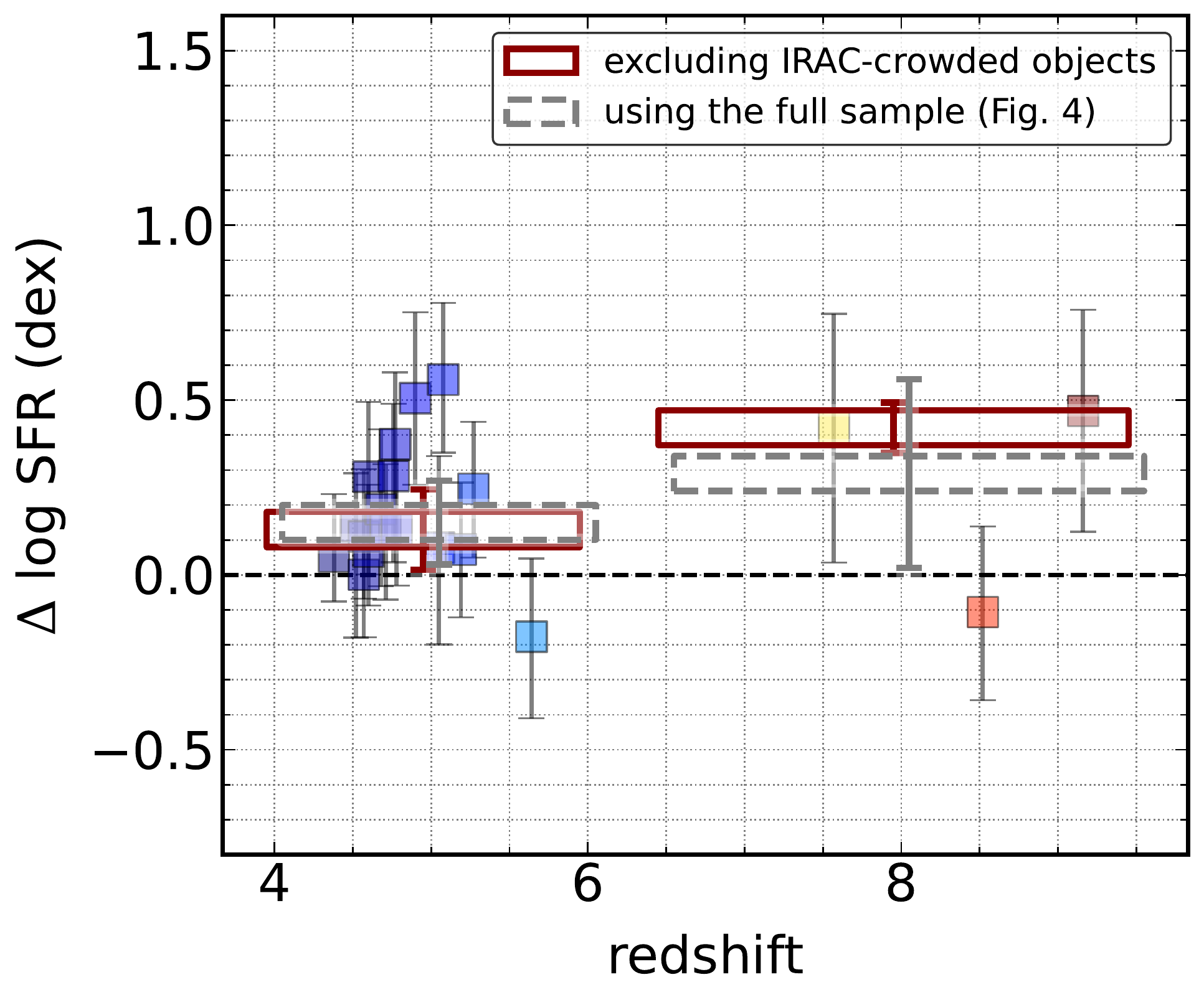}
      }
  \end{center}
\caption{ Testing the impact of sources with ``crowded'' IRAC
photometry.   The plots in this figure are similar to those in
Figure~\ref{fig:mass-sfr_miri_vs_nomiri}, and compare the
stellar masses and SFRs derived from the SED modeling for galaxies
including the MIRI F560W and F770W data and without the MIRI data.  In
both panels the results show the difference between the mass (SFR)
derived without MIRI data and the mass (SFR) deriving including the
MIRI data.   In this figure, we have excluded objects that have a
neighbor with $r \leq 3$\arcsec\ and MIRI [5.6] $\leq$26.7~mag.   This
eliminates 11 objects, and allows us to test if crowding in the IRAC
data (which has lower angular resolution) impacts object photometry in
the IRAC bands.   We do not observe any significant offset compared to
the results in Figure~\ref{fig:mass-sfr_miri_vs_nomiri}: the
median offsets in stellar mass and SFR change by $\approx$0.1~dex.
Therefore we conclude that blended IRAC photometry does not
significantly impact the results here.  \label{fig:mass-sfr_miri_vs_nomiri_unblended}}
\end{figure*}

From our sample, we identified 11 galaxies that have a neighbor within
3\arcsec\ in the
MIRI 5.6~\micron\ image.   To estimate their effect on our study we
removed these objects from the sample and recomputed the offsets in
stellar mass and SFR for the results that include the MIRI 5.6 and
7.7~\micron\ data versus the results that exclude the MIRI data.
These results are shown in
Figure~\ref{fig:mass-sfr_miri_vs_nomiri_unblended}.  Contrasting this
figure with the one for the full sample
(Figure~\ref{fig:mass-sfr_miri_vs_nomiri}) shows there is little
change in the median offsets in stellar mass and SFR.   The galaxy
sample used in this Appendix is obviously smaller, but the median
values do not change appreciably.   For the sample that excludes
blended objects, the offsets in stellar mass are $\Delta
\log M_\ast = 0.21$~dex for $4 < z < 6$ and 0.53 for $6 < z < 9$
(though the \addedtwo{latter} now includes only three galaxies).   The offsets in
SFR are $\Delta \log \mathrm{SFR} $ = 0.13 dex for $4 <
z < 6$ and 0.42 dex for $6 < z < 9$.  These are within $\approx$0.1
dex of the values reported for the full sample (in
Figure~\ref{fig:mass-sfr_miri_vs_nomiri}).

Similarly, we also investigated how the IRAC data for sources with
close neighbors impact our finding that the stellar populations of
the galaxies in our sample are generally ``bluer'' when the MIRI data
are included in the analysis (see
Section~\ref{section:rest-frame_colors} and
Figure~\ref{fig:rel_colors}).   We repeated our analysis of the
rest-frame colors in Section~\ref{section:rest-frame_colors} with our
sample of galaxies that excludes those objects with a neighboring MIRI
5.6~\micron\ source with $[5.6] < 26.7$~mag and within $r \leq
3\arcsec$.   We find that in this case the relative rest-frame colors
change only slightly. The rest-frame far-UV--$I$ color become
\textit{bluer}  by 0.015~mag (to have a total rest-frame (blue) color of $\Delta (m_{1600~\AA} - I) \approx 0.42$~mag)
when the objects with crowded IRAC photometry are excluded.

Therefore, we conclude that our results are not dominated by photometry from sources
crowded in the IRAC data.   Obviously, future studies using
\jwst/NIRCam will be valuable to testing the IRAC photometry (see,
e.g., \citealt{Bagley_2022}).

\bibstyle{aasjournal}
\bibliography{ceers_miri}{}


\end{document}

%% file: defs.tex
\def\lesssim{\mathrel{\hbox{\rlap{\hbox{%
 \lower4pt\hbox{$\sim$}}}\hbox{$<$}}}}
\def\gtrsim{\mathrel{\hbox{\rlap{\hbox{%
 \lower4pt\hbox{$\sim$}}}\hbox{$>$}}}}
\def\I.{\kern.2em{\sc i}} 
\def\II.{\kern.2em{\sc ii}} 
\def\III.{\kern.2em{\sc iii}} 
\def\IV.{\kern.2em{\sc iv}} 
\newcommand{\hst}{\textit{HST}}
\newcommand{\spitzer}{\textit{Spitzer}}

\newcommand{\mirifive}{\hbox{$[5.6]$}}
\newcommand{\miriseven}{\hbox{$[7.7]$}}

\newcommand{\lsim}{\lesssim}

\newcommand{\hb}{\hbox{H$\beta$}}
\newcommand{\ha}{\hbox{H$\alpha$}}

\newcommand{\msol}{\hbox{$M_\odot$}}

\newcommand\myRoman[1]{\@Roman{#1}\relax}%

\newcommand{\mone}{\hbox{$[3.6]$}}
\newcommand{\mtwo}{\hbox{$[4.5]$}}

\newcommand{\jwst}{\textit{JWST}}

\renewcommand{\mone}{\hbox{$[3.6]$}}
\renewcommand{\mtwo}{\hbox{$[4.5]$}}
\newcommand{\mthree}{\hbox{$[5.8]$}}
\newcommand{\mfour}{\hbox{$[8.0]$}}

\definecolor{aggiemaroon}{HTML}{500000}

\newcommand{\neiii}{\hbox{[Ne\,{\sc iii}]}}
\newcommand{\oii}{\hbox{[O\,{\sc ii}]}}
\newcommand{\oiii}{\hbox{[O\,{\sc iii}]}}
\newcommand{\nii}{\hbox{[N\,{\sc ii}]}}

\newcommand{\bagpipes}{\hbox{\texttt{BAGPIPES}}}
\newcommand{\sextractor}{\hbox{\texttt{SE}}}

\definecolor{aggiemaroon}{HTML}{500000}

%% file: figset_montage.tex
\figsetgrpnum{\thefigure.1}
\figsetgrptitle{ID 5090}
\figsetplot{montage_figures/slf_cutout_id5090.pdf}
\figsetgrpnote{Montage of images for galaxy ID 5090. All images are 6 arcsec by 6 arcsec.}
\figsetgrpnum{\thefigure.2}
\figsetgrptitle{ID 5742}
\figsetplot{montage_figures/slf_cutout_id5742.pdf}
\figsetgrpnote{Montage of images for galaxy ID 5742. All images are 6 arcsec by 6 arcsec.}
\figsetgrpnum{\thefigure.3}
\figsetgrptitle{ID 6811}
\figsetplot{montage_figures/slf_cutout_id6811.pdf}
\figsetgrpnote{Montage of images for galaxy ID 6811. All images are 6 arcsec by 6 arcsec.}
\figsetgrpnum{\thefigure.4}
\figsetgrptitle{ID 7364}
\figsetplot{montage_figures/slf_cutout_id7364.pdf}
\figsetgrpnote{Montage of images for galaxy ID 7364. All images are 6 arcsec by 6 arcsec.}
\figsetgrpnum{\thefigure.5}
\figsetgrptitle{ID 7600}
\figsetplot{montage_figures/slf_cutout_id7600.pdf}
\figsetgrpnote{Montage of images for galaxy ID 7600. All images are 6 arcsec by 6 arcsec.}
\figsetgrpnum{\thefigure.6}
\figsetgrptitle{ID 7818}
\figsetplot{montage_figures/slf_cutout_id7818.pdf}
\figsetgrpnote{Montage of images for galaxy ID 7818. All images are 6 arcsec by 6 arcsec.}
\figsetgrpnum{\thefigure.7}
\figsetgrptitle{ID 11329}
\figsetplot{montage_figures/slf_cutout_id11329.pdf}
\figsetgrpnote{Montage of images for galaxy ID 11329. All images are 6 arcsec by 6 arcsec.}
\figsetgrpnum{\thefigure.8}
\figsetgrptitle{ID 12514}
\figsetplot{montage_figures/slf_cutout_id12514.pdf}
\figsetgrpnote{Montage of images for galaxy ID 12514. All images are 6 arcsec by 6 arcsec.}
\figsetgrpnum{\thefigure.9}
\figsetgrptitle{ID 12773}
\figsetplot{montage_figures/slf_cutout_id12773.pdf}
\figsetgrpnote{Montage of images for galaxy ID 12773. All images are 6 arcsec by 6 arcsec.}
\figsetgrpnum{\thefigure.10}
\figsetgrptitle{ID 13179}
\figsetplot{montage_figures/slf_cutout_id13179.pdf}
\figsetgrpnote{Montage of images for galaxy ID 13179. All images are 6 arcsec by 6 arcsec.}
\figsetgrpnum{\thefigure.11}
\figsetgrptitle{ID 13389}
\figsetplot{montage_figures/slf_cutout_id13389.pdf}
\figsetgrpnote{Montage of images for galaxy ID 13389. All images are 6 arcsec by 6 arcsec.}
\figsetgrpnum{\thefigure.12}
\figsetgrptitle{ID 14178}
\figsetplot{montage_figures/slf_cutout_id14178.pdf}
\figsetgrpnote{Montage of images for galaxy ID 14178. All images are 6 arcsec by 6 arcsec.}
\figsetgrpnum{\thefigure.13}
\figsetgrptitle{ID 15445}
\figsetplot{montage_figures/slf_cutout_id15445.pdf}
\figsetgrpnote{Montage of images for galaxy ID 15445. All images are 6 arcsec by 6 arcsec.}
\figsetgrpnum{\thefigure.14}
\figsetgrptitle{ID 18441}
\figsetplot{montage_figures/slf_cutout_id18441.pdf}
\figsetgrpnote{Montage of images for galaxy ID 18441. All images are 6 arcsec by 6 arcsec.}
\figsetgrpnum{\thefigure.15}
\figsetgrptitle{ID 18449}
\figsetplot{montage_figures/slf_cutout_id18449.pdf}
\figsetgrpnote{Montage of images for galaxy ID 18449. All images are 6 arcsec by 6 arcsec.}
\figsetgrpnum{\thefigure.16}
\figsetgrptitle{ID 19180}
\figsetplot{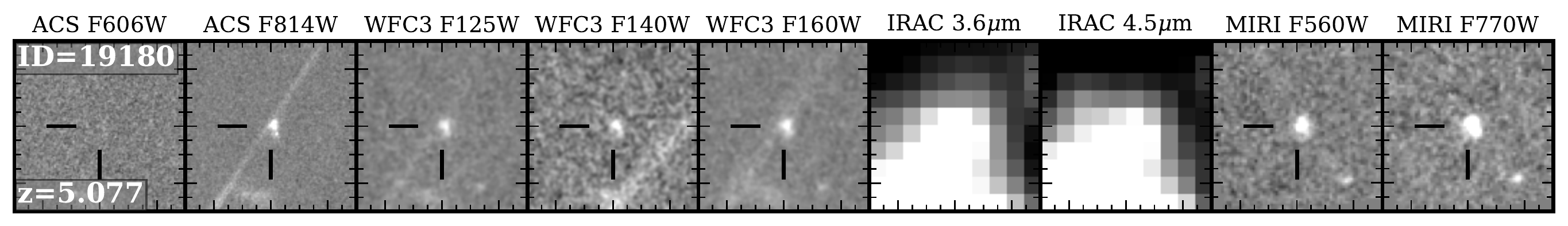}
\figsetgrpnote{Montage of images for galaxy ID 19180. All images are 6 arcsec by 6 arcsec.}
\figsetgrpnum{\thefigure.17}
\figsetgrptitle{ID 24007}
\figsetplot{montage_figures/slf_cutout_id24007.pdf}
\figsetgrpnote{Montage of images for galaxy ID 24007. All images are 6 arcsec by 6 arcsec.}
\figsetgrpnum{\thefigure.18}
\figsetgrptitle{ID 26890}
\figsetplot{montage_figures/slf_cutout_id26890.pdf}
\figsetgrpnote{Montage of images for galaxy ID 26890. All images are 6 arcsec by 6 arcsec.}
\figsetgrpnum{\thefigure.19}
\figsetgrptitle{ID 34813}
\figsetplot{montage_figures/slf_cutout_id34813.pdf}
\figsetgrpnote{Montage of images for galaxy ID 34813. All images are 6 arcsec by 6 arcsec.}
\figsetgrpnum{\thefigure.20}
\figsetgrptitle{ID 35896}
\figsetplot{montage_figures/slf_cutout_id35896.pdf}
\figsetgrpnote{Montage of images for galaxy ID 35896. All images are 6 arcsec by 6 arcsec.}
\figsetgrpnum{\thefigure.21}
\figsetgrptitle{ID 37653}
\figsetplot{montage_figures/slf_cutout_id37653.pdf}
\figsetgrpnote{Montage of images for galaxy ID 37653. All images are 6 arcsec by 6 arcsec.}
\figsetgrpnum{\thefigure.22}
\figsetgrptitle{ID 37703}
\figsetplot{montage_figures/slf_cutout_id37703.pdf}
\figsetgrpnote{Montage of images for galaxy ID 37703. All images are 6 arcsec by 6 arcsec.}
\figsetgrpnum{\thefigure.23}
\figsetgrptitle{ID 39096}
\figsetplot{montage_figures/slf_cutout_id39096.pdf}
\figsetgrpnote{Montage of images for galaxy ID 39096. All images are 6 arcsec by 6 arcsec.}
\figsetgrpnum{\thefigure.24}
\figsetgrptitle{ID 41375}
\figsetplot{montage_figures/slf_cutout_id41375.pdf}
\figsetgrpnote{Montage of images for galaxy ID 41375. All images are 6 arcsec by 6 arcsec.}
\figsetgrpnum{\thefigure.25}
\figsetgrptitle{ID 41545}
\figsetplot{montage_figures/slf_cutout_id41545.pdf}
\figsetgrpnote{Montage of images for galaxy ID 41545. All images are 6 arcsec by 6 arcsec.}
\figsetgrpnum{\thefigure.26}
\figsetgrptitle{ID 41564}
\figsetplot{montage_figures/slf_cutout_id41564.pdf}
\figsetgrpnote{Montage of images for galaxy ID 41564. All images are 6 arcsec by 6 arcsec.}
\figsetgrpnum{\thefigure.27}
\figsetgrptitle{ID 42638}
\figsetplot{montage_figures/slf_cutout_id42638.pdf}
\figsetgrpnote{Montage of images for galaxy ID 42638. All images are 6 arcsec by 6 arcsec.}
\figsetgrpnum{\thefigure.28}
\figsetgrptitle{ID 45145}
\figsetplot{montage_figures/slf_cutout_id45145.pdf}
\figsetgrpnote{Montage of images for galaxy ID 45145. All images are 6 arcsec by 6 arcsec.}
\figsetgrpnum{\thefigure.29}
\figsetgrptitle{ID 49365}
\figsetplot{montage_figures/slf_cutout_id49365.pdf}
\figsetgrpnote{Montage of images for galaxy ID 49365. All images are 6 arcsec by 6 arcsec.}

%% file: figset_spec_plots_burst.tex
\figsetgrpnum{\thefigure.1}
\figsetgrptitle{ID 5090}
\figsetplot{spec_plots_burst/spec_burst_both_5090.pdf}
\figsetgrpnote{SED fits including a burst at $z_f=100$ for galaxy ID 5090.}
\figsetgrpnum{\thefigure.2}
\figsetgrptitle{ID 6811}
\figsetplot{spec_plots_burst/spec_burst_both_6811.pdf}
\figsetgrpnote{SED fits including a burst at $z_f=100$ for galaxy ID 6811.}
\figsetgrpnum{\thefigure.3}
\figsetgrptitle{ID 7364}
\figsetplot{spec_plots_burst/spec_burst_both_7364.pdf}
\figsetgrpnote{SED fits including a burst at $z_f=100$ for galaxy ID 7364.}
\figsetgrpnum{\thefigure.4}
\figsetgrptitle{ID 7600}
\figsetplot{spec_plots_burst/spec_burst_both_7600.pdf}
\figsetgrpnote{SED fits including a burst at $z_f=100$ for galaxy ID 7600.}
\figsetgrpnum{\thefigure.5}
\figsetgrptitle{ID 7818}
\figsetplot{spec_plots_burst/spec_burst_both_7818.pdf}
\figsetgrpnote{SED fits including a burst at $z_f=100$ for galaxy ID 7818.}
\figsetgrpnum{\thefigure.6}
\figsetgrptitle{ID 11329}
\figsetplot{spec_plots_burst/spec_burst_both_11329.pdf}
\figsetgrpnote{SED fits including a burst at $z_f=100$ for galaxy ID 11329.}
\figsetgrpnum{\thefigure.7}
\figsetgrptitle{ID 12514}
\figsetplot{spec_plots_burst/spec_burst_both_12514.pdf}
\figsetgrpnote{SED fits including a burst at $z_f=100$ for galaxy ID 12514.}
\figsetgrpnum{\thefigure.8}
\figsetgrptitle{ID 12773}
\figsetplot{spec_plots_burst/spec_burst_both_12773.pdf}
\figsetgrpnote{SED fits including a burst at $z_f=100$ for galaxy ID 12773.}
\figsetgrpnum{\thefigure.9}
\figsetgrptitle{ID 13179}
\figsetplot{spec_plots_burst/spec_burst_both_13179.pdf}
\figsetgrpnote{SED fits including a burst at $z_f=100$ for galaxy ID 13179.}
\figsetgrpnum{\thefigure.10}
\figsetgrptitle{ID 13389}
\figsetplot{spec_plots_burst/spec_burst_both_13389.pdf}
\figsetgrpnote{SED fits including a burst at $z_f=100$ for galaxy ID 13389.}
\figsetgrpnum{\thefigure.11}
\figsetgrptitle{ID 14178}
\figsetplot{spec_plots_burst/spec_burst_both_14178.pdf}
\figsetgrpnote{SED fits including a burst at $z_f=100$ for galaxy ID 14178.}
\figsetgrpnum{\thefigure.12}
\figsetgrptitle{ID 15445}
\figsetplot{spec_plots_burst/spec_burst_both_15445.pdf}
\figsetgrpnote{SED fits including a burst at $z_f=100$ for galaxy ID 15445.}
\figsetgrpnum{\thefigure.13}
\figsetgrptitle{ID 18441}
\figsetplot{spec_plots_burst/spec_burst_both_18441.pdf}
\figsetgrpnote{SED fits including a burst at $z_f=100$ for galaxy ID 18441.}
\figsetgrpnum{\thefigure.14}
\figsetgrptitle{ID 18449}
\figsetplot{spec_plots_burst/spec_burst_both_18449.pdf}
\figsetgrpnote{SED fits including a burst at $z_f=100$ for galaxy ID 18449.}
\figsetgrpnum{\thefigure.15}
\figsetgrptitle{ID 19180}
\figsetplot{spec_plots_burst/spec_burst_both_19180.pdf}
\figsetgrpnote{SED fits including a burst at $z_f=100$ for galaxy ID 19180.}
\figsetgrpnum{\thefigure.16}
\figsetgrptitle{ID 24007}
\figsetplot{spec_plots_burst/spec_burst_both_24007.pdf}
\figsetgrpnote{SED fits including a burst at $z_f=100$ for galaxy ID 24007.}
\figsetgrpnum{\thefigure.17}
\figsetgrptitle{ID 26890}
\figsetplot{spec_plots_burst/spec_burst_both_26890.pdf}
\figsetgrpnote{SED fits including a burst at $z_f=100$ for galaxy ID 26890.}
\figsetgrpnum{\thefigure.18}
\figsetgrptitle{ID 34813}
\figsetplot{spec_plots_burst/spec_burst_both_34813.pdf}
\figsetgrpnote{SED fits including a burst at $z_f=100$ for galaxy ID 34813.}
\figsetgrpnum{\thefigure.19}
\figsetgrptitle{ID 35896}
\figsetplot{spec_plots_burst/spec_burst_both_35896.pdf}
\figsetgrpnote{SED fits including a burst at $z_f=100$ for galaxy ID 35896.}
\figsetgrpnum{\thefigure.20}
\figsetgrptitle{ID 37653}
\figsetplot{spec_plots_burst/spec_burst_both_37653.pdf}
\figsetgrpnote{SED fits including a burst at $z_f=100$ for galaxy ID 37653.}
\figsetgrpnum{\thefigure.21}
\figsetgrptitle{ID 37703}
\figsetplot{spec_plots_burst/spec_burst_both_37703.pdf}
\figsetgrpnote{SED fits including a burst at $z_f=100$ for galaxy ID 37703.}
\figsetgrpnum{\thefigure.22}
\figsetgrptitle{ID 39096}
\figsetplot{spec_plots_burst/spec_burst_both_39096.pdf}
\figsetgrpnote{SED fits including a burst at $z_f=100$ for galaxy ID 39096.}
\figsetgrpnum{\thefigure.23}
\figsetgrptitle{ID 41375}
\figsetplot{spec_plots_burst/spec_burst_both_41375.pdf}
\figsetgrpnote{SED fits including a burst at $z_f=100$ for galaxy ID 41375.}
\figsetgrpnum{\thefigure.24}
\figsetgrptitle{ID 41545}
\figsetplot{spec_plots_burst/spec_burst_both_41545.pdf}
\figsetgrpnote{SED fits including a burst at $z_f=100$ for galaxy ID 41545.}
\figsetgrpnum{\thefigure.25}
\figsetgrptitle{ID 41564}
\figsetplot{spec_plots_burst/spec_burst_both_41564.pdf}
\figsetgrpnote{SED fits including a burst at $z_f=100$ for galaxy ID 41564.}
\figsetgrpnum{\thefigure.26}
\figsetgrptitle{ID 42638}
\figsetplot{spec_plots_burst/spec_burst_both_42638.pdf}
\figsetgrpnote{SED fits including a burst at $z_f=100$ for galaxy ID 42638.}
\figsetgrpnum{\thefigure.27}
\figsetgrptitle{ID 45145}
\figsetplot{spec_plots_burst/spec_burst_both_45145.pdf}
\figsetgrpnote{SED fits including a burst at $z_f=100$ for galaxy ID 45145.}
\figsetgrpnum{\thefigure.28}
\figsetgrptitle{ID 49365}
\figsetplot{spec_plots_burst/spec_burst_both_49365.pdf}
\figsetgrpnote{SED fits including a burst at $z_f=100$ for galaxy ID 49365.}